\documentclass[fleqn,usenatbib]{mnras}

\usepackage{mathptmx}
\usepackage[T1]{fontenc}

\DeclareRobustCommand{\VAN}[3]{#2}
\let\VANthebibliography\thebibliography
\def\thebibliography{\DeclareRobustCommand{\VAN}[3]{##3}\VANthebibliography}
\usepackage{graphicx}	% Including figure files
\usepackage{amsmath}	% Advanced maths commands

\title{Radio AGN selection in LoTSS DR2}

\author[M.J. Hardcastle et al.]{M.J.\ Hardcastle,$^{1}$\thanks{E-mail: m.j.hardcastle@herts.ac.uk}
J.C.S.\ Pierce,$^{1}$, K.J.\ Duncan$^2$, G.\ G\"urkan$^{1,3}$, Y.\ Gong$^4$,
M.A.\ Horton$^5$, B. Mingo$^{1}$,\newauthor
H.J.A.
R\"ottgering$^6$, and D.J.B.\ Smith$^{1}$
\\
$^{1}$Department of Physics, Astronomy and Mathematics, University of
Hertfordshire, College Lane, Hatfield AL10 9AB, UK\\
$^2$ Institute for Astronomy, Royal Observatory, Blackford Hill,
Edinburgh, EH9 3HJ, UK\\
$^3$ Australia Telescope National Facility, CSIRO Space and Astronomy,
PO Box 1130, Bentley WA 6102, Australia\\
$^4$ School of Physical Sciences, The Open University, Walton Hall, Milton Keynes MK7 6AA, UK\\
$^5$ Astrophysics Group, Cavendish Laboratory, University of Cambridge, JJ Thomson Ave, Cambridge CB3 0HE\\
$^6$ Leiden Observatory, Leiden University, PO Box 9513, NL-2300 RA Leiden, the Netherlands\\
}

\date{Accepted 2025 April 7. Received 2025 April 1; in original form 2024 December 9}

\pubyear{2025}

\begin{document}
\label{firstpage}
\pagerange{\pageref{firstpage}--\pageref{lastpage}}
\maketitle

\begin{abstract}
The wide-area component of the LOFAR Two-Metre Sky Survey (LoTSS) is
currently the largest radio survey ever carried out, and a large
fraction of the 4.5 million radio sources it contains have been
optically identified with galaxies or quasars with spectroscopic or
photometric redshifts. Identification of radio-luminous AGN from this
LoTSS source catalogue is not only important from the point of view of
understanding the accretion history of the universe, but also enables
a wide range of other science. However, at present the vast majority of the
optical identifications lack spectroscopic information or well-sampled
spectral energy distributions. We show that
colour and absolute magnitude information from the {\it Wide-Field
  Infrared Survey Explorer (WISE)} allows for the robust and efficient
selection of radio AGN candidates, generating a radio AGN candidate sample
of around 600,000 objects with flux density $> 1.1$ mJy, spanning
  144-MHz luminosities between $10^{21}$ and $10^{29}$ W Hz$^{-1}$. We use the catalogue to constrain the total
sky density of radio-luminous AGN and the evolution of their
luminosity function between $z=0$ and $z\approx 1$, and show that the typical
mass of their host galaxies, around $10^{11} M_\odot$, is essentially independent of
radio luminosity above around $L_{144} \approx 10^{24}$ W Hz$^{-1}$.
Combining with Very Large Array Sky Survey (VLASS) data, we show that
the core prominences, radio spectral indices and variability of
extended sources from the sample are qualitatively consistent with the
expectations from unified models. A catalogue of the radio AGN
candidates is released with this paper.
\end{abstract}


%%%%%%%%%%%%%%%%%%%%%%%%%%%%%%%%%%%%%%%%%%%%%%%%%%

%%%%%%%%%%%%%%%%% BODY OF PAPER %%%%%%%%%%%%%%%%%%

\section{Introduction}
\subsection{Why select AGN?}
\label{sec:intro}

Although the central supermassive black holes of galaxies must grow by
accretion and merger so as to maintain the observed black hole
mass/galaxy mass relation \citep{Reines+Volonteri15}, direct evidence
for accretion comes only from the observation of active galactic
nucleus (AGN) activity at one or more wavelengths. However,
observational selection methods are essentially all biased in terms of
the population of AGN that they can find. For example, selection of
radiatively efficient AGN by their infrared (IR) or X-ray emission
produces largely disjoint sets of galaxies in many studies
\citep[e.g.][]{Hickox+09} despite the fact that we expect AGN to be
luminous in both bands. Some of the selection effects in these
populations are clearly expected in the context of unified models of
AGN \citep{Antonucci93}, in which anisotropic extinction (due to the
`torus') affects the optical and X-ray emission of AGN from certain
lines of sight. Other selection effects remain poorly understood.

One key selection effect in most wavebands is that the requirement for
luminous radiation {\it directly} produced by the accretion onto the black
hole selects, both theoretically \citep{Narayan+Yi95-2} and
observationally, for objects accreting at a rate above approximately 1
per cent of their Eddington rate, defined as
\[
\dot M_{\mathrm{Edd}} = \frac{4\pi G M m_p }{\eta c \sigma_T}
\]
where $G$ is the gravitational constant, $M$ is the black hole mass,
$m_p$ the mass of a proton, $\eta$ a radiative efficiency factor, $c$
the speed of light and $\sigma_T$ the Thomson cross-section. It
follows that most AGN selection methods cannot even in principle
select for accretion that is taking place at lower mass accretion
rates than $\dot M \approx 10^{-2} \dot M_{\mathrm{Edd}}$. This
imposes biases against detecting the accretion onto the most slowly
growing or most massive black holes.

A crucial realization of the past couple of decades is that radio
selection, while of course still biased, is not limited by an
accretion rate bias in the same way. Radio emission from active
galaxies is produced (dominantly in the high-luminosity
  population, among other possible mechanisms that we will discuss in
more detail later) by synchrotron emission from particles accelerated
as a result of the evolution of relativistic jets generated close to
the central supermassive black hole. It has long been known
\citep{Hine+Longair79} that only a fraction of radio-luminous AGN
(hereafter radio AGN or RLAGN) show strong optical emission
lines of the type associated with quasars or Seyfert 1 and 2 galaxies
which participate in standard unified models. A consensus has now
emerged
\citep{Hardcastle+07-2,Antonucci11,Best+Heckman12,Hardcastle+Croston20}
that some RLAGN have all the apparatus of optical, IR and X-ray
emission directly from the accretion flow, and an obscuring torus,
that we see with other AGN selection methods: these are the
radiatively efficient \citep{Hardcastle18b} or thermal
\citep{Antonucci11} RLAGN, and are observationally described as
radio-loud quasars or high-excitation radio galaxies (HERGs). Others,
which are generally the majority in samples selected without a strong
bias to high radio luminosities, do not have any of the standard AGN
apparatus but do still possess powerful jets: these are radiatively
inefficient \citep{Hardcastle18b} or non-thermal RLAGN
\citep{Antonucci11} and are observationally described as
low-excitation radio galaxies (LERGs). Crucially, the Eddington ratios
estimated for the different classes do appear to show a division at
the expected level around 1 per cent Eddington
\citep{Best+Heckman12,Mingo+14,Drake+24}, although it is difficult to
estimate these Eddington ratios accurately in the absence in many
cases of accurate black hole mass and jet power estimates. Selection
of RLAGN thus offers a view of accretion onto black holes at low
Eddington rates that cannot be achieved in any other way, although at
the cost of losing the ability to infer easily what those accretion
rates actually are, since jet power does not necessarily relate
closely to accretion rate \citep{Mingo+14,Hardcastle+Croston20}, is
not easy to infer accurately from radio data alone
\citep{Hardcastle+19} and is almost certainly variable over the
lifetime of a typical extended radio source.

If we choose to proceed in the face of these difficulties, it then
becomes vital to understand how to select RLAGN from large
extragalactic radio surveys. Such surveys are numerically dominated,
at the faint end, by star-forming galaxies whose radio emission is
largely due to cosmic rays accelerated in the intergalactic medium by
supernovae related to recent star formation. In principle the gold
standard for selection of {\it radiatively efficient} AGN is spectral
energy distribution (SED) fitting, which allows the fraction of
radiation due to an AGN to be estimated from broad-band photometry
\citep{CalistroRivera+16,Boquien+19,Best+23,Das+24} and therefore is
not in principle biased against, for example, dust-obscured AGN. It
could be argued that these codes will still not select faint AGN in
bright galaxies, but in fact the lower Eddington limit for radiatively
efficient AGN and the black hole mass/galaxy mass relation require a
minimum bolometric luminosity for an AGN in a given
galaxy\footnote{Very roughly, this fraction is simply given by the
normalization of the black hole mass/galaxy mass relation together
with the specific bolometric luminosity of a galaxy due directly or
indirectly to stars, which is of order the solar luminosity. For the
values quoted by \cite{Reines+Volonteri15}, with a black hole mass of
0.025 per cent of the galaxy mass, the ratio between the AGN and
galaxy bolometric luminosity can thus range between about 0.1 and 10
for Eddington ratios between 0.01 and 1. Fainter radiative AGN
relative to their host galaxies can only exist in this model if their
black hole masses fall substantially below the typical value measured
by \citeauthor{Reines+Volonteri15}.} and so at least in principle with
good enough broad-band photometry all radiatively efficient AGN can be
selected in this way. Of course, in practice, and particularly for
wide-area surveys, the high-quality broad-band photometry needed for
SED fitting may not exist, and then it is necessary to resort to
proxies of AGN activity such as optical spectroscopy, X-ray emission
or radio or optical colours, which may not be reliable in all cases.

In some senses the situation is far worse in the radio. Many, probably
most, radio AGN are not radiatively efficient, and orecisely
because the accretion rate is not limited, here the broad-band or
monochromatic radio luminosity can take almost any value, limited only
at the top end by the requirement that the accretion powering the jets
should not be super-Eddington \citep{Mingo+14}. Arbitrarily low
accretion rates and thus jet powers appear to be able to exist. In
addition, the radio emission from jets is not necessarily spectrally
distinguishable from that due to star formation, and even when it is,
good-quality broad-band radio spectral information is rarely
available. The current most widely used method for selection of radio
AGN is the `radio excess' method, where the star-formation rate of the
host galaxy is estimated, directly or indirectly via some
well-calibrated proxy, to allow an inference of the expected radio
luminosity due to star formation; then the radio emission from an
object that is significantly more luminous than this can be assigned
to AGN activity
\citep[e.g.][]{Yun+01,Kauffmann+08,Hardcastle+16,Gurkan+18,Drake+24}.
This, however, clearly excludes AGN activity that produces emission
that is similarly luminous to or less luminous than what is generated
by star formation. These non-excess radio AGN can in general only be
distinguished from star formation spatially, or on a statistical basis
within a given population \citep[e.g.][]{Yue+24}, and in many cases
spatial decomposition would require observations of much higher
resolution than are generally available in surveys
\citep[although see][]{Morabito+21,Morabito+25}. Moreover, the radio-excess method
requires a good star-formation rate estimator from SED fitting or
spectroscopy and, as with radiatively efficient AGN, these are not
always available from wide-area surveys \cite[see][]{Smith+16}.

As we have argued above, though, a census of accretion in the Universe
is severely incomplete without including RLAGN, despite the fact that
we currently cannot even in principle find all of them. RLAGN selection
is not only important in itself, but permits a large amount of other
studies of, among other things, AGN feedback and life cycles. The current
best wide and deep radio surveys are those generated by the LOFAR
Two-metre Sky Survey (LoTSS: \citealt{Shimwell+17}), which has
released observations both of individual deep fields and of wide areas
of the northern sky. This paper focuses on RLAGN selection in LoTSS.

\subsection{RLAGN selection in the LoTSS wide-area survey}

The current most sensitive LoTSS radio images are the first data
release of the deep fields, covering the ELAIS-N1, Lockman Hole and
Bo\"otes areas, \citep{Sabater+21,Tasse+21} and the second data
release, DR2, of the wide-area survey \citep{Shimwell+22}. The deep
fields benefit from excellent multi-wavelength data and almost
complete (95 per cent) optical identification \citep{Kondapally+21} which has
permitted SED fitting methods to be used to identify radio-excess RLAGN
\citep{Best+23,Das+24}. The wide-area DR2 contains many more radio sources (over 4.4
million) but the optical data for this survey,
where the {\it WISE} surveys and the DESI Legacy Survey are the
deepest available in the mid to near-IR and optical bands
respectively, are much shallower. Only 85 per cent of sources have any
kind of optical identification, with the redshift fraction being even
smaller \citep{Hardcastle+23}, while the far-IR data needed for
high-quality SED fitting are completely lacking over most of the DR2
area. The key advantage of the wide-area survey is that it has much
better statistics, particularly for the rare luminous radio AGN, and
so gives us a much more accurate overall view of the radio AGN
population, at least in the local universe in which host galaxies and
optical identifications can be found given our ancillary data. The
wide-area LOFAR survey also benefits from synergies with other
wide-area surveys, notably including the Very Large Array Sky Survey,
VLASS \citep{Lacy+20}, which is less sensitive to steep-spectrum
emission or extended sources but provides high sensitivity to
flat-spectrum cores and higher angular resolution over all of the
LoTSS coverage. The largest wide-area catalogue of RLAGN in general to
date was provided by \cite{Hardcastle+19}, based on the much smaller
LoTSS DR1 release which contained only $\sim 325,000$ sources in
total.

In the present paper we aim to carry out radio AGN selection for the
DR2 optical catalogue based on the data currently available. Our
optical identifications are taken from \cite{Hardcastle+23}, hereafter
H23, who obtained optical/near-IR (hereafter simply `optical')
counterparts by a mixture of likelihood-ratio crossmatching for
compact sources, more sophisticated algorithms for extended sources,
and visual inspection. A minority of spectroscopic redshifts come from
SDSS, DESI and HETDEX, with the bulk of the redshifts being the
photometric ones derived by \cite{Duncan22}. \cite{Drake+24},
hereafter D24, have recently cross-matched the optical identifications
to sources from the SDSS spectroscopic sample, which allows
spectroscopic classifications of that subset of our objects to be
made. We begin by defining a radio sample and developing methods to
separate RLAGN and SFG in our dataset in the absence of complete
spectroscopic data, using the D24 classifications as a starting point
(Section \ref{sec:selection}). We then go on to check the consistency
of the sources in this catalogue with expectations and with previous
work in Section \ref{sec:properties}, before demonstrating its
application to a number of science questions in Section
\ref{sec:discussion}. A summary and prospects for future work are
given in Section \ref{sec:summary}.

Throughout the paper we use a concordance cosmology in which $H_0 =
70$ km s$^{-1}$ Mpc$^{-1}$, $\Omega_M = 0.3$ and $\Omega_\Lambda =
0.7$. Spectral index $\alpha$ is defined in the sense $S \propto
\nu^{-\alpha}$ and for calculation of luminosities we take $\alpha =
0.7$ unless otherwise stated.

\section{Sample definition and selection}
\label{sec:selection}

The process for RLAGN selection in LoTSS DR2 has steps in common with
the DR1 AGN catalogue created by \cite{Hardcastle+19}, hereafter H19,
but is based on the DR2 optical ID catalogue from H23. A step-by-step description is
given in this section of the paper.

\subsection{Completeness and ID cuts}

Initially we selected all sources with total flux density, after
  any source association, greater than 1.1
mJy and with an optical identification and redshift (corresponding to
the FCOZ sub-sample in H19). The 1.1 mJy flux limit was taken from the
95 per cent completeness level given by \cite{Shimwell+22} and should
ensure that we miss almost no point sources at this flux level that
have optical IDs provided by H23. By chance, this total flux cut is
almost exactly $10^4$ times lower than that used by the widely used 3CRR
survey \citep{Laing+83}. It is important to note, of course, that there
is also a strong surface brightness selection for this dataset and
that the completeness threshold only applies to point sources: we may
well be missing significant numbers of extended sources that should
have been selected above our flux limit. We return to a discussion of
this below, Section \ref{sec:pdd}.

The DR2 catalogue of H23 lists 1,776,977 sources with {\tt Total\_flux}
greater than or equal to 1.1 mJy, of which 966,323 (54.4 per cent)
have an optical ID and a redshift estimate either from spectroscopy or
from a good photometric redshift\footnote{As described by H23,
  photometric redshifts are said to be good if the {\tt flag\_qual}
  column of \cite{Duncan22} is set to 1, i.e. the photometry is free
  from blending and artefacts, the object is not star-like in optical images, and $\sigma_z / (1+z) < 0.2$.} (i.e. the {\tt z\_best} column is not
blank). We further impose a redshift cut $z>0.01$, which ensures that
host galaxies are not completely resolved by the {\it WISE} photometry
that we will use in subsequent analysis, and exclude objects that do
not have any {\it WISE} information (including upper limits) in the
{\it WISE} 1, 2 or 3 bands. {\it WISE} magnitudes in band 3 in our
catalogue come from Neo{\it WISE} by way of the DESI Legacy Survey
catalogue \citep{Duncan22} and so both this and the requirement of a
photometric redshift means that we require a Legacy survey detection
for a source to be selected. Hereafter this flux-complete catalogue of
963,764 radio sources with redshift information, corresponding to the
`FCOZ' sample of H19, is referred to as the `parent catalogue' or
`parent sample'. We take the sky area covered by this the catalogue to be 5,200
deg$^2$, based on the area of the full-band Legacy Survey coverage
which is needed for photometric redshifts.

% selection.py

\subsection{Merger with emission-line catalogue}

The parent catalogue is then merged with the emission-line catalogue
from D24, with duplicate columns stripped
out. D24's catalogue is based on the H23 catalogue and so can be
matched exactly by LoTSS source identifier. This gives probabilistic
emission-line classifications for 154,044 sources, of which 110,638
are in the parent catalogue; the remainder are below the flux density
or redshift limits that we use here.

D24 provide probabilistic classifications for all objects. In
particular, they give an estimate of the probability that an object is
a star-forming galaxy or a radio-excess AGN based on its position in a
Baldwin-Philips-Terlevich (BPT) diagram \citep{Baldwin+81}, and its
measured H$\alpha$ and radio luminosities. In what follows we use a
0.95 threshold on this probability calculated by D24 to select
high-confidence radio-excess and star-forming galaxies (hereafter RXG
and SFG) to help us to define selection regions for these classes
  of objects where spectroscopic data are lacking. The intersection
of the D24 catalogue with the parent catalogue lists 28,065 RXG and
27,127 SFG at this confidence level.

% class_cm.py

\subsection{Merger with the SDSS DR16 quasar catalogue}

We finally merge the parent catalogue with the SDSS DR16
(`quasar-only') quasar catalogue \citep{Lyke+20} with a simple optical
positional crossmatch with offset $<1$ arcsec. Objects that have a
match in the DR16 catalogue are very likely to be quasars: however,
the converse is not true since not all quasars will have been observed
spectroscopically, and in addition there is a part of the DR2 sky area
that is not covered by SDSS. Moreover, the eBOSS survey, which adds
significantly to the total number of high-$z$ quasars, does not
uniformly cover the SDSS footprint \citep{Dawson+16}. There are 26,567 matches to DR16 quasars
in the table.

% absmag.py

\subsection{The {\it WISE} colour-colour plot}

\begin{figure}
\includegraphics[width=\linewidth]{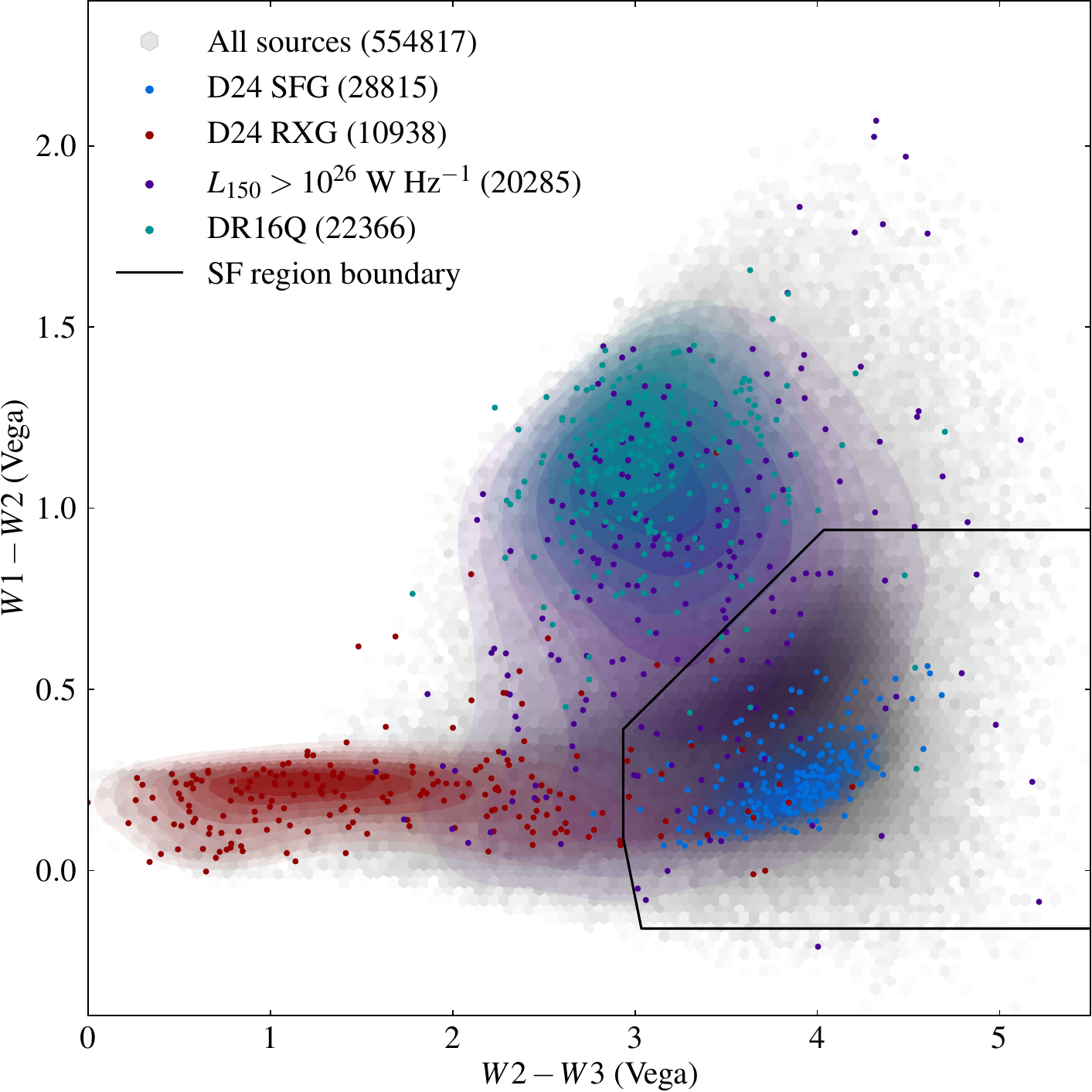}  
\caption{The observational {\it WISE} colour-colour diagram for the
  DR2 sample. Overlaid on the grey density plot showing the full
  parent sample with detections in all three {\it WISE} bands are the
  locations of D24 RXGs, D24 SFGs, DR16 quasars and luminous radio
  sources. To allow visualization of the areas occupied by each
  population, these are plotted as contours from the KDE estimate of
  the source density for the different classes, with colour contours
  representing intervals in source density on a square root scale.
  Overplotted are a small number of representative points from each
  subsample to give a sense of the scatter: the total number of points
  in the subsamples is indicated in the legend. Lines indicate the
  locus populated by SFG and avoided by many RLAGN discussed in the
  text.}
% wisecc_class_density.py
\label{fig:wisecc}
\end{figure}

The selections from these two catalogues can then be overplotted on
the {\it WISE} colour-colour plot from H23 (Fig.\ \ref{fig:wisecc}).
This shows clearly that the D24 classifications correspond, like those
of \cite{Sabater+19}, to largely distinct regions in colour-colour
space, as expected (although the D24 analysis made no use of the {\it
  WISE} colours). The approximate (hand-drawn) locus that could be
used to exclude objects with SFG colours in the manner described by
H19 is shown as a solid line on this plot. Also shown are the luminous
($L_{144} > 10^{26}$ W Hz$^{-1}$) sources, which are likely to be AGN
in all circumstances, since this value exceeds any plausible level of
emission from star formation: as in H19, luminous sources occupy largely
distinct locations in colour space, with many lying in the region
where we expect to find quasars, and the addition of the DR16 quasars
confirms this.

\begin{figure*}
\includegraphics[width=0.48\linewidth]{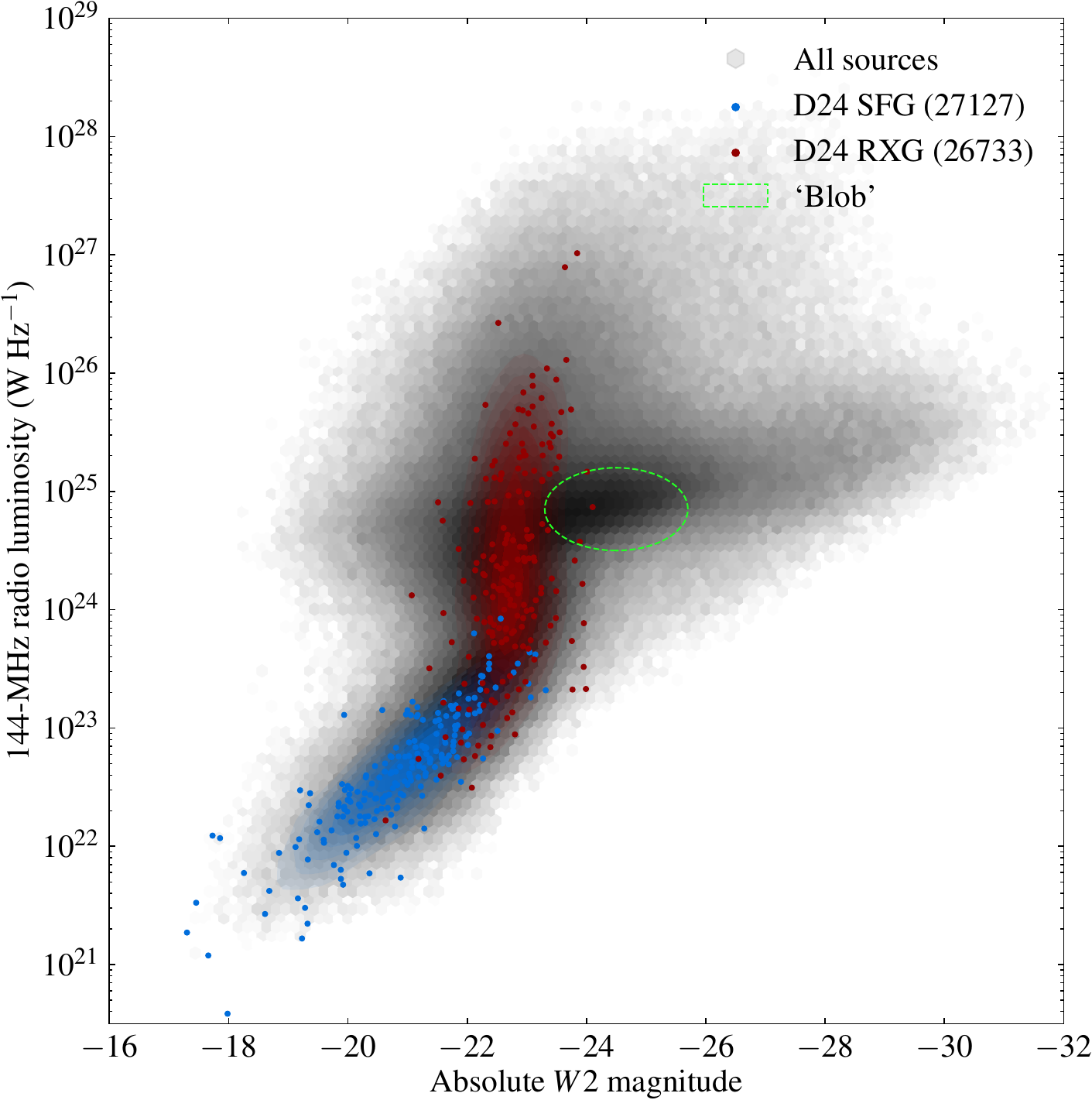}  
\includegraphics[width=0.48\linewidth]{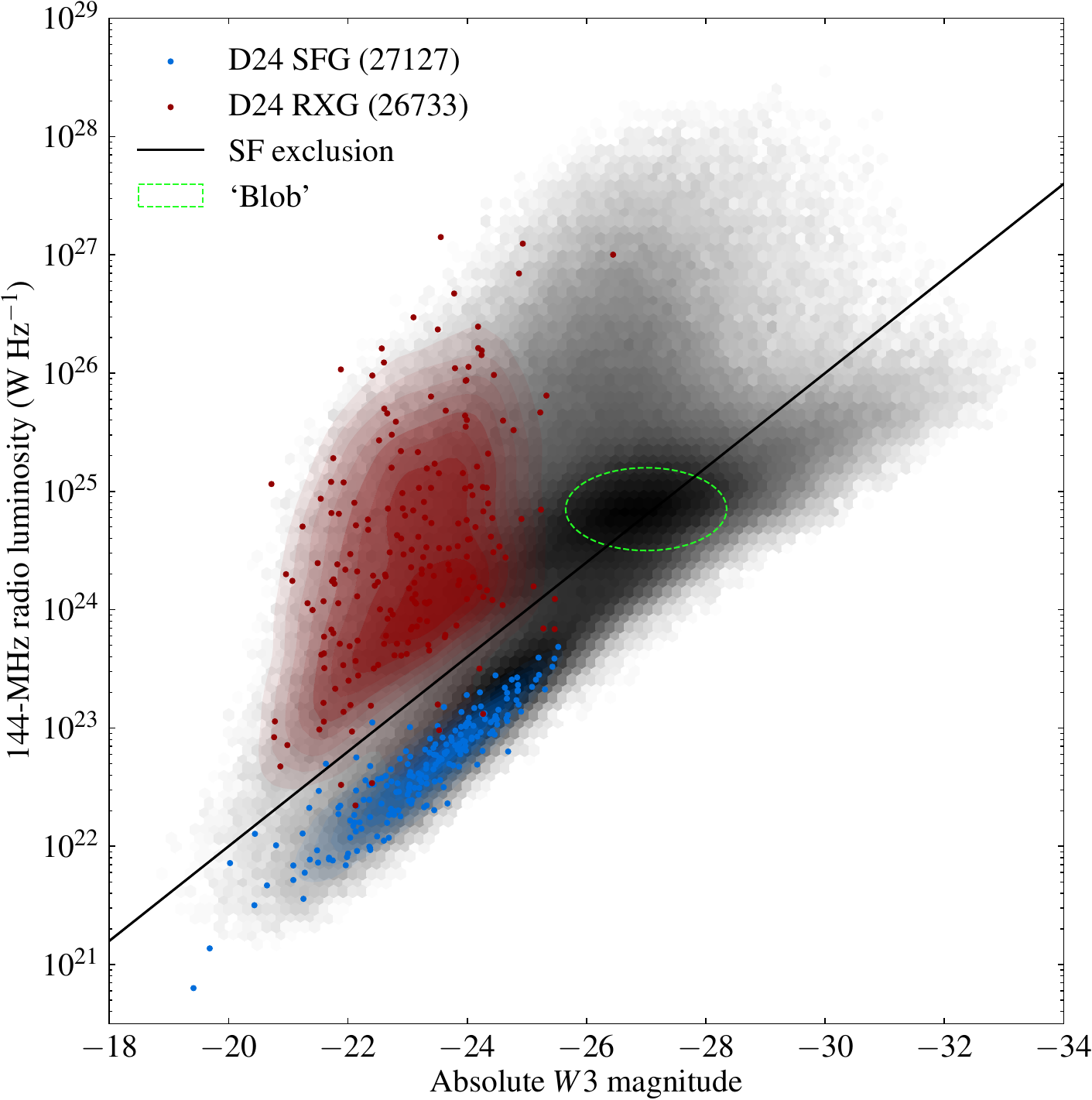}
\caption{The absolute magnitude/radio luminosity relation for the
  parent sample and the two emission-line classified samples from D24.
  Left panel shows $W2$ magnitude and right panel $W3$ magnitude. Note
  that in the right-hand panel a significant number of upper limits (407,657/963,754)
  are plotted as though they were detections. The solid line in the
  right-hand panel ($\log_{10}(L_{144}/\mathrm{W Hz}^{-1}) = 14-M_{W3}/2.5$) indicates a possible dividing line below which
  objects can be rejected as being SFG, while the dashed ellipse shows
  the location of the `blob' of objects at high {\it WISE}
  luminosities and $L_{144} \approx 10^{25}$ W Hz$^{-1}$. Colour
  contours are as in Fig.\ \ref{fig:wisecc}.}
\label{fig:absmag}
% absmag_lum_w2_density.py
% absmag_lum_w3_density.py
\end{figure*}

An important point here is that we exclude from the plot objects that
are not detected in any of the {\it WISE} bands. In practice this
mostly excludes objects without a detection in {\it WISE} band 3 (i.e.
the error {\tt magerr\_w3} in the catalogue is not listed). Only
555,200 objects in the parent catalogue have a {\it WISE} 3 detection.
For the others, the {\it WISE} magnitude is a lower limit. This has
the useful feature that selection in colour space would be
conservative -- if we excluded from our RLAGN sample objects that lay
in the SFG colour locus, we might also exclude some true RLAGN which
should actually lie to the left of their position in the plot, but we
would not expect to be contaminated by objects with SFG colours. The
problem is that this cut would be likely to exclude a number of true
RLAGN with colours in the SFG locus as well \citep{Gurkan+14}, and
this is clearly seen in the figure, where a number of objects with
$L_{144} > 10^{26}$ W Hz$^{-1}$ would lie in the exclusion region.
Although this was the approach of H19, we prefer to take a different
approach here to avoid excluding radio-excess AGN with colours in the
star-forming locus. Accordingly, colour selection for SFG/AGN
selection is not used in the rest of the paper.

\subsection{The mid-IR/radio relation}
\label{sec:SFcut}

Another way of visualizing the relationship between radio power and
galaxy properties is to look at the relationship between radio
luminosity and IR luminosity or absolute magnitude, as noted by
e.g. \cite{Mingo+16} and H23. {\it WISE}
magnitudes are particularly interesting here as they trace different
combinations of SF-heated dust and old stellar population. We form the
{\it WISE} absolute magnitude here by converting {\it WISE} magnitudes
to flux density\footnote{See
\url{https://wise2.ipac.caltech.edu/docs/release/allsky/expsup/sec4_4h.html}.},
computing a power-law spectral index between the flux density values,
and then $K$-correcting the magnitudes based on this power-law
extrapolation (e.g. for the {\it WISE} band 2 magnitude the spectral
index between bands 1 and 2 is used). This naive approach has the
advantage that it can be used without relying on templates, as the
tabulated absolute magnitude values in the catalogue do; however it
clearly breaks down even in principle for $z>0.35$ for bands 1 and 2
and $z>1.6$ for bands 2 and 3, after which we would be extrapolating
rather than interpolating between the two values. In addition only
about half the sample has a secure detection in band 3.

Nevertheless it is instructive to look at the positions of the D24 SFG
and RXG samples on this plot (Fig.\ \ref{fig:absmag}) as they show
strikingly different behaviour in the two {\it WISE} bands. In $W2$,
we see that the SFGs trace a linear correlation which merges
seamlessly into a vertical stripe for the RXGs. This is because $W2$
is simply a proxy of mass for these objects (which can be verified by
plotting it against the stellar mass estimates provided by H23) and we
are seeing the well-known effect that RLAGN appear in large numbers
above some stellar mass threshold, corresponding to around
$\log_{10}(M_*/M_\odot) \approx 10.7$. However, in the $W3$ plot there
is an almost complete separation between the D24 RXG and SFG. Here it
is important to bear in mind that the fact that the RXGs are mostly
non-detections in $W3$ would only increase the real separation. The
$W3$ band (observer-frame broad-band 7-17 $\mu$m) is dominated by hot
dust from star formation and so the strong correlation between $W3$
luminosity and H$\alpha$ emission means that the D24 RXG and SFG,
which are by construction well separated on a plot of $L(H\alpha)$ vs
$L_{144}$, are well separated here as well. Moreover, the
main-sequence line defined by the SFGs clearly continues beyond the
low-luminosity SFGs from D24 (which run out around $10^{24}$ W
Hz$^{-1}$ or star formation rates $\sim 100 M_\odot$ yr$^{-1}$). We
conclude that we can conservatively separate RXGs and SFGs in the
$W3/L_{144}$ plot with a simple linear dividing line which, unlike the
approach of H19, does not exclude objects that have SFG colours but a
strong radio excess over what is expected from star formation. This
approach mirrors that of \cite{Mingo+16} but we have the advantage of
being able to calibrate the line independently using the D24
spectroscopic classifications. Our chosen line could exclude true RXG
that have weak radio emission but strong radiative AGN-related (torus) emission
in the {\it WISE} band, but if such objects exist, they are not
present in any significant numbers in the D24 sample.

\subsection{Quasars}
\label{sec:quasars}

The appearance of Fig.\ \ref{fig:absmag} is complicated by the
existence of a quasar population which gives rise to $W2$ absolute
magnitudes brighter than $-24$ and also to bright $W3$ magnitudes,
though in this case not necessarily brighter than the maximal $W3$
magnitude for a starburst galaxy, which might be $-29$ or so for a
radio luminosity of $10^{25}$ W Hz$^{-1}$. This population will
consist of both traditional type 1 quasars and also radio-loud type 2
quasars, or narrow-line radio galaxies (NLRG), since both can have
excess emission over the expectation from starlight at $W2$ and $W3$
(see e.g. \citealt{Hickox+17}). $W3$ excess is likely to be more
reliable as an indicator of RE AGN activity in general if SFG are
excluded, since $W2$ is more affected by obscuration. By construction,
all these objects are radio-detected, and they have radio luminosities
significantly larger than the typical SFG, but they would be
traditionally described either as `radio-quiet' (RQQ) or `radio-loud'
(RLQ) based on the ratio of their radio to optical or IR luminosity.

In H19 we excluded on a similar plot (based on the $Ks$ band at 2.2
$\mu$m) a branch of what we took to be radio-quiet quasars, which can
be seen in both panels of Fig.\ \ref{fig:absmag} extending to the right at
$L_{144}$ values of a few $\times 10^{25}$ W Hz$^{-1}$. The naive
star-formation exclusion line in Fig.\ \ref{fig:absmag}, right panel,
goes straight through the bulk of this population, and in particular
bisects a `blob' of objects with $L_{144} \approx 10^{25}$ W Hz$^{-1}$
and $-25 \la W3 \la -27$. A similar group of objects is at $-24 \la W2
\la -26$.

The nature of the radio emission from RQQ is not well understood but
it remains possible that some or all of it is from strong star
formation \citep{Gurkan+19}: in that case one approach would just be
to apply the naive line to the quasar population as well. However,
that neglects the fact that the quasar must contribute to the emission
in W3, and it now seems plausible that, although some quasars may
  be radio-silent \citep{Radcliffe+21}, the radio emission in
radio-detected quasars is at least partly AGN-related
\citep{CalistroRivera+24,Yue+24,Njeri+25}. If we wished to reproduce the H19
analysis, we could cut out the `quasar branch' seen in the plots more
completely by modifying the selection line to exclude the branch and
the `blob' at its base at the cost of introducing significant (and
physically unmotivated) structure into the luminosity distribution of
the sample.

\begin{figure*}
\includegraphics[width=0.48\linewidth]{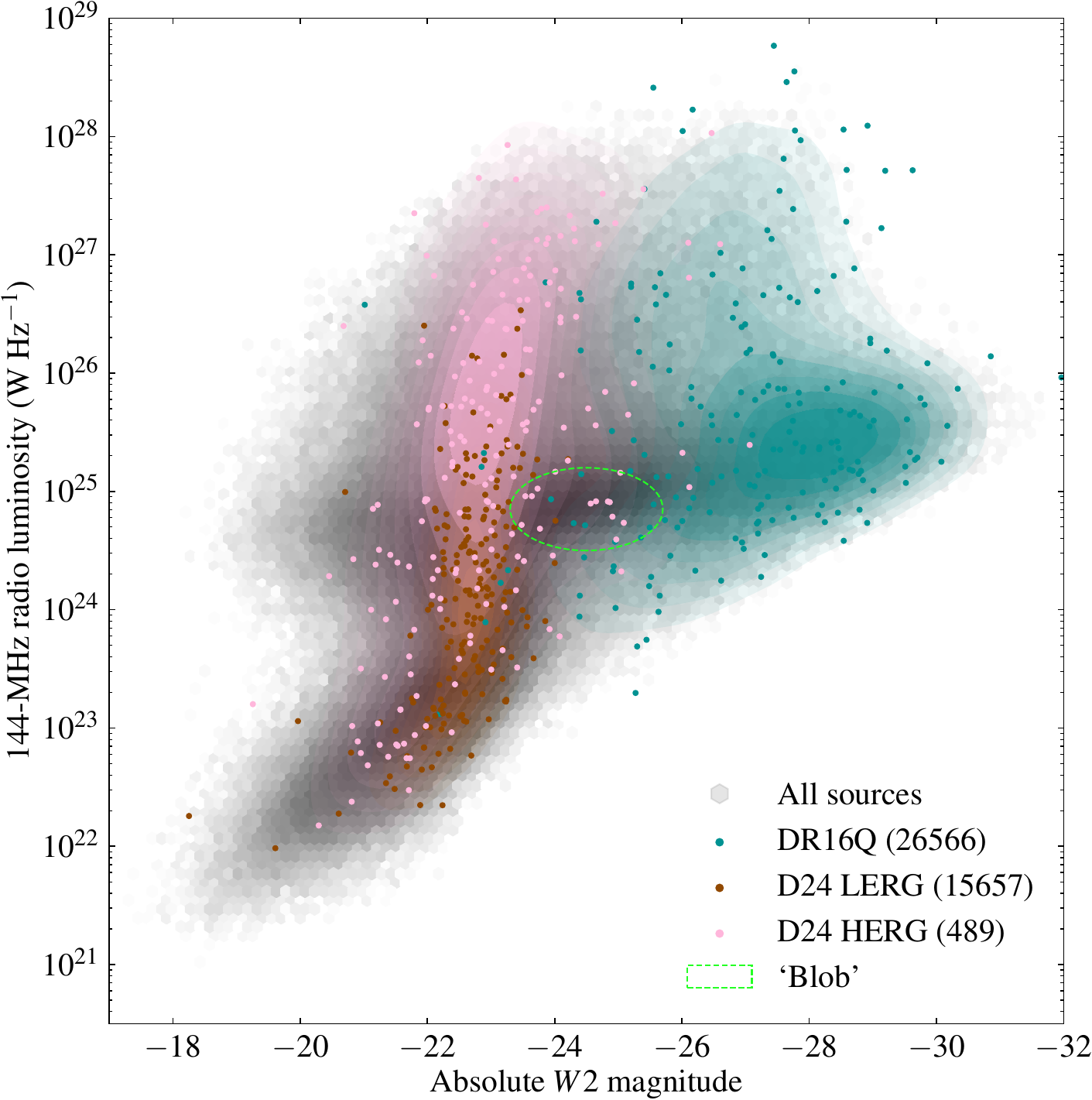}  
\includegraphics[width=0.48\linewidth]{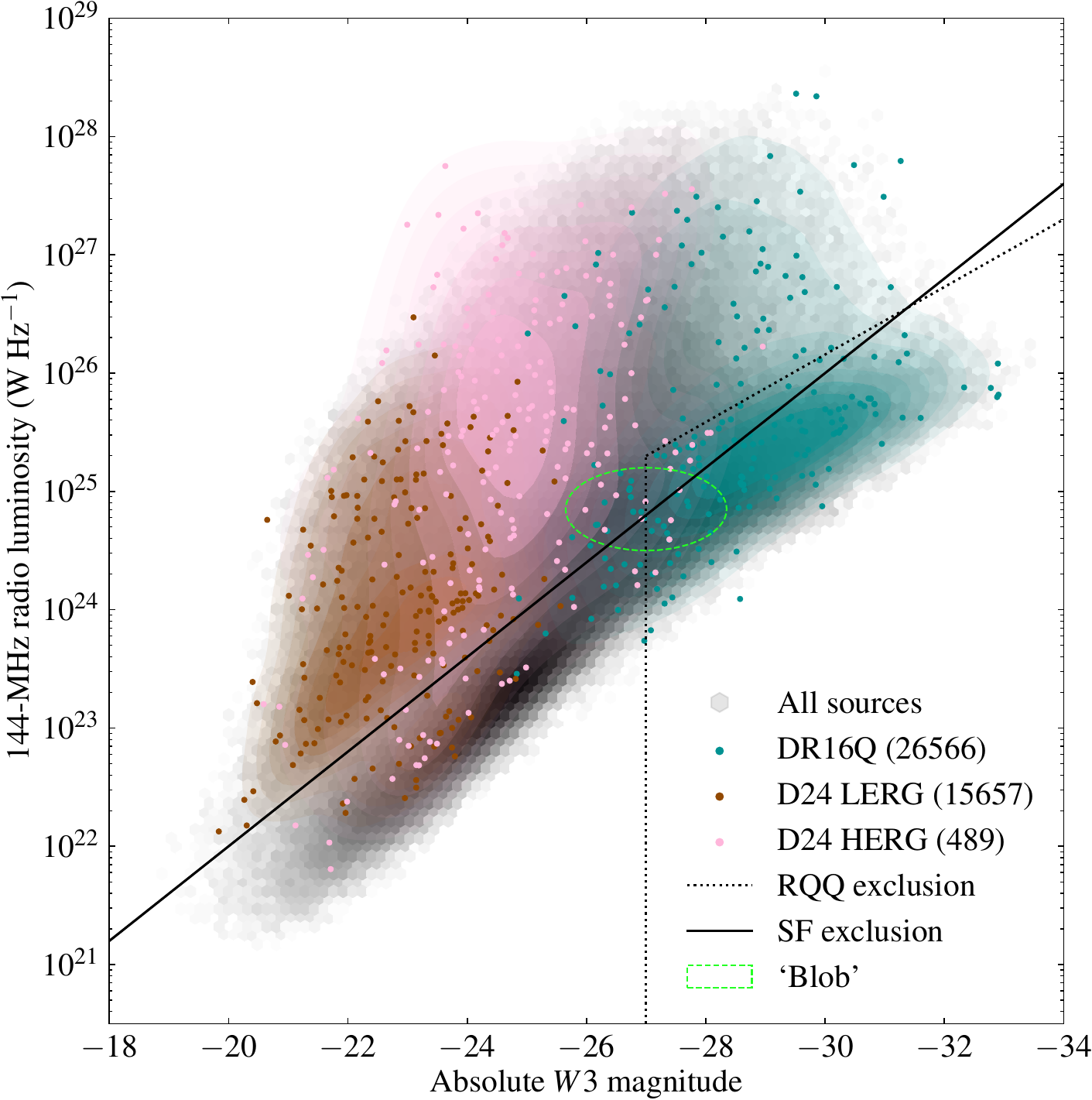}
\caption{As Figure \ref{fig:absmag}, but with emission-line classes
  for RLAGN and DR16 quasars labelled. Note that there are many fewer
  HERGs than the other types of object plotted here: the coloured
  density plots represent the locations where they are
most likely to be found rather than the total numbers of objects.}
\label{fig:wise_spc}
% absmag_lum_w2_spc_density.py
% absmag_lum_w3_spc_density.py
\end{figure*}

However, SDSS (type 1) quasars do
not seem to be the dominant population of objects in the `blob'. In
Fig. \ref{fig:wise_spc} we show the positions on the {\it WISE}/LOFAR
luminosity plot of D24 LERGs and HERGs together with the positions of
objects identified as SDSS quasars. The SDSS quasars turn out to avoid the
`blob' region; many of these objects seem actually to be galaxies or
non-SDSS quasars in
the $1 \la z \la 2$ redshift range with $W3$ non-detections. Some may
of course be quasars or extreme star-forming objects (e.g.
ultra-luminous infra-red galaxies or ULIRGs at high $z$), but in general
this part of parameter space cannot be excluded. DR16Q objects are
almost all detected in W3.

\subsection{Selection of RLAGN}
\label{sec:agn_selection}

%apply_cuts.py

From the preceding two subsections, conservatively we can cut out
clear SFGs, and thus restrict the sample to objects where the
  radio emission is presumptively due to RLAGN activity, by taking objects that:
\begin{enumerate}
\item Are detected in $W3$, AND
\item Lie below and to the right of the SF exclusion line on
 the right-hand panel of Fig.\ \ref{fig:wise_spc}, AND
\item Have $\log_{10}(L_{144}/\mathrm{W Hz}^{-1}) < 24.8$ (above which the SFG population
  merges into the `blob' discussed above and then the RQQs).
\end{enumerate}

This cut removes 331,221 objects whose radio emission is
  plausibly due to star formation from the parent sample,
leaving only objects where the radio emission is above the SF
  exclusion line. We cannot exclude the possibility that we are
  excluding some objects in this way which should be included in the
  AGN selection (e.g. objects where the radio emission is faint but on scales
  much larger than the host galaxy and so cannot be due to star formation) but this is unavoidable in a
  simple radio excess selection and in general we lack the resolution
  to investigate this distinction for our parent sample.

If we wish to remove the apparent `RQQ' branch we can make further cuts to remove sources that:

\begin{enumerate}
\item Are detected in $W3$, AND
\item Lie below and to the right of the RQQ exclusion line on
  the right-hand panel of Fig.\ \ref{fig:wise_spc}, AND
\item Have $W3$ brighter than $-27$
\end{enumerate}

This would exclude a further 66,718 objects. Whether this is the
correct thing to do depends partly on the science case of interest,
and partly on the nature of radio emission from the `RQQ' population \citep[e.g.][]{Panessa+19}.
In what follows we exclude these objects for consistency with H19, and
because they are generally treated separately in other analyses
\citep[e.g][]{Mingo+16}; see further discussion below (Section \ref{sec:nature_rl}).

This approach errs on the side of inclusivity for the RLAGN
sample that remains after these cuts are made since
we consider only objects detected in W3, which, as discussed above, is
only a fraction of our sample, though most known SFGs from the D24
analysis and most DR16Q are detected in W3. We could err in the
direction of having a cleaner (though less complete) RLAGN-only sample
by also excluding objects that meet these criteria with an upper limit
in W3. A comparison of the two approaches is shown in Table
\ref{tab:agnselect}. In what follows we use the inclusive sample of
565,821 objects for further analysis.

Our sample selection leads
to a sky surface density of candidate RLAGN (above our completeness
flux density limit) of $\sim 110$ deg$^{-2}$, exceeding
the sky densities often estimated for infrared selection
\citep[e.g.][]{Stern+12} and notably around twice the sky density of
the DR1 sample from H19. Given
that we require an optical ID and a redshift, it seems very likely
that this sky density is a very conservative lower limit even for sources that
would meet our selection criteria. We comment further on this below,
Section \ref{sec:howmany}.

% typecount.py
\begin{table}
  \caption{Results of inclusive and exclusive RLAGN selection as
    described in the text}
  \label{tab:agnselect}
  \begin{center}
    \begin{tabular}{lrr}
    \hline
    Name&Inclusive&Exclusive\\
    \hline
    Original sample&963,764&\\
    Star formation exclusion&331,225&357,424\\
    RQQ exclusion&66,718&130,844\\
    Remaining AGN&565,821&475,496\\
    \hline
    \end{tabular}
    \end{center}
\end{table}

\subsection{Selection of radiatively efficient and inefficient objects}
\label{sec:re_selection}

In the absence of emission-line classification, which will be provided
for many of these objects by the WEAVE-LOFAR survey \citep{Smith+16},
detailed broad-band SED fitting is expected to be the best way of
establishing the presence or absence of an energetically significant
radiatively efficient (RE) AGN \citep[e.g.][]{Best+23,Das+24}. Our
data are not good enough to do this: in particular, they lack
far-infrared measurements which help to tie down the contribution of
star formation to dust heating.

\begin{figure*}
  \includegraphics[width=\linewidth]{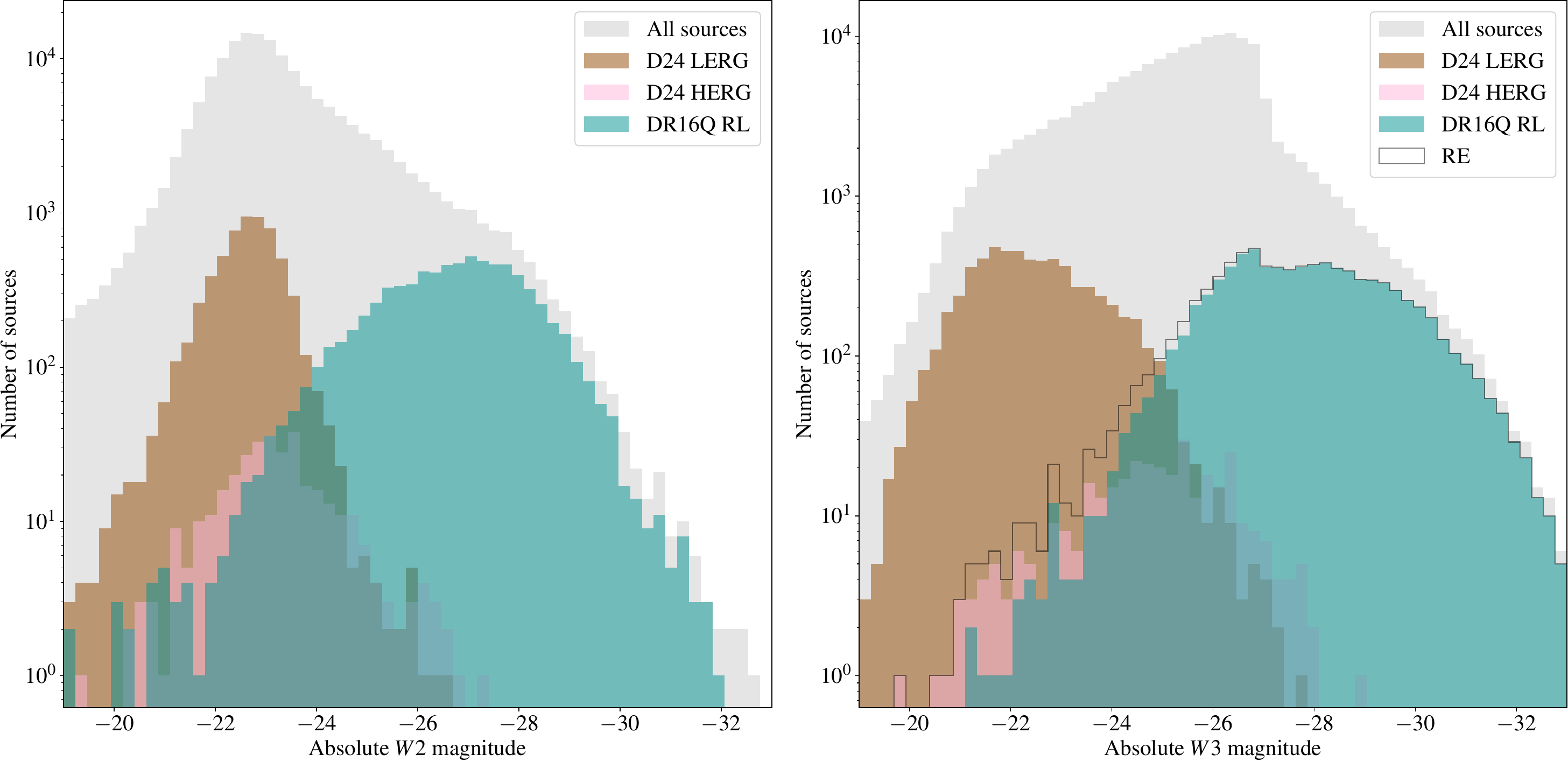}
  \caption{{\it WISE} magnitude distribution for RLAGN detected in
  $W3$ together with the distributions of emission-line classified
  sources. Left: $W2$ distribution. Right: $W3$ distribution, with
  combined D24 HERGs and DR16 quasars indicated with the RE line.}
  \label{fig:wisehist}
\end{figure*}

There are some alternative routes to an approximate classification,
however. In the $W2$/radio luminosity relation of
Fig.\ \ref{fig:wise_spc}, the quasars lie generally to the
luminous side of $W2 \approx -24$, while there is very little
difference between D24 HERGs and LERGs\footnote{D24 required the
emission lines needed for a BPT diagram to classify an object
spectroscopically, and so were only able to classify LERGs that had
these emission lines, while many absorption-line-only systems would
also normally be classified as LERGs. For simplicity we
refer to the `LINELERG' class of D24 as LERGs, while noting that there
will be other objects in the population studied by D24 that would
typically be classified in that way.} in this quantity (which for
normal galaxies is just driven by the old stellar population); HERGs
lie at typically higher radio luminosities than LERGs, with a good
deal of scatter, as is well known. But in the relation with $W3$ there
is a clear offset between the typical LERG and HERG locations in the
{\it WISE} magnitude, with both D24 HERGs and DR16 quasars being to
the luminous side of the distribution. Histograms of the {\it WISE}
absolute magnitudes for the RLAGN sample of the previous section confirm
this basic picture (Fig.\ \ref{fig:wisehist}). We could attempt to use
cuts in $W2$ and $W3$ (requiring detections in both bands) to select
and classify radiative AGN, corresponding to the suggestion of
\cite{Gurkan+14}, but there remains significant overlap
between the HERG and LERG populations in these plots. 

% absmag_elc.py

\begin{figure*}
  \includegraphics[width=\linewidth]{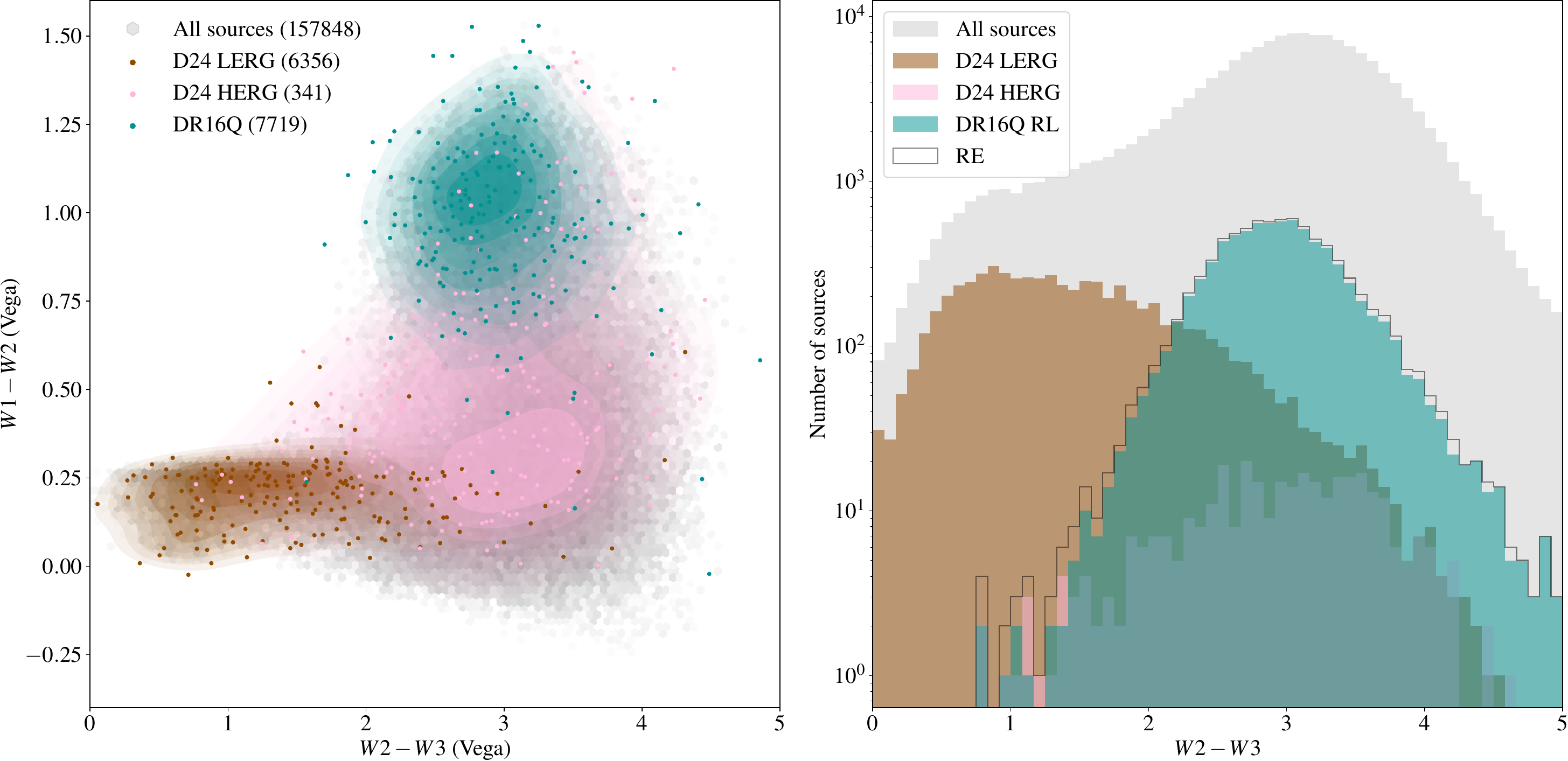}
  \caption{(Left) {\it WISE} colour-colour plot and (right) histogram
    of the $W2-W3$ colour for the RLAGN sample, with
    emission-line classes from D24 and the DR16 quasars indicated. Note the removal of the
    previously dominant population in the SFG locus by the RLAGN
    selection, as expected. Only 157,848 sources with $W3$ detections
    are plotted.}
  \label{fig:wisecc_class}
\end{figure*}

Alternatively we can look at colour-colour plots
(Fig.\ \ref{fig:wisecc_class}). Once SFG are removed, D24 LERGs,
(non-quasar) HERGs and DR16 quasars do sit in reasonably distinct
locations on these plots, as has been seen in earlier studies \citep{Gurkan+14,Prescott+18}. Again, the problem is that there is no sharp
boundary between the LERG and HERG populations in $W2-W3$, and that a
$W3$ detection is needed for the use of any line in that colour space.
Selecting a cut around $W2-W3 > 2$ would minimize the
cross-contamination between RE and RI objects for our sample but it
would clearly still exist. We can also note that there is a reasonably
sharp cutoff for $W1-W2 > 0.4$ above which few LERGs exist. Taking
cuts in colour space where RE objects have $W1-W2 > 0.4$ or $W2-W3 >
2$ (requiring a $W3$ detection for the latter) and quasars have $W1-W2
> 0.75$ (corresponding to standard `AGN' selection criteria, e.g.
\citealt{Stern+12}) leads to classifying 305,078 objects as RE (of
which 108,580 would be quasars and 196,498 would be non-quasar RE
objects, i.e. NLRG), 57,232 as RI and 203,511 as unclassifiable (because they
have $W1-W2<0.4$ and an upper limit on $W2-W3$ which does not allow
them to be placed on either side of the $W2-W3=2$ boundary line). It
is important to note that there will be substantial
cross-contamination with this method across the boundaries of the
RE/RI sample (12 per cent of D24 LERGs are classified as RE by this
method, for example) and also the calibration we use is best at low
$z$, whereas we know that the positions of particular types of source
on these plots are likely to evolve with $z$ \citep{Assef+13}.
Reliable and complete classification of the sample will have to await
more spectroscopic information from WEAVE-LOFAR and DESI and/or better
broad-band optical SEDs from {\it Euclid}. We will, however, use these
colour-colour classifications later in the paper, while bearing in
mind the substantial caveats on their application.

% wisecc_class_density_elc.py

\section{Properties of the catalogued sources}
\label{sec:properties}
In the preceding section we described the process of generating a RLAGN
catalogue. The catalogue is released with this
paper\footnote{\url{https://lofar-surveys.org/dr2_release.html}} and
a description of the columns it contains is given in Appendix
\ref{app:catalogue}. In this section we carry out some checks on the
properties of the catalogued objects to establish their consistency
with expectations from earlier work.

\subsection{Comparison with H19}

The H19 catalogue made from LoTSS DR1 data cannot be compared directly
with the current one because H19 used a lower (and somewhat
optimistic) completeness threshold of 0.5 mJy because of the position
of their sources at the location of best sensitivity for LOFAR on the
sky. However, if we cut the H19 RLAGN catalogue of 23,244 objects at our
current completeness threshold of 1.1 mJy, we get 14,379 objects, of
which 12,114 (84 per cent) have a matching ID position in the new parent
catalogue -- a match rate comparable to what we reported for the
catalogues as a whole in H23. Of these 12,114, 10,979 (91 per cent) would
also be classed as RLAGN in the new catalogue, with almost all the
remainder being excluded as SFG: these would be objects whose location
on the {\it WISE} colour-colour plot used by H19 is not consistent
with star formation, but which lie close to the $W3$-$L_{144}$
relation we use here. On the other hand, 1,461 (10 per cent) of the objects
excluded by H19 as SFG on the basis of their {\it WISE} colours would
be classed by us as RLAGN, which is as expected given the discussion of
colour selection above. We can conclude that the H19 selection is
broadly picking out the same classes of object as we do here, with
some scatter across the AGN/SFG boundary as expected.

\subsection{Source counts}
\label{sec:sourcecounts}
\begin{figure*}
  \includegraphics[width=\linewidth]{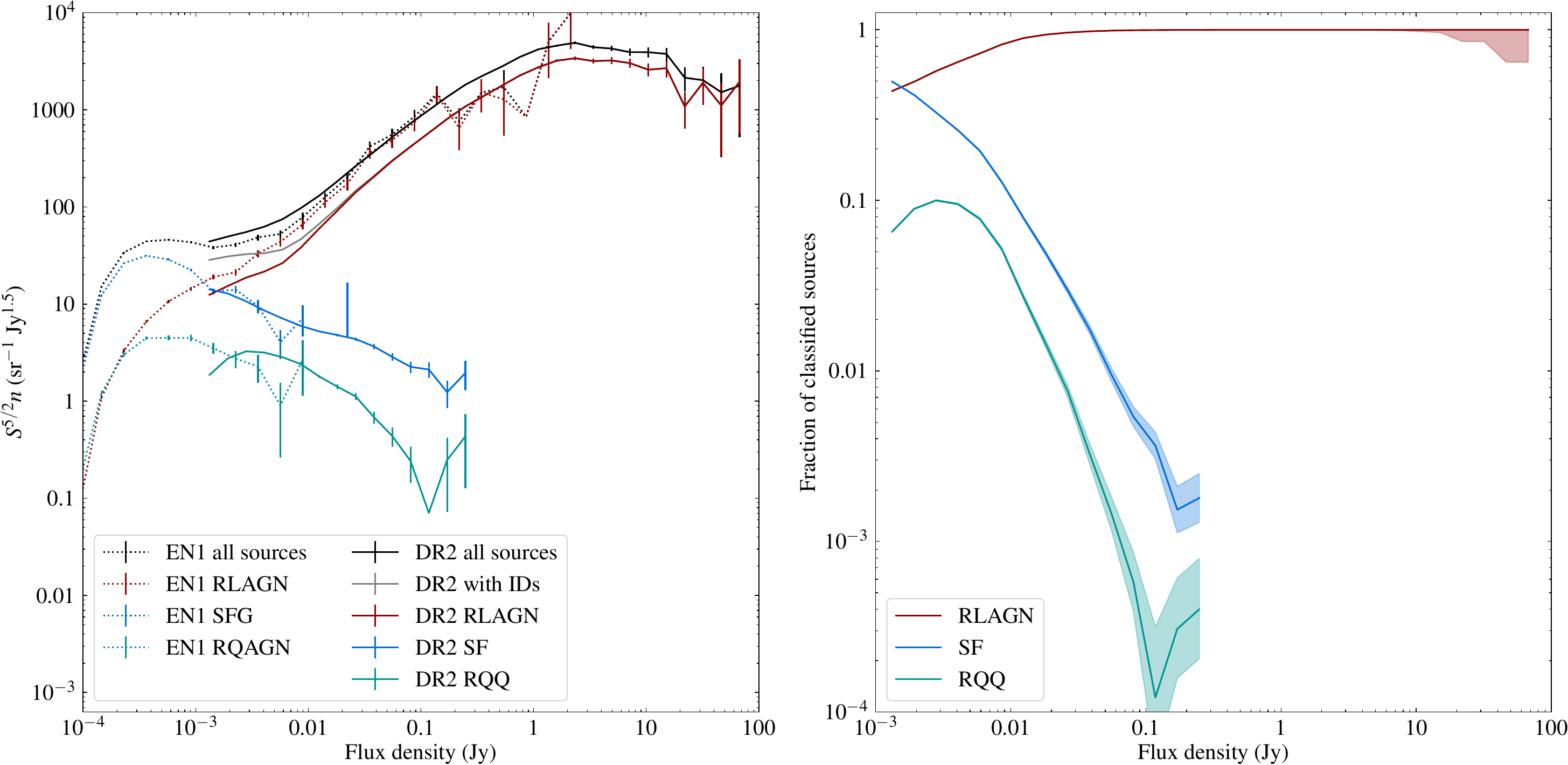}
\caption{Left: Conventional Euclidean-normalized
  differential source counts as a function of flux density for the
  overall LoTSS catalogue (after component association through visual
  inspection) above the flux density limit, the full parent catalogue,
  and the sources classified as RLAGN, SFGs and RQQ, together with the
  ELAIS-N1 source counts classified by \protect\cite{Best+23} which
  are shown as dashed lines. Error bars are Poissonian. The ELAIS-N1
  counts are not corrected for incompleteness at the faint end. Right:
  the fraction of the parent catalogue classified as RLAGN, SFG or RQQ
  as a function of flux density, with Jeffreys interval binomial
  uncertainties plotted as the shaded area.}
\label{fig:sourcecounts}
\end{figure*}

Extragalactic radio source counts are expected to be dominated at the bright
end by RLAGN and at the faint end by star-forming galaxies and RQQ, and
as a sanity check we present the source counts broken down by
classification in Fig.\ \ref{fig:sourcecounts}. No completeness
correction is carried out, since we believe that the lower flux cut of
our parent sample ensures good completeness. The expected picture
is seen, with SFG dominating the numbers of sources only at the very
lowest flux density levels. RQQ do not dominate at any flux density in
these classifications but are most important at flux densities of a
few mJy. These results are very consistent with those of, e.g.
\cite{Hardcastle+16} or \cite{Siewert+20}, with the only clear difference being the
significantly better statistics in the present work, and qualitatively
in agreement with models such as those of \cite{Wilman+08}, although there
are differences in detail (see \citealt{Siewert+20} or
\citealt{Mandal+21} for a more detailed discussion). We note that we may be incomplete to SFGs
at the bright end of the number counts because of the lower redshift
cut imposed as part of our selection, and of course we are still only
able to classify the 58 per cent of the sample that have optical IDs
and redshift information. Source counts for the LoTSS deep fields DR1
field ELAIS-N1 \citep{Sabater+21} are also plotted, broken down
according to the SED-fitting classifications of \cite{Best+23}, and
these show encouragingly consistent results with our much simpler
classification scheme, with the exception that almost all the
unclassified sources in DR2 (those that do not meet our ID and
redshift selection criteria) are classified as AGN in the deep fields.
We return to this point below, Section \ref{sec:howmany}.
  
\subsection{Distributions of key quantities}

% make_agn_sf_hists.py
\begin{figure*}
  \includegraphics[width=\linewidth]{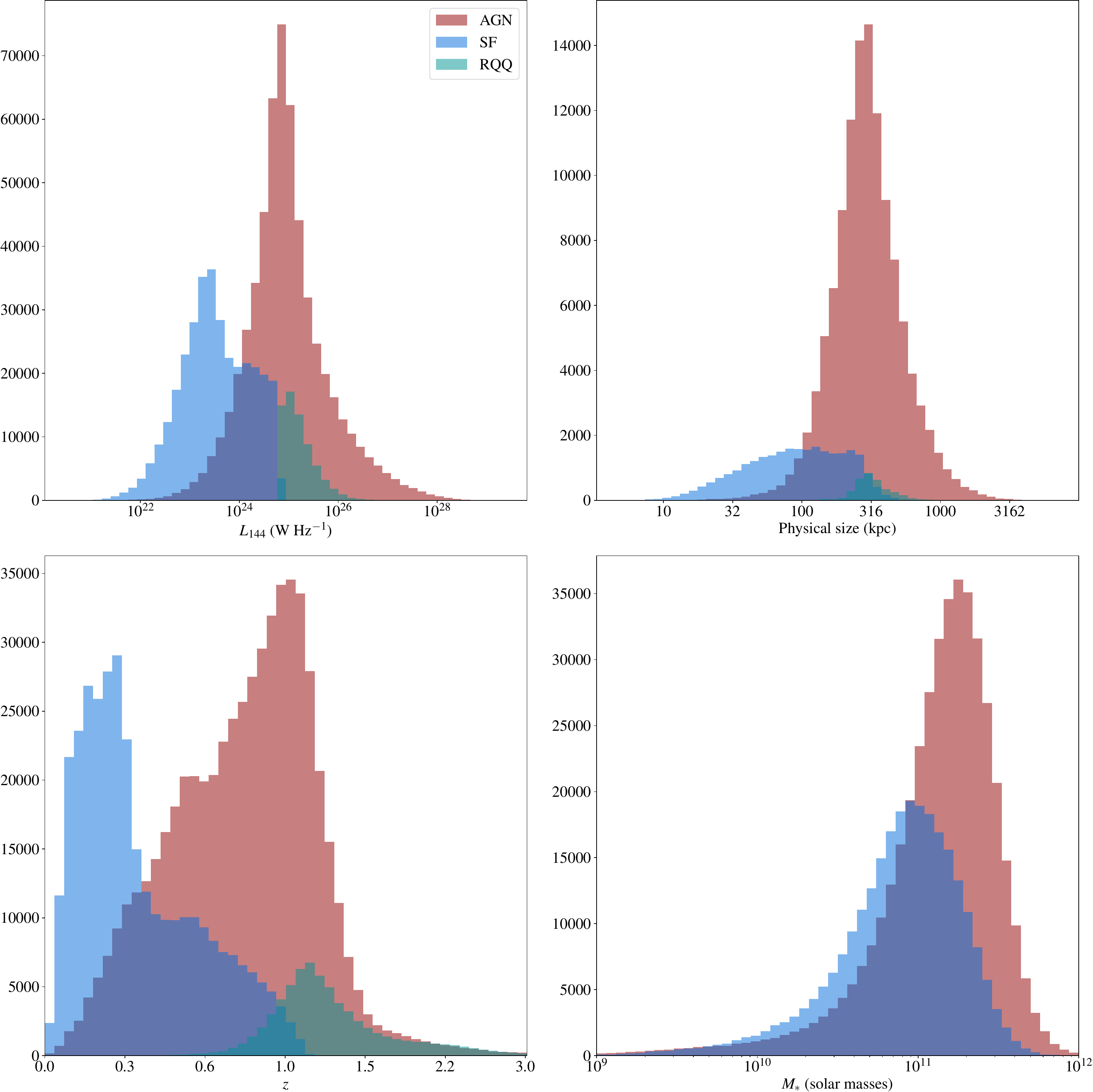}
  \caption{Luminosity (top left), size (top right), redshift (bottom
    left) and host galaxy stellar mass (bottom right) distributions of the three
    catalogues. In the size histogram, only resolved sources as
    defined by H23 are
    plotted, while the RQQ sample is excluded from the mass histogram and
    only objects with {\tt flag\_mass==True} are shown.}
  \label{fig:prophist}
  \end{figure*}

Fig.\ \ref{fig:prophist} shows the distribution of luminosity, size,
redshift and stellar mass for the RLAGN, SFG and RQQ samples. These show
the expected separation between the two sets of quantities, although a
small number of SFG with physical size $>100$ kpc is unexpected and
may indicate from contamination of the SFG sample by RLAGN, or, perhaps
more likely, poorly measured physical sizes for SFG which may be
represented by large Gaussians in PyBDSF fitting. Of note is
the steep decline of RLAGN above $z \sim 1.2$ -- massive galaxies are no
longer detected by the Legacy survey above this redshift. SFG cannot
have $\log_{10}(L_{144}/\mathrm{W Hz}^{-1}) > 24.8$ in our separation scheme and so a
sharp feature appears in their luminosity distribution as well, but
otherwise the luminosity and redshift distributions are not too
different to those of our earlier work with different AGN/SF
separation methods \citep{Hardcastle+16,Hardcastle+19}: the main
difference is that we are sensitive to sources at higher redshifts and
so have more RLAGN overall and in particular more luminous RLAGN than
in previous work. The stellar mass distribution shows strong overlap between the
SFG and RLAGN hosts, as expected from e.g. Fig.\ \ref{fig:absmag}, but a
clear tendency for RLAGN hosts to be more massive overall. We miss
low-mass SFGs in general because they will not be detected in the
radio other than at very low redshift, thus biasing the SFG mass
distribution high. As noted by
H23, not all galaxies in the sample have reliable mass estimates, and
of course no quasars do: overall 73 per cent of RLAGN and 80 per cent of
SFG are plotted here.

\subsection{Power-linear size diagram}
\label{sec:pdd}
% make_pdd.py
\begin{figure*}
  \includegraphics[width=\linewidth]{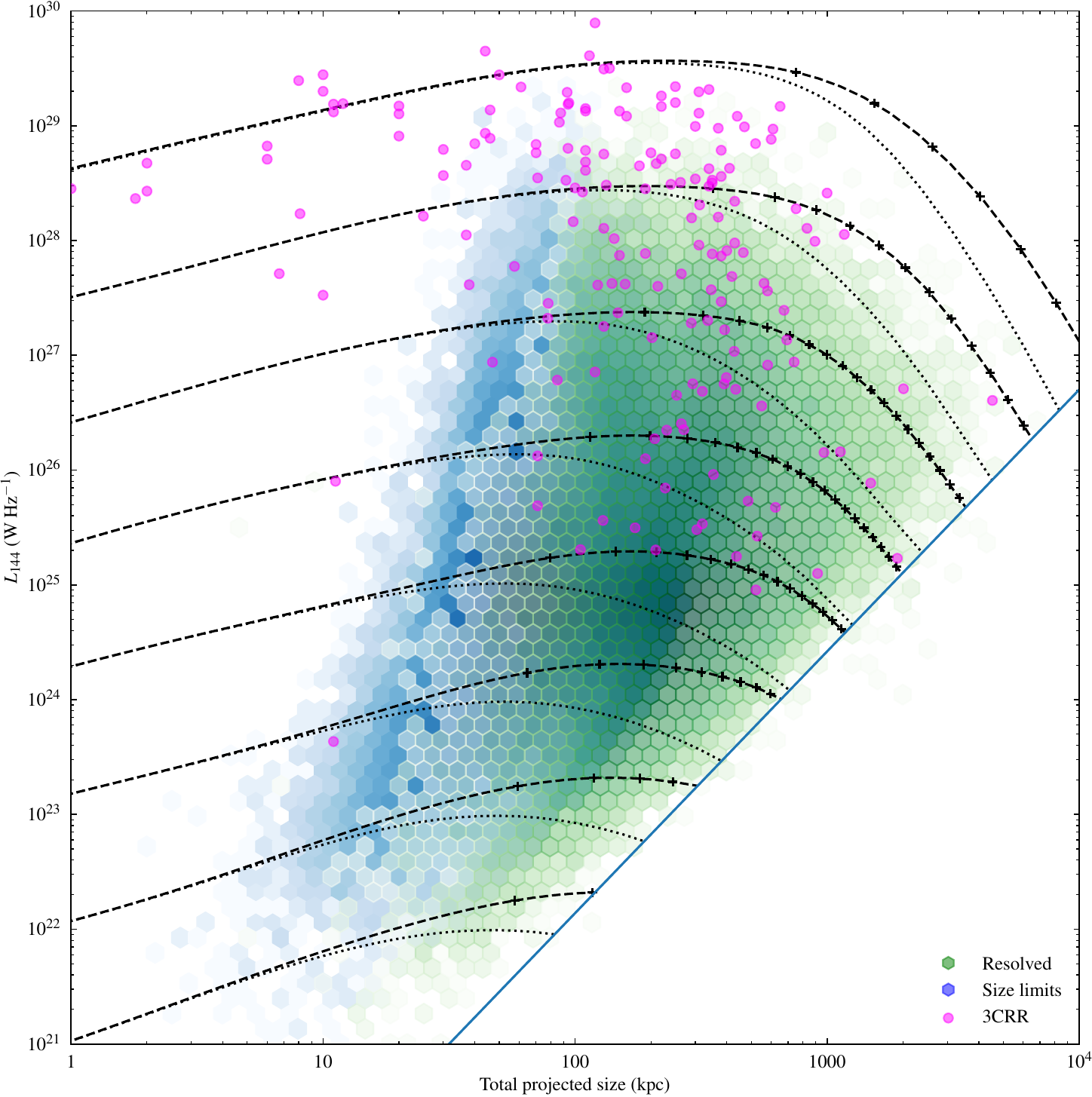}
  \caption{Power/linear size plot for the RLAGN sample. All 565,821 RLAGN
    are plotted: the green density plot indicates size estimates for
    sources that are resolved by LOFAR as defined by H23 while the blue points are for
    sources that are unresolved, in which case the size should be
    taken as an upper limit. We also show the positions of 3CRR
    sources \citep{Laing+83} for comparison. A solid blue line
    indicates the approximate region that cannot be populated due to
    the surface brightess limit of LoTSS. We overplot $z=0$ (dashed) and $z=1$ (dotted)
    theoretical evolutionary tracks in this space \citep{Hardcastle18} for sources lying in
  the plane of the sky in a group environment ($M_{500} = 2.5 \times
  10^{13} M_\odot$, $kT = 1$ keV) for two-sided jet powers (from
  bottom to top) $Q=10^{33}, 10^{34}, \dots, 10^{40}$ W; see the text
  for details. Crosses on the $z=0$ tracks are plotted at intervals of 50
  Myr, with linear size increasing monotonically with time.}
  \label{fig:pdd}
  \end{figure*}

Fig. \ref{fig:pdd} shows the power-linear size ($P$-$D$) diagram for the
RLAGN sample. Relative to the similar figure shown by H19, this plot contains far more
  sources and probes deeper into the low-surface-brightness regime to
  the right hand side of the plot due to the improved sensitivity of
  LoTSS DR2. Individual sources' evolution causes
them to describe tracks in this plane: representing the whole
population on such a plot gives us a sense of the range of properties
that objects in the sample have, particularly when combined with
theoretical models. As the plot shows, if we (unsafely) ignore the
wide range of environments and redshifts that are represented within
the sample, the objects that we select span 7 orders of magnitude in
kinetic power and have maximum ages ranging from $10^7$ yr to several Gyr.
Due to the size of the sample and the sensitivity of LoTSS DR2, we select an unprecedentedly large number of physically very large
sources, as reported by \cite{Mostert+24}, and we can see that these
must probe the extreme high-age end of the visible source lifetime
distribution.

What is also apparent from this Figure is two key limitations of our
selection that will affect future analysis using the sample. The first
is the large number of unresolved sources. Only 118,498 objects in the
sample (21 per cent) have reliable size measurements. As shown by
Sweijen et al.\ (submitted), a much larger fraction of sources have size
measurements when 6-arcsec images made using the Dutch LOFAR baselines
are combined with the full 0.3 arcsec available using the
International LOFAR Telescope, but we are still some way away from
having these high-resolution images for wide sky areas like those of
LoTSS DR2. Size information for a sample provides key information
about the age distribution, as discussed by H19, and in
particular could help us to determine the distribution of lengths of
accretion event that drive jet activity (the lifetime function). A multimodal lifetime
function would be implied, for example, by the idea that a substantial
fraction of young radio sources selected as Compact Symmetric Objects
(CSOs) are actually triggered by tidal disruption events (TDEs) as
proposed by \cite{Readhead+24}. Accurate size measurements of a large
sample like the present one, selected in as unbiased a way as possible
directly from low-frequency surveys, are crucial to making these
inferences across the population. These will be available in the
future through processing of the long-baseline LoTSS data and through
the proposed ILoTSS project\footnote{\url{https://lofar-surveys.org/ilotss.html}}. Since the constraints on the lifetime
function from the existing data have an influence on the inference of
radio AGN feedback effects, we will discuss them in more detail in a
future paper (Pierce et al in prep).

The second limitation of our selection is the surface brightness
limit, which manifests itself as a sharp cutoff in the sampling of
$P$-$D$ space below and to the right of the diagonal line in
Fig.\ \ref{fig:pdd}. Physically large sources can be detected but only
if they are also luminous: we have no way of knowing whether
there are, say, Mpc-scale RLAGN with $L_{144} = 10^{23}$ W Hz$^{-1}$,
since they are simply undetected in our sample. More subtly, some
large objects will be detected in the LoTSS images but will not be
included in our selection because it has not been possible to identify
them optically; for example, remnant sources with no compact structure
may be particularly difficult to associate with a particular host
galaxy \citep[e.g.][]{Brienza+17}. \cite{Oei+23} and \cite{Mostert+24} have shown that
dedicated methods of searching for diffuse sources and/or of their
optical identifications can increase the numbers of detections, but
these methods only somewhat shift the boundaries of the area of the
$P$-$D$ plane that we are not sensitive to rather than removing it
altogether. Analysis of the size (and therefore, as discussed,
age/lifetime) distribution of RLAGN, or of the populations of special
types of RLAGN such as remnant sources, needs to take account of this
effect since even next-generation surveys will not fully remove the
bias for the faintest objects. We estimate that in the current sample
only the most luminous sources ($>10^{26}$ W Hz$^{-1}$) are unaffected
by surface brightness limits, and even here the completeness is
likely to be reduced by the challenges of optically identifying all of
these physically large sources.

With these caveats, the population of the $P$-$D$ plot and the
comparison with models still gives
some insights into RLAGN physics. For example, there is a region in the
top right of the plot (powerful, physically large sources with nominal
jet powers $10^{39} < Q < 10^{40}$ W) which is almost completely
unpopulated, although these objects would be easily detected if they
existed. This tells us that the most powerful jets, whose powers are
of the order of the Eddington luminosity for a $10^9$ solar mass black
hole, almost never live beyond $\sim 100$ Myr. Less powerful sources,
with $10^{38} < Q < 10^{39}$, clearly can persist for much longer and
(because of the surface brightness selection effect) most of our
extreme giant sources are in this category \citep[cf.][]{Oei+24}.

\subsection{Colour-colour classification compared with other methods}
\label{sec:ccc_check}

\begin{figure*}
  \includegraphics[width=\linewidth]{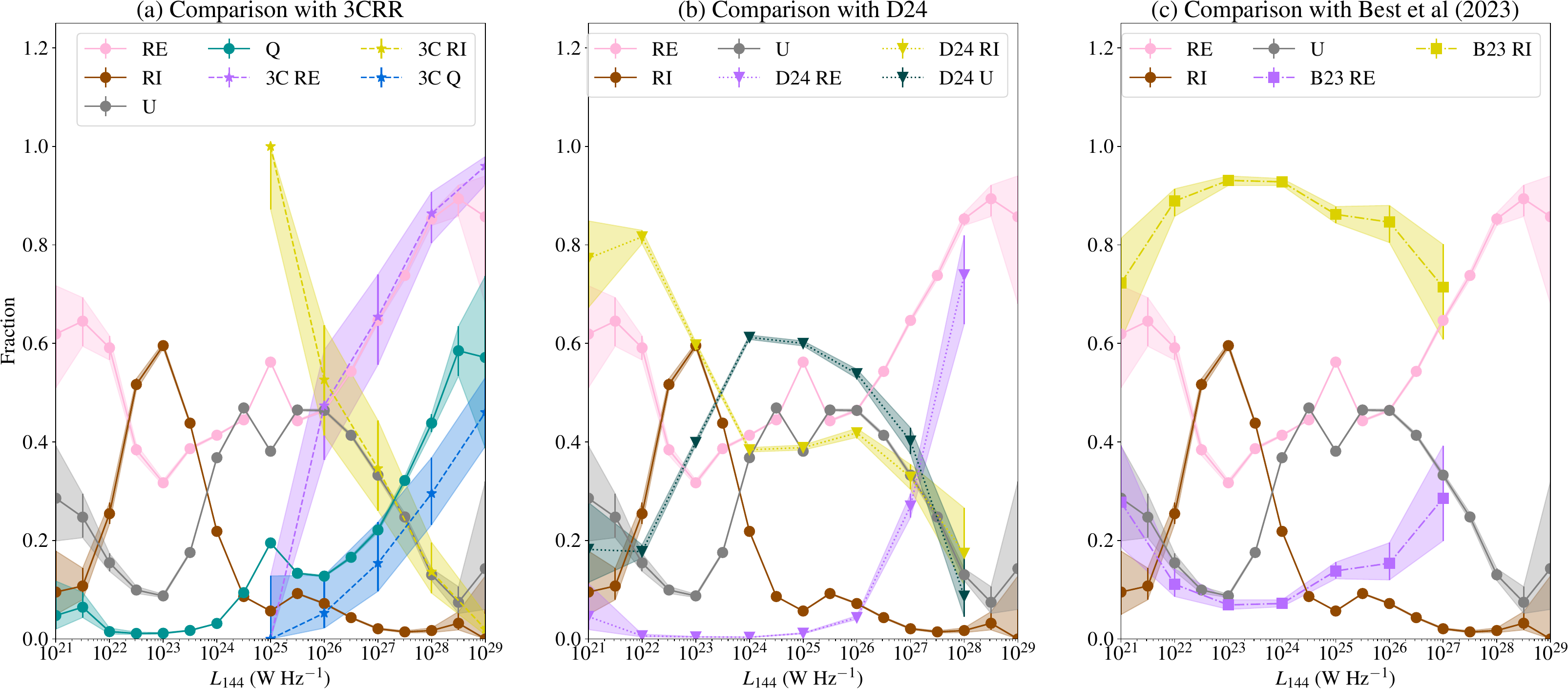}
  \caption{A comparison of the {\it WISE} colour-colour plot
    classifications with those from other work. All panels show the
    proportion of objects with particular classifications as a function of the
    total RLAGN at a given $L_{144}$ with $z<1.2$. From left to right, we
    compare our classifications with (a) the 3CRR catalogue, (b) D24
    and (c) \protect\cite{Best+23}. Shaded regions show the binomial
    credible intervals. For clarity, luminosity bins where there would be
  fewer than 4 sources in total are not plotted.}
  \label{fig:ccc_compare}
\end{figure*}

Since our colour-colour classification of RLAGN will be used later in the paper
we here compare its results briefly with that from other work. Panel
(a) of Fig \ref{fig:ccc_compare} shows a comparison with the 3CRR
catalogue \citep{Laing+83} which has complete spectroscopic classifications
for luminous sources. The agreement here is remarkably good, including
with the proportion of quasars, if we assume that almost all our
unclassified objects are RI at $L_{144}>10^{26}$ W Hz$^{-1}$. The
proportion of RE objects that we obtain is much higher at the lowest
luminosities, though. Panel (b) compares with D24 and we see that
again the agreement is reasonably good at high luminosities,
particularly as some of D24's unclassified objects will certainly be
RE. However at low luminosities we obtain many more RE objects and we
suggest that this is likely due to contamination of our sample by lower-mass
galaxies that are intrinsically RI but have star-forming colours, or
of course by some objects that should have been excluded as SFGs but are
not removed by our selection criterion. This
is particularly clear around $10^{25}$ W Hz$^{-1}$ as there is a
localized peak in RE objects that must be related to the SFG maximum
luminosity at $\log_{10}(L_{144}/\mathrm{W Hz}^{-1}) = 24.8$. Finally, the comparison
with \cite{Best+23} (B23) shows strong disagreement at all luminosities,
with B23 having a much higher proportion of RI objects throughout;
however, this means that B23 also strongly disagree at moderate to high luminosity with
the spectroscopic classifications of 3CRR and D24, which is
surprising. If we ignore the discrepancy with B23, we can conclude
that our classification should give consistent results with previous
{\it spectroscopic} classification at $L_{144}>10^{26}$ W Hz$^{-1}$, with
the unclassified objects at high luminosity plausibly being mostly RI. 

\section{Discussion}
\label{sec:discussion}

\subsection{How many radio AGN are there?}
\label{sec:howmany}

We noted above (Section \ref{sec:agn_selection}) that our RLAGN
selection gives a sky density of $\sim 100$ deg$^{-2}$. We know on the
basis of the source counts (Section \ref{sec:sourcecounts}) that there
must be a large number of additional RLAGN in the parent sample not
identified as such by us: at a minimum, all of the sources with flux
density larger than a few mJy that do not have an optical counterpart
and a secure redshift seem likely to be RLAGN with a high-$z$ host
galaxy. Given that the redshift fraction in the sample stands at 58
per cent at the moment, it is plausible that the true sky density of
RLAGN with flux density $>1.1$ mJy is a factor 1/0.58 = 1.7 times larger
than our estimate. Indeed, the catalogues and SED-based classification
of \cite{Best+23}, using the LoTSS deep fields where the optical
identifications are almost complete, give a sky density of 160 deg$^{-2}$ RLAGN
above our flux cut in the ELAIS-N1 field, consistent with this
estimate, and this can be seen in the source counts plot of
Fig.\ \ref{fig:sourcecounts}, where the RLAGN line for ELAIS-N1 lies
consistently above ours in the range where they overlap. While the
detailed classification of sources through SED fitting and
spectroscopy may change, the number count arguments imply that this
conclusion is likely to be robust.

We can also see from the number counts that a substantial number of
radio-excess RLAGN are likely to exist at flux densities below our cut
of 1.1 mJy. As noted above, the optical data become insensitive to
massive galaxies above $z\approx 1.2$, at which point, by chance, an
$\alpha = 0.7$ radio source with $S_{144} = 1.1$ mJy has a luminosity
almost exactly at our adopted cutoff for star formation, around $7
\times 10^{24}$ W Hz$^{-1}$. Thus, not only are we insensitive to
high-$z$ AGN because of our optical data, but also as we move to high
$z$ our flux density cut prevents us from including in our sample
objects that would be unambiguously AGN by our selection criteria even
if their host galaxies were detected (see further discussion of this
in the following subsection). Again, the \cite{Best+23} catalogue can
be used to get a sense of the order of magnitude of this effect: in
ELAIS-N1 there are 670 deg$^{-2}$ objects classed as radio-excess AGN
in total, so that the objects that are below our completeness flux cut
actually dominate numerically over these that are above it. As the
deep-fields catalogue samples down to the point where SFG start to
dominate the number counts of radio sources, a number of the order of
1000 deg$^{-2}$ is likely to be an upper limit on the sky density of
true radio-excess AGN (taking account also of the effects imposed by
surface brightness limits as discussed in Section \ref{sec:pdd}).
Hence we can conclude that we have found around a tenth of all of the
RLAGN by this definition in the DR2 sky area. Even so, the sky densities
of RLAGN even in our current wide-area survey are already
comparable with those of optically selected AGN expected from
next-generation wide-area surveys such as {\it Euclid}
\citep{Euclid+25}, consistent with the result of \cite{Sabater+19}
that all massive galaxies have AGN-related radio emission at some level. Deeper
wide-area radio sky surveys such as the proposed ILoTSS survey, which
will cover a similar sky area to LoTSS DR2 but at twice the depth and
0.3 arcsec resolution, or deep and wide surveys with the forthcoming
SKA, should be expected to select all the radio-excess AGN in their
sky coverage.

We finally return to a point made in Section \ref{sec:intro}: even if
it can be done with complete accuracy, {\it radio-excess} selection
does not in practice capture all sources that have non-negligible
radio emission due to AGN activity. Many star-forming galaxies,
including arguably the Milky Way, show radio emission that is the
result of past or present activity of the central supermassive black
hole but that does not currently dominate over the radio emission due
to star formation. A complete census of the black-hole activity in the
Universe would need to account for these low-level but potentially
very numerous sources with both star-formation and AGN-related radio
emission, and doing so would almost certainly give a significantly
larger radio AGN sky density \citep{Morabito+25}.

\subsection{The nature of the RLAGN population}
\label{sec:nature_rl}

As noted in Section \ref{sec:intro}, there are in principle a number
of different mechanisms that can produce AGN-related radio emission,
not limited to the powerful jets that dominate the high-luminosity
radio AGN population \citep{Hardcastle+Croston20}. Since we select as
a RLAGN any source whose radio emission exceeds the prediction from
the mid-IR/radio relation (Section \ref{sec:SFcut}) we are potentially
including different physical types of object in our sample.
\cite{Panessa+19} list some of the different mechanisms for generation of
radio emission from what they describe as `radio-quiet' AGN, but it is
important to understand that an object that is `radio-quiet' by their
definition -- which can only consistently be applied to radiatively
efficient classical AGN, since it relies on optical or X-ray emission
from the accretion disk -- could also be `radio-luminous' by ours.
Thus our sample could be `contaminated' by objects whose radio
emission in excess of star formation is generated by e.g. disk winds
or coronal emission rather than by a jet. If the RQQ are removed as
discussed above (Section \ref{sec:agn_selection}), then such objects
in our sample would be limited to objects that do not lie in the RQQ
exclusion region (Fig.\ \ref{fig:wise_spc}), perhaps including some of
the `blob' population discussed in Section \ref{sec:quasars}. We expect this
population to be most important at low radio luminosities, and there
is considerable evidence for a group of low-luminosity compact
RLAGN in the LOFAR population (H19) including some that are clearly
radiatively efficient AGN \citep{Chilufya+24}. At present there is no
way of separating these objects out of our sample (or any other) in
the absence of detailed information about the radiative AGN
contribution to our sample from spectroscopy, and so their presence in
the sample should be borne in mind in what follows.

\subsection{The evolution of the RLAGN luminosity function}
\label{sec:lf}

The wide-area LOFAR observations give us an unrivalled view of the
rare, luminous RLAGN in the universe out to the point where we start to
run out of optical identifications at $z\approx 1$. One way of
parameterizing this population and its cosmological evolution is to
construct the radio luminosity function as a function of redshift
\citep[e.g.][]{Condon+84,Dunlop+Peacock90,Willott+01,Mauch+Sadler07,Prescott+16,Smolcic+17,Williams+18,Novak+18,Kondapally+22,Wang+24}
and in this section we use our RLAGN/SFG separation to investigate the
cosmological evolution of the overall RLAGN population. The very large
number of objects in the sample means that we can get a much more
precise view of the evolution of the luminosity function, particularly
at the high-luminosity end, than has previously been
possible. 

Luminosity functions are constructed following the standard $\rho =
\sum_{i}1/V_i$ method (as described by \citealt{Schmidt68,Condon89}),
where $V_i = V_{\rm max} - V_{\rm min}$ is the volume within which a
given radio galaxy could have been observed with the available
observations and sample criteria (following \citealt{Williams+18}).
$V_{\rm min}$ is the volume enclosed within the observed sky area out
to the minimum distance at which a target could have been observed,
derived from the minimum redshift limit in the given bin. $V_{\rm
  max}$ is the volume enclosed out to the maximum available distance
for the target, considering the imposed {\it WISE} Band 1 flux limit
($W1< 21.2$ mag) and the upper redshift limit for the bin: we
  discuss the details of the optical $V_{\rm max}$ calculation in
  Appendix \ref{app:optvmax}. Error bars are Poissonian.

\begin{figure}
  \includegraphics[width=\linewidth]{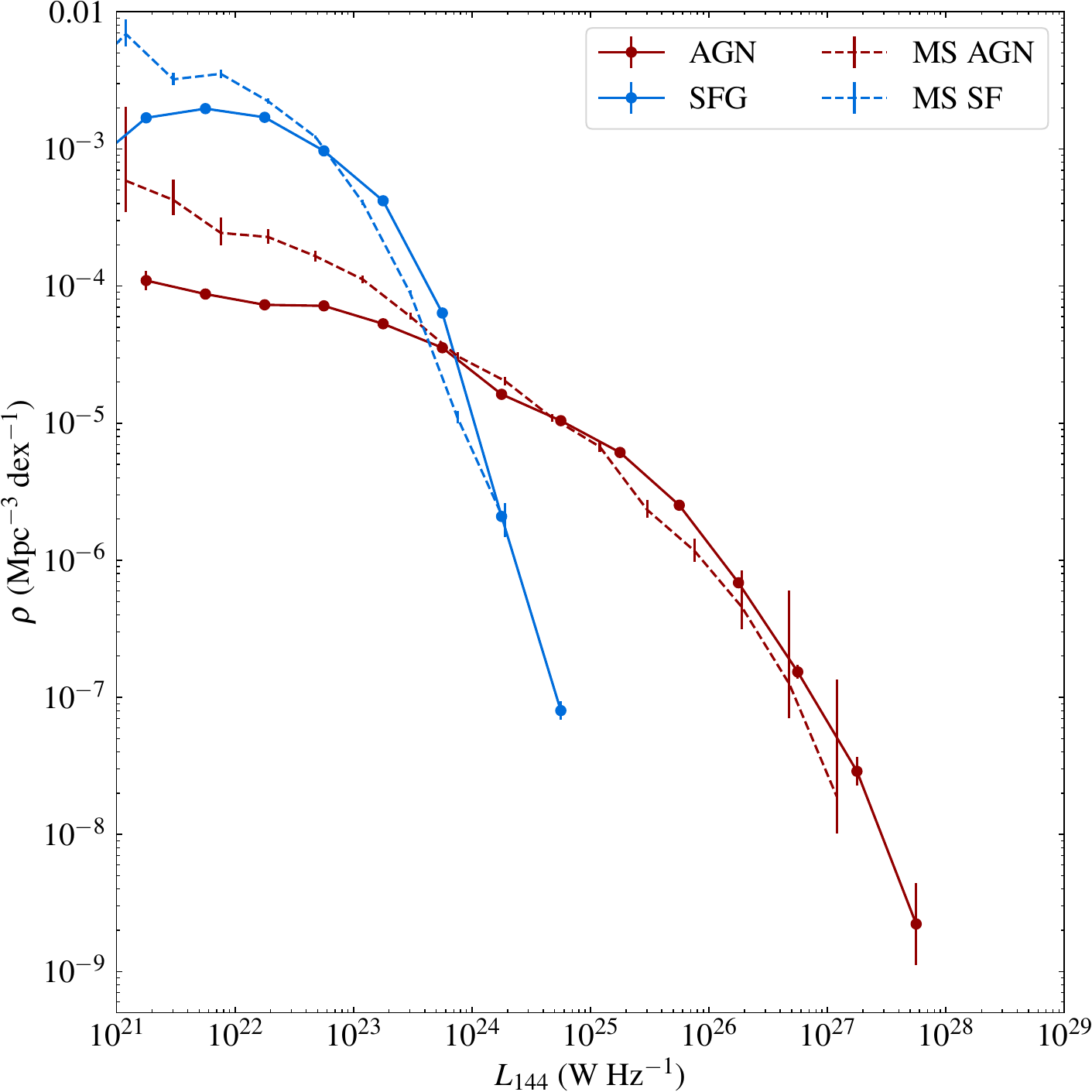}
  \caption{The luminosity function of LOFAR sources (left) in the range $0.01 < z < 0.3$,
    with both RLAGN and SFGs shown and the results of
    \protect\cite{Mauch+Sadler07} plotted for comparison.}
  \label{fig:lf_full}
\end{figure}

Fig.\ \ref{fig:lf_full} shows the overall luminosity function for the
RLAGN in the sample for $z<0.3$, along with the local luminosity
function of \cite{Mauch+Sadler07}, hereafter MS07, for comparison, converted to
144 MHz assuming $\alpha = 0.7$. Unsurprisingly our LF is in
reasonable agreement with that of MS07, but we have significantly
fewer objects at the lowest luminosities, which we attribute to
missing optical IDs differentially at the low-flux end (H23). We know
from the work of e.g. \cite{Kondapally+22} that the local LF from the
LoTSS deep fields, with essentially complete optical IDs, is more in
line with the MS07 results --- our results are more in line with the
wide-field work of \cite{Sabater+19}. The lowest-luminosity bins are
populated by relatively low numbers of objects, so small changes in
the optical ID fraction can make a large difference. We may also be
classifying slightly more objects as SFG than MS07 did in the
luminosity range $10^{23}$ to $10^{24}$ W Hz$^{-1}$, but a direct
comparison is hard given that the SFG population shows strong
cosmological evolution.

\begin{figure*}
  \includegraphics[width=0.48\linewidth]{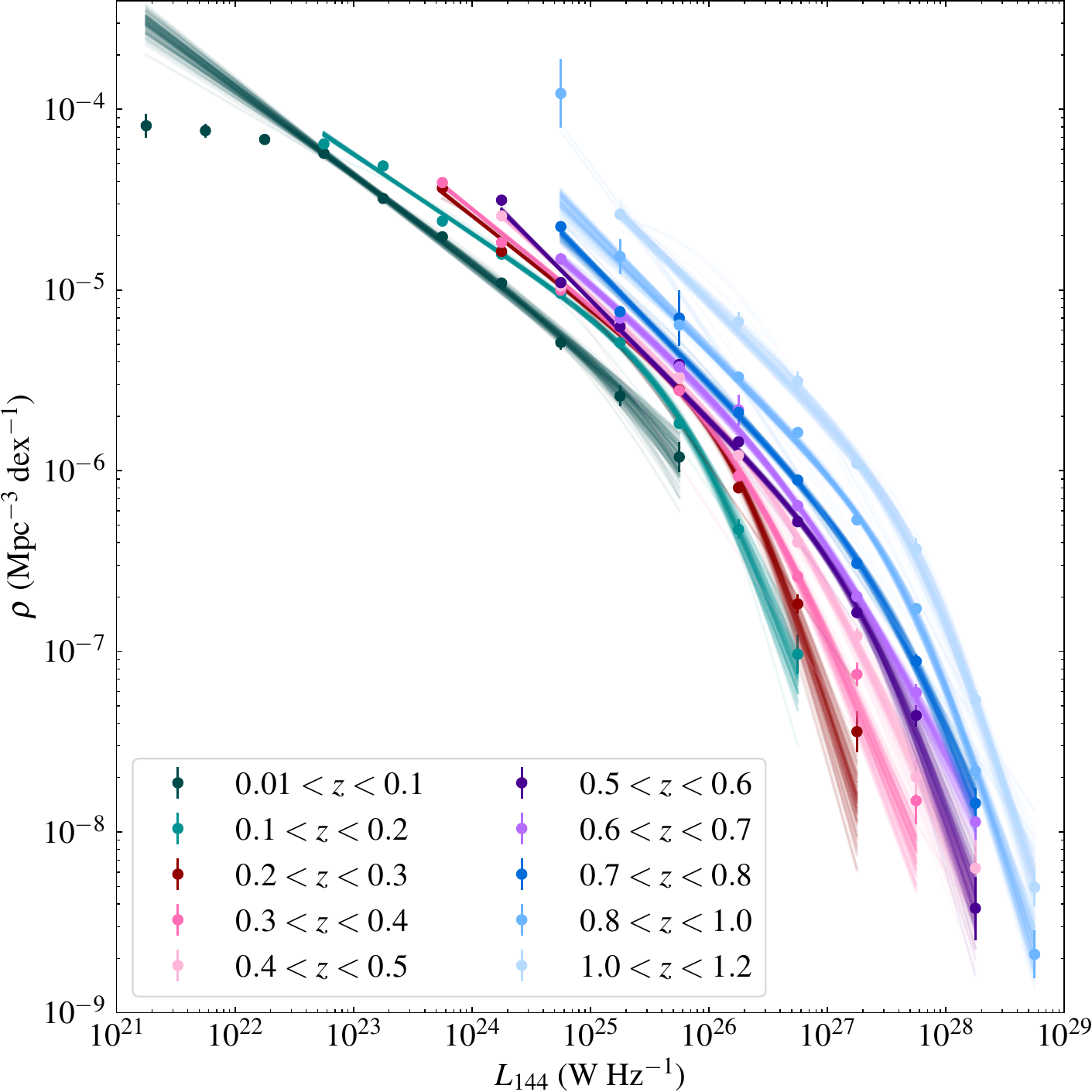}
  \includegraphics[width=0.48\linewidth]{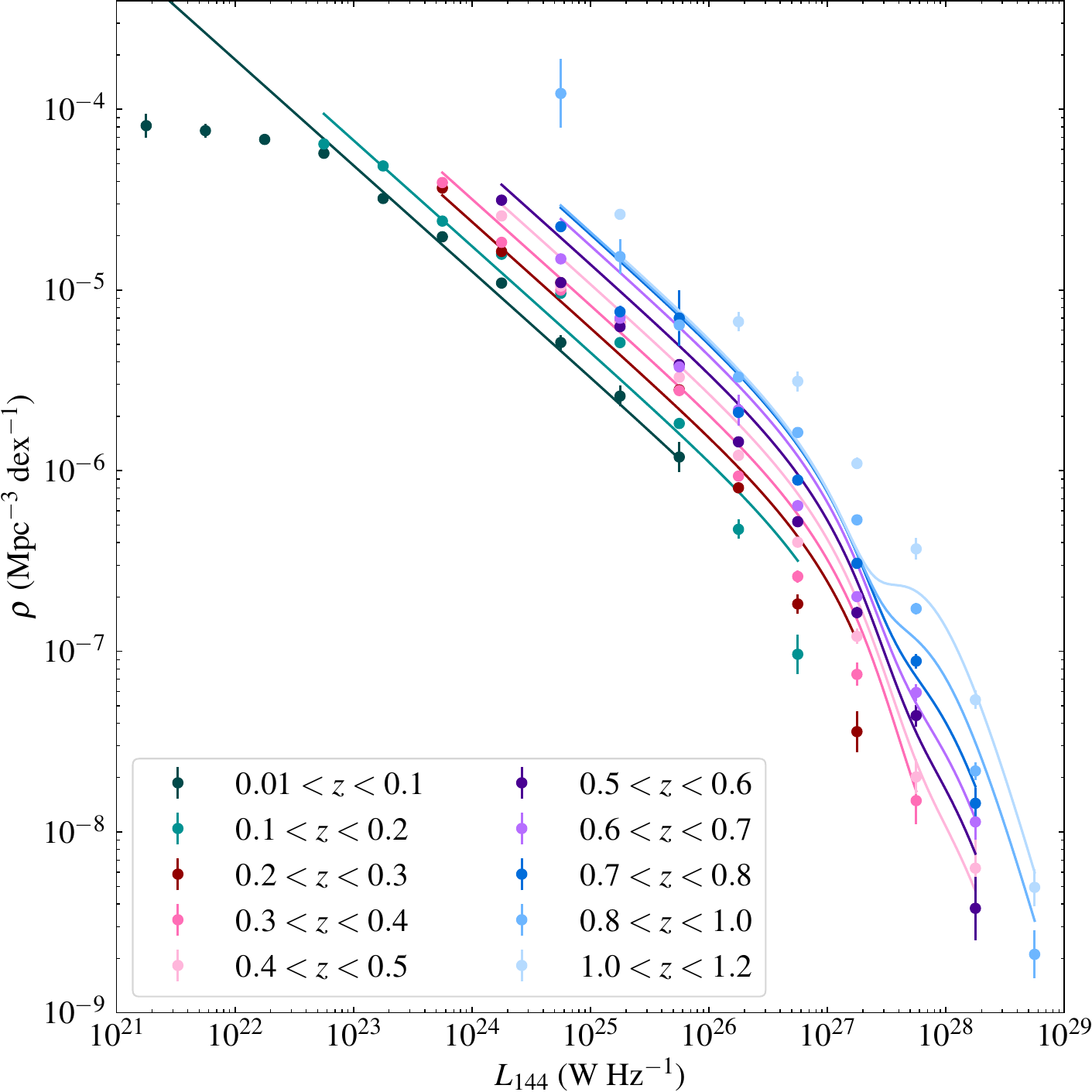}
  \caption{The luminosity function of LOFAR RLAGN as a function of
    redshift. Colours show different redshift bins: points with error
    bars are the measured values. Left: the lines show samples from the
    MCMC inference of the parameters of a dual power-law model. Right: the lines show the predictions of the models of \protect\cite{Willott+01}
    converted to the cosmology and luminosity units of this paper.}
  \label{fig:lf_evolution}
\end{figure*}

More interesting is the redshift evolution of the whole RLAGN population,
seen in Fig.\ \ref{fig:lf_evolution}. As we have such large numbers of
sources, we can construct precise luminosity functions in very small
redshift bins compared to previous studies, and hence can carry
  out a much more model-independent analysis of the cosmic evolution
  of the population than is generally the case in earlier studies. We binned the RLAGN sample in redshift out to
$z=1.2$ in steps of $\Delta z = 0.1$ out to $z=0.8$ and $\Delta z =
0.2$ thereafter, reflecting the increased uncertainties on the
photometric redshifts at the high-$z$ end. We then fitted each luminosity
function with the dual power-law model used by \cite{Dunlop+Peacock90}
and MS07,
\begin{equation}
  \rho(L) = \frac{C}{(L/L_*)^{\alpha} + (L/L_*)^{\beta}}
  \label{eq:dpl}
\end{equation}
where $C$ gives the normalization of the luminosity function at $L
\approx L_*$, $L_*$ is the characteristic break luminosity and
$\alpha$ and $\beta$ are the power-law indices at low and high
luminosity respectively. We used the {\tt emcee} sampler
\citep{Foreman-Mackey+13} to carry out a Markov-chain Monte Carlo
(MCMC) fit to the data above $L_{144} = 3 \times 10^{22}$ W Hz$^{-1}$ for each redshift bin with a $\chi^2$
likelihood function, leaving all four parameters free: $C$ and $L_*$ have uninformative uniform priors
in log space, while we require $0<\alpha<1$ and $1<\beta<2.5$ to
ensure that the power-law parameters are not degenerate. Credible
intervals on the fitted parameters, and their best value estimates,
were then derived from the (16, 50, 84)th percentiles of samples on
the marginalized distribution for each parameter. 100 random samples
from the {\tt emcee} run (after the burn-in samples where the MCMC
code has not yet converged are discarded) are overplotted on the data
in Fig.\ \ref{fig:lf_evolution}, where it can clearly be seen that the
higher-redshift bins have systematically higher-luminosity cutoffs and
higher normalization.

\begin{figure*}
    \includegraphics[width=\linewidth]{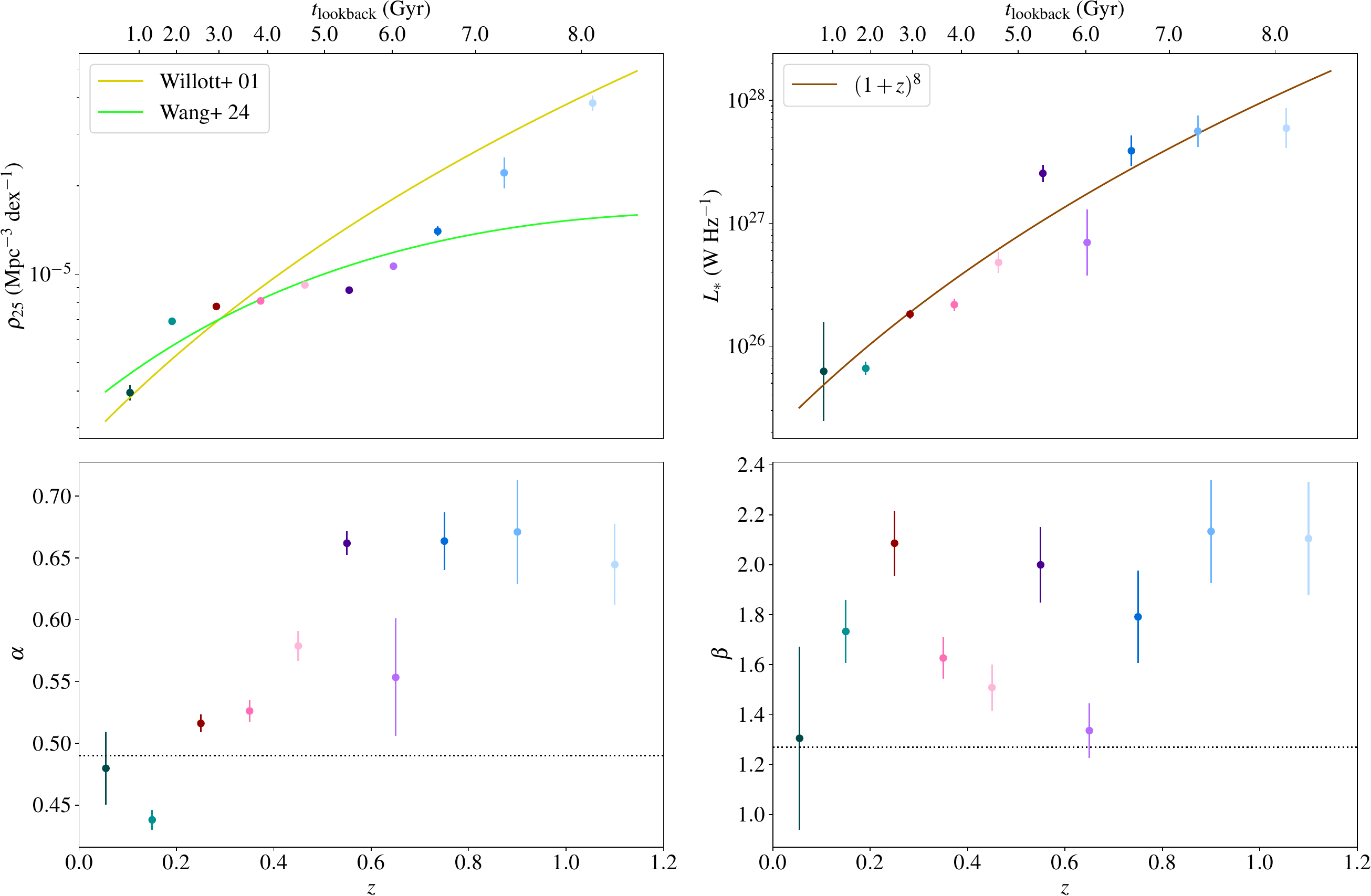}
    \caption{Evolution of the fitted parameters of a dual power-law
      model as a function of redshift. Top left: $\rho_{25}$, the normalization of
      the luminosity function at $10^{25}$ W Hz$^{-1}$, with overlaid
      curve of $(1+z)^{3.48}$ as expected from the models of
      \protect\cite{Willott+01}, together with the fitted pure density
      evolution model of \protect\cite{Wang+24}, both normalized to
      align with the data. Top right:
    the characteristic turnover luminosity of the dual power-law
    luminosity function, $L_*$, with overlaid curve of $(1+z)^{8}$ to show the strong redshift dependence of this parameter. Bottom left and right: the parameters
    of the two power laws $\alpha$ and $\beta$ respectively, where the
    dotted lines show the MS07 values of these parameters. The
    $x$-axes show redshift and lookback time (light travel time).
    Colours of the points are as in Fig.\ \ref{fig:lf_evolution}.
    Error bars show the one-dimensional 68 per cent confidence
      credible intervals on the derived parameters, and thus do not account
      for correlations between parameters.}
    \label{fig:lf_parameters}
    \end{figure*}

The evolution of the population is made quantitative in
Fig.\ \ref{fig:lf_parameters} where we plot the parameters of the fit
as a function of redshift: as $L_*$ and $C$ are correlated, we
illustrate the evolution of normalization by plotting $\rho_{25}$, the
comoving source density at a fixed luminosity of $10^{25}$ W
Hz$^{-1}$, inferred from the model fits. It can be seen that the
inferred $L_*$ increases by nearly three orders of magnitude over the
redshift range sampled here. It is hard to see how redshift/optical ID
incompleteness could bias this evolution of $L_*$ in this positive
direction, and, viewed as pure
luminosity evolution, it is very strong
($\log(L_{*,\mathrm{max}}/L_{*,\mathrm{min}})/\log(1+z_\mathrm{max})
\approx 8$ out to $z=1$). Although it is possible that there is contamination from
luminous SFGs that are not excluded from our RLAGN selection because
they fall above the luminosity limit, we do not expect them or
  contaminating `radio-quiet AGN' to be
important at the very high break luminosities seen at high $z$.
$\rho_{25}$ also evolves positively with redshift.

A useful comparison
is with the models of \cite{Willott+01}, who use a two-population
model for low-luminosity and high-luminosity RLAGN (intended to
represent RI and RE objects in our terminology respectively). Taking
their model C and converting to our cosmology, we find reasonable agreement
with the data at low $z$, as shown in
  Fig.\ \ref{fig:lf_evolution}, but relative to us they overpredict the density
evolution of low-luminosity objects at low $z$ and underpredict it at
the highest $z$, while coming close to the
observed evolution of high-luminosity objects\footnote{The slight
kink in the predicted LF at high luminosity and redshift is a
consequence of the two-population model used by \cite{Willott+01} and
is present in the original paper.}. This is perhaps not
surprising as low-luminosity objects at high $z$ would have been
poorly constrained by the data they used, which had a much higher flux
density limit.

More recent work \citep[e.g]{Smolcic+17,Wang+24} has used deep fields
to study cosmological evolution of the low-luminosity population out
to high redshift. Relative to our work, these studies generally have
smaller numbers of AGN sources and do not probe the most luminous
objects. There is therefore a degeneracy between density and
luminosity evolution in these samples that is not present in our work,
where the wide area ensures that we have a wide range of radio
luminosities for every redshift bin. On the other hand, the deeper
optical and radio data available for deep fields means that they can
probe to considerably higher redshift than is possible for us. To
limit the numbers of degrees of freedom, deep field studies have
tended to fix the parameters of Eq.\ \ref{eq:dpl} to those determined
in the local universe, e.g. by MS07, giving a local luminosity
function $\rho_0(L)$, and then fit for luminosity or density evolution
of the form
  \begin{equation}
    \rho(z,L) = (1+z)^{\alpha_D}
    \rho_0\left(\frac{L}{(1+z)^{\alpha_L}}\right)
  \end{equation}
  where $\alpha_D$ and $\alpha_L$ parametrize the density and
  luminosity evolution, respectively, and can be constants or
  functions of $z$. A key difference between these analyses and ours
  is that our fits are strongly inconsistent with constant values for
  the parameters $\alpha$ and $\beta$ in Eq.\ \ref{eq:dpl}, while they
  agree with MS07 at $z=0$. Indeed, the evolution of our break
  luminosity (Fig.\ \ref{fig:lf_parameters}) is largely driven by the
  apparent steepening in $\alpha$ with redshift, though there is some
  positive evolution even if we fix $\alpha$ to its local value. The
  evolution of number density at fixed luminosity is relatively
  independent of these choices, though, and as seen in
  Fig.\ \ref{fig:lf_parameters}, we are in reasonable agreement with
  the parametrization of \cite{Wang+24} out to $z\approx 0.8$, but
  then infer stronger evolution of the AGN number density than they
  do, more in line with that of \cite{Willott+01}. This difference is
  even stronger if we force the parameters $L_*$, $\alpha$ and $\beta$
  to have their local values. Given the relatively coarse redshift
  binning necessarily used in the deep-field work, it seems possible
  that the discrepancy may indicate a more complex evolution of the
  RLAGN population, with a less smooth dependence on cosmic time, than
  is visible in small-sample studies.
      
We can also compare with the work of \cite{Kondapally+22} using the
LoTSS deep fields, who find
only modest positive luminosity evolution and negative density
evolution in what they class as the LERG population in bins of $0.5 <
z < 1.0$ and $1.0 < z < 1.5$ --- at the highest redshifts it seems
possible that the discrepancy between this and our results is due to
the higher fraction in our sample of rare but luminous HERGs, which
have been found to evolve strongly in earlier studies
\citep{Best+14,Pracy+16}. Again, detailed comparison with earlier work that
looks at the HERG/LERG difference will need to await spectroscopic
source classifications. With deeper optical/IR data for
identifications and more spectroscopic or photometric redshifts, as
will be provided by {\it Euclid} among other facilities, it should be
possible to push an analysis of this type in the wide fields, with
excellent statistics, out to the point where the space density of
RLAGN starts to decline again.

\subsection{Host galaxy masses and radio luminosity dependence}
\label{sec:masses}

Host galaxy stellar mass estimates\footnote{Stellar mass estimates are as described by
H23 and \cite{Duncan22}, and are based on the optical and {\it WISE}
band 1 and 2 photometry, making use of SED fitting of a set of
parametric star-formation histories derived using the
\cite{Bruzual+Charlot03} stellar population synthesis models and a \cite{Chabrier03} initial mass function.} are available for around 73 per cent of
the RLAGN sample, excluding any quasars; for the sample with $z<1.2$ 81
per cent of sources have mass estimates that we treat as reliable,
following H23. The mass
estimate success rate is nearly 100 per cent for $z\la 0.7$ and falls
off with redshift above that, as expected as the signal to noise of
the optical data deteriorates. While a more complete set of mass
estimates would be ideal, the data we have allow us to explore the
relationship between radio properties, source classification, and galaxy mass.

\begin{figure*}
  \includegraphics[width=0.48\linewidth]{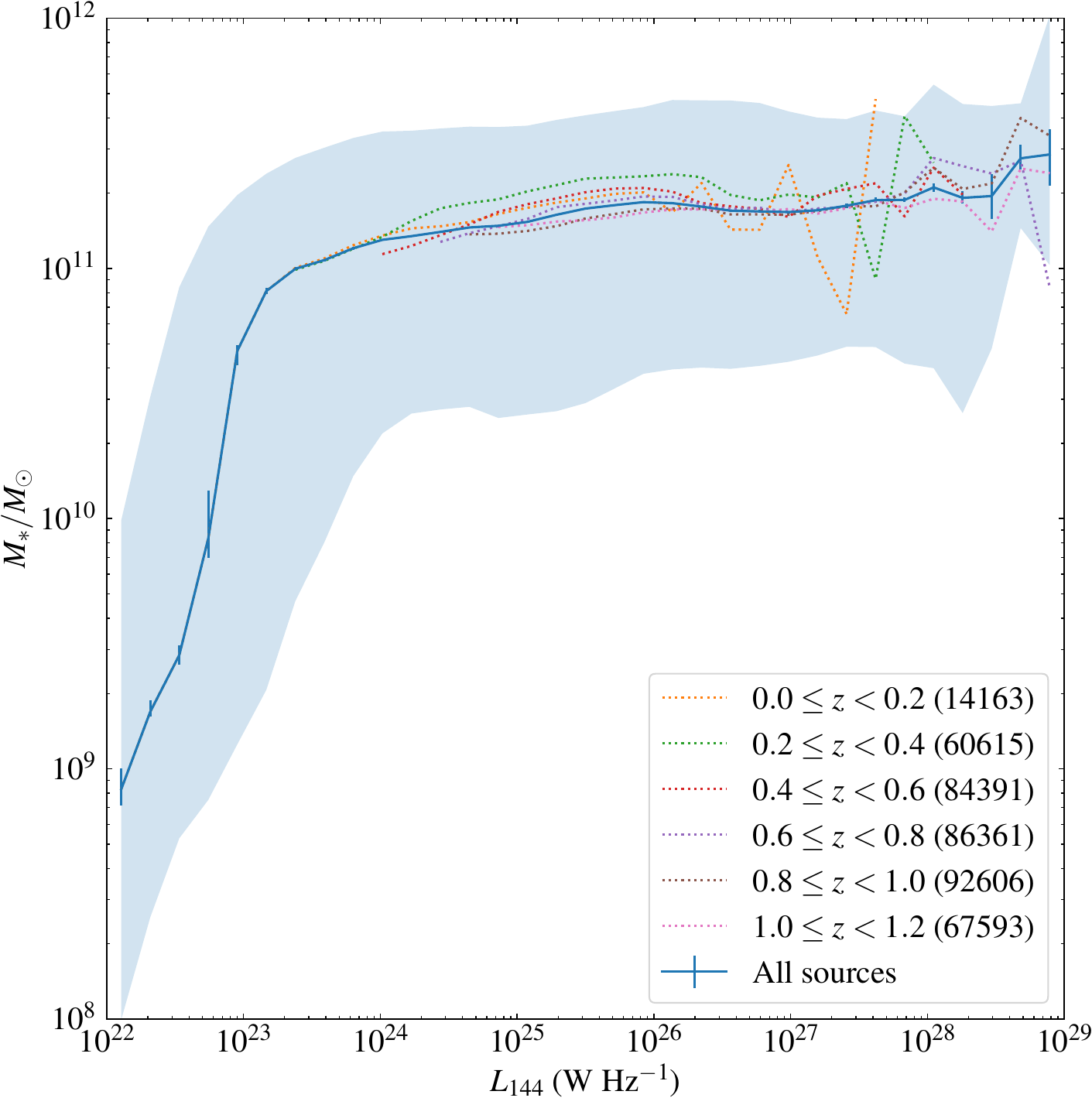}
  \includegraphics[width=0.48\linewidth]{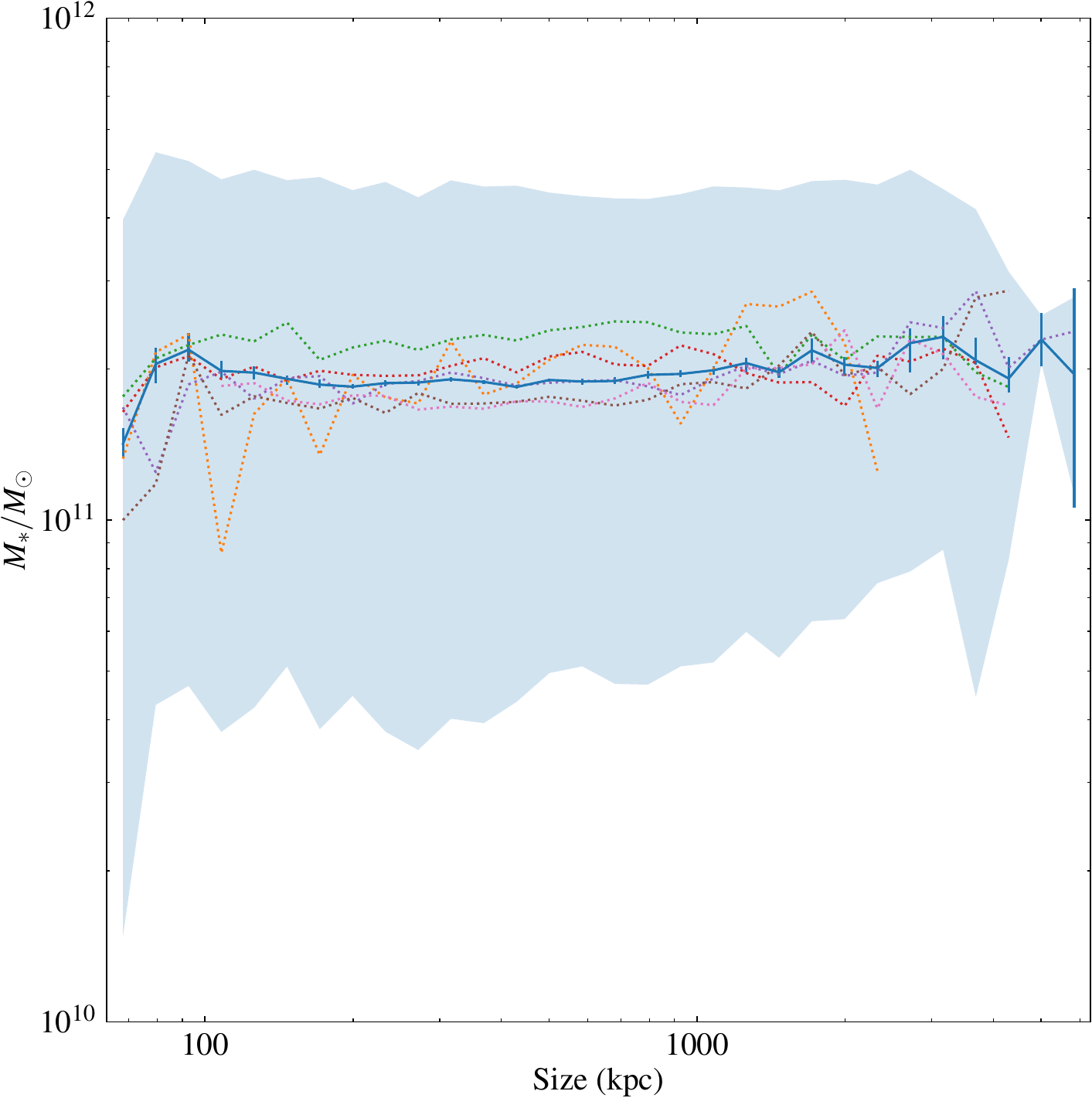}
  \caption{Host galaxy stellar mass estimates as a function of radio
    luminosity (left) and source linear size (right). Only sources
    with {\tt flag\_mass==True} are plotted in both figures, and a
    redshift cut $z<1.2$ is imposed. Both panels show binned median
    (logarithmic) masses together with their $1\sigma$ bootstrap
    uncertainty (line and error bars) together with the 5-95
    percentile range of the mass estimates to give a sense of the
    breadth of the distribution (shaded area). Both panels also
    present a breakdown of the population into redshift bins (coloured
    lines: error bars not shown for clarity). The left-hand figure
    shows all 405,729 RLAGN that meet our selection criteria. In the
    right-hand figure only the 49,961 resolved sources with $L_{144} >
    10^{25}$ W Hz$^{-1}$ are shown. Note the different scales on the
    $y$-axes of the two plots.}
  \label{fig:massls}
\end{figure*}

The left panel of Fig.\ \ref{fig:massls} shows the mean mass of host
galaxies as a function of radio luminosity. Our large sample gives
excellent statistics, with the median being very accurately measured
in spite of the large dispersion in masses for any given radio
luminosity. We see that there is a very strong dependence of mean mass
on luminosity for low luminosities, below $L_{144} \approx 10^{23}$ W
Hz$^{-1}$ --- it is possible that this is an effect of incompleteness
at the low luminosity end, since only objects above $7 \times 10^{24}$
W Hz$^{-1}$ can be seen over the full redshift range shown here.
Alternatively, it may be an effect of the presence of non-jetted
  radio AGN in the population at low radio luminosity as discussed in
  Section \ref{sec:nature_rl}.
However, the characteristic mass flattens out above this radio
luminosity, and thereafter the mean mass is between 1 and $2 \times
10^{11}$ $M_\odot$ over 4-5 orders of magnitude in radio luminosity,
with the scatter being broad but consistent over this range. This of
course is another view of the well-known $K$-$z$ relation for radio
galaxies \citep[e.g.][]{Lilly+Longair84,Willott+03,Rocca-Volmerange+04} but here seen in
physical terms. The trend in the median, in the sense that galaxies
with $L_{144} = 10^{28}$ W Hz$^{-1}$ are around a factor 2 more
massive on average than galaxies with $L_{144} = 10^{24}$ W Hz$^{-1}$,
is shallow but clearly significant, and is not a redshift effect since
it is seen over a broad range in redshift. Host galaxies are also
typically more massive, for a given luminosity, at lower redshift, but
the difference here is less than a factor 2 across the entire redshift
range we can sample (i.e. about half of cosmic time). It is
interesting that the highest masses are seen at $0.2<z<0.4$ and not
$z<0.2$, but this may be a selection effect in the sense that the most
massive galaxies are simply rarer in the lowest-redshift bins. The
differences that we see are in line with the differences in $K$-band absolute
magnitude seen by \cite{Willott+03}, assuming that the magnitude
simply scales with stellar mass, and with the results of
\cite{Williams+Rottgering15}. Considering these typical host galaxy
masses to correspond to black hole masses as in the relationship
derived for nearby ellipticals \citep{Kormendy+Ho13} then we can say
from the range of jet powers in our sample (Section \ref{sec:pdd})
that the radio galaxy hosts in the region where the mass is
independent of radio luminosity span a broad range of Eddington
ratios, between $\sim 10^{-5}$ and $\sim 1$. We will return to this
point in a future paper when individual per-source jet power estimates
are available.

% plot_agn_mass_lum.py
% plot_agn_mass_size.py

The right-hand panel of Fig.\ \ref{fig:massls} also shows that for
resolved sources there is essentially no dependence of host galaxy
mass on size. Sources that have a well-measured length $>100$ kpc tend
to be a little more massive (a factor $\sim 1.5$) than the population
above $L_{144} = 10^{25}$ W Hz$^{-1}$ as a whole, but there
is no evidence that the host galaxies of the extreme Mpc-scale giants
are in any way different from their smaller counterparts. This is also
not a redshift effect. Similar conclusions were reached on the basis
of the K-band absolute magnitude distribution by H19, but our sample
is much larger.

%\begin{figure}
%  \includegraphics[width=1.0\linewidth]{agn_mass_lum_ccc-crop.pdf}
%  \caption{As Fig.\ \ref{fig:massls}, but showing objects with $z<0.5$
%    classified using the {\it WISE} colour-colour plots.}
%\label{fig:masslccc}
%\end{figure}

%Finally, Fig.\ \ref{fig:masslccc} shows the mass-luminosity relation
%for objects with $z<0.5$ (to minimize the effects of the different
%redshift distributions) and classifications from the colour-colour
%diagram as described in Section \ref{sec:re_selection}. We see that
%the median masses of the objects in the RI class are systematically high with
%respect to the RE objects across the luminosity range, though by no
%more than about a factor 2 in the region above $L_{144} \approx
%10^{24}$ W Hz$^{-1}$. The mass difference is in the sense and of the
%magnitude of what has been seen in other studies using absolute
%magnitudes \citep[e.g.][]{Hardcastle+13}. Interestingly, the sharp
%ramp up in mass with luminosity below $10^{24}$ W Hz$^{-1}$ appears to
%be driven by the RE objects, with the RI objects' distribution being
%smoother. We note that the large fraction of unclassified objects
%most closely follow the RE rather than the RI trend, which would be
%unexpected if we think that the unclassified objects are mostly RI as
%discussed in Section \ref{sec:ccc_check}. There may be a
%selection effect here, in that lower-mass galaxies are less likely to
%have the {\it WISE} detection needed for classification; a full
%investigation again requires emission-line classifications from WEAVE.

\subsection{Flat-spectrum cores in extended sources}
\label{sec:cores}

\begin{figure*}
  \includegraphics[width=0.48\linewidth]{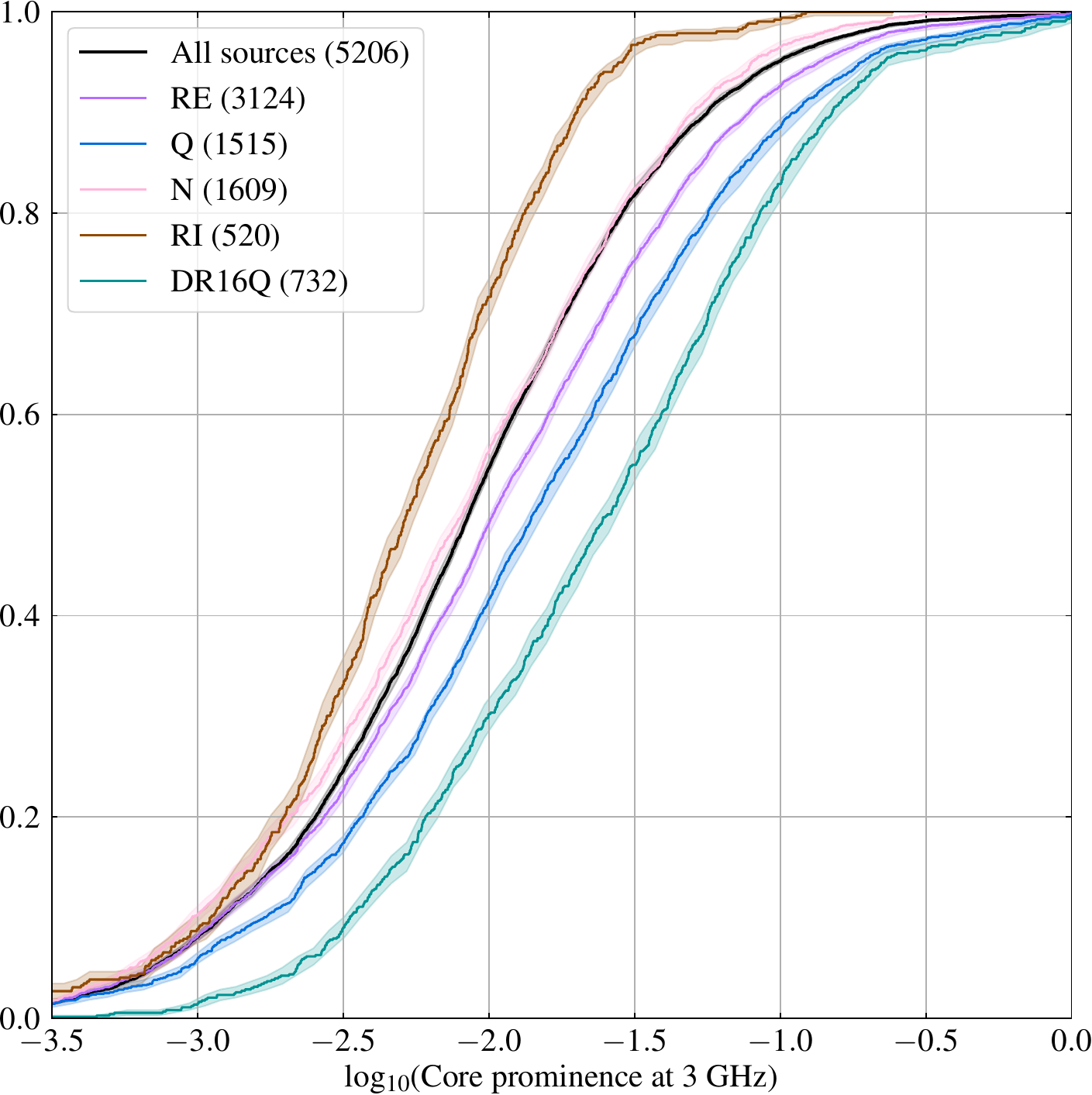}
  \includegraphics[width=0.48\linewidth]{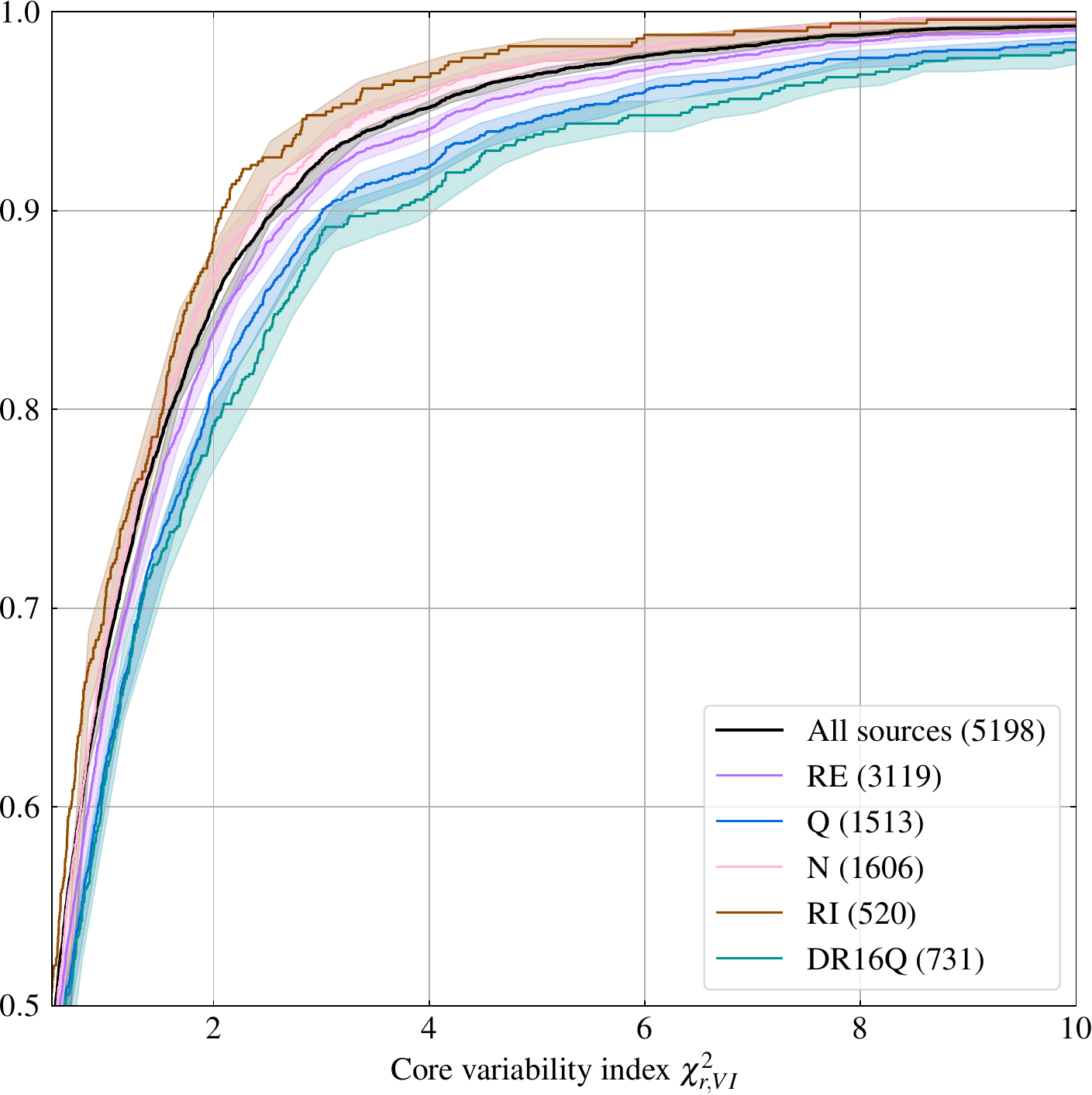}
  \includegraphics[width=0.48\linewidth]{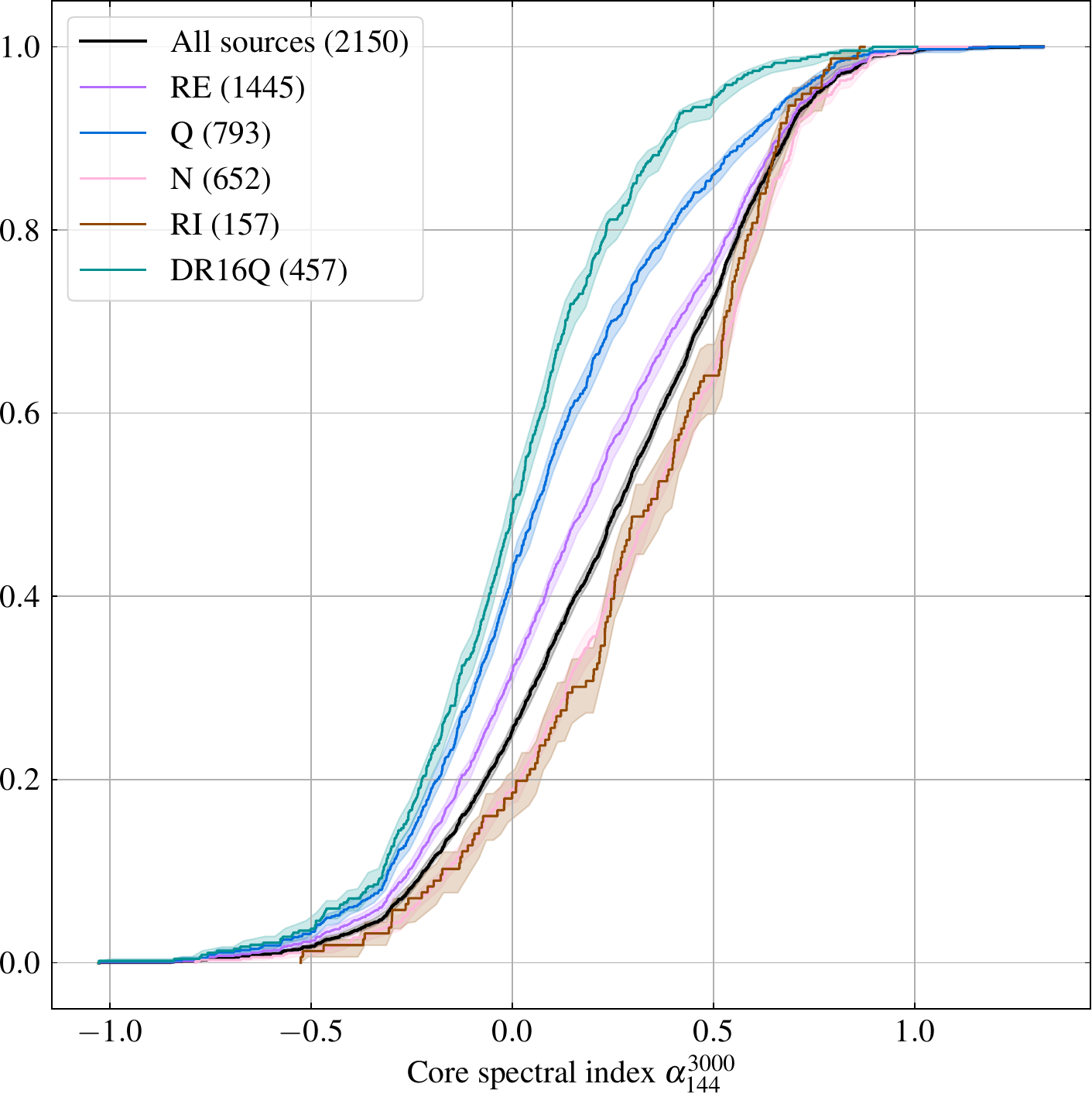}
  \includegraphics[width=0.48\linewidth]{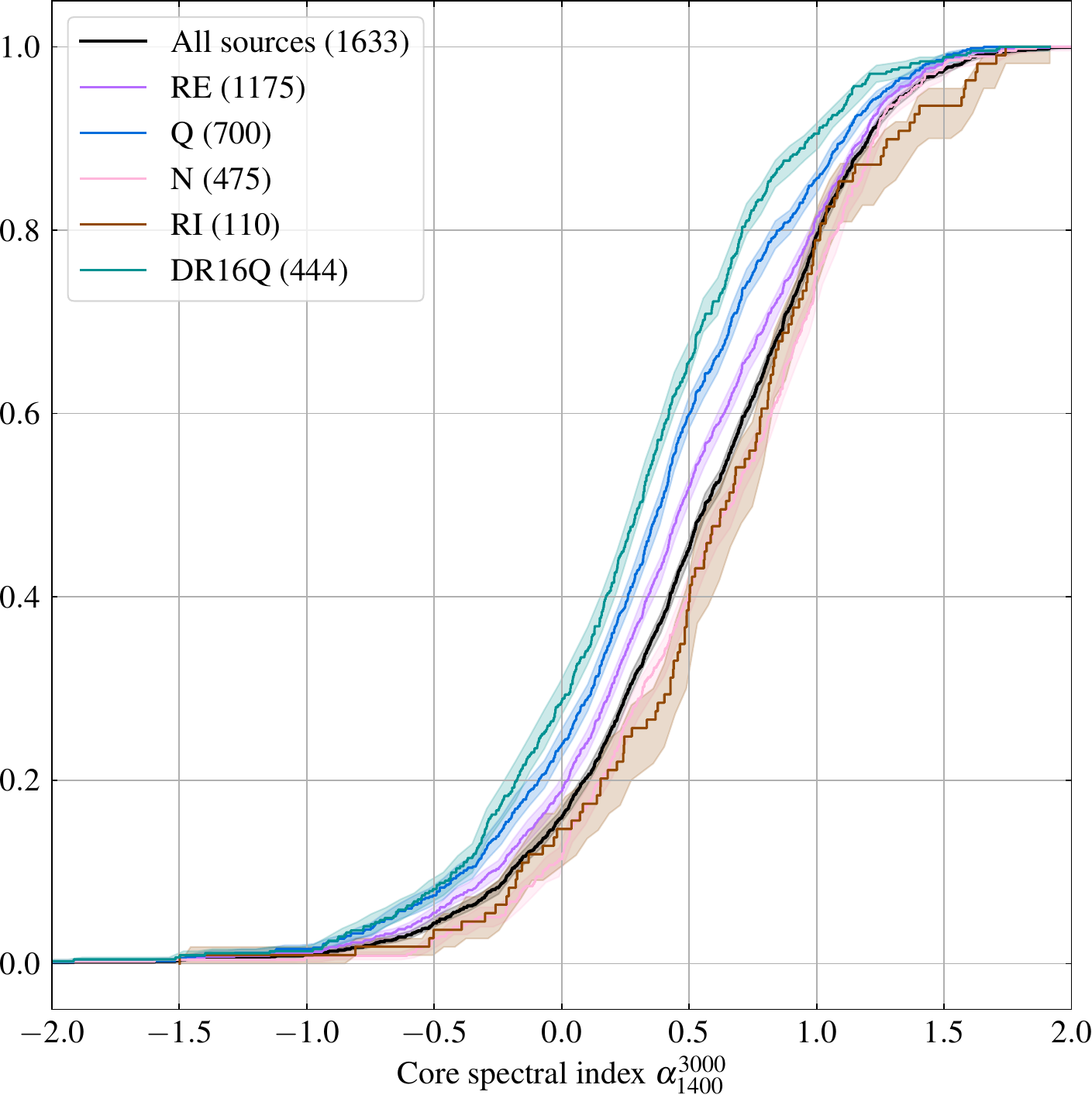}
\caption{Cumulative distribution functions of core-related quantities
  separated by classification. Each line shows a CDF for a
  quantity (in some cases restricted to the region of interest) and
  colours indicate different subsamples classified from the {\it WISE}
  colour-colour diagram as described in Section
  \ref{sec:re_selection}. Subsamples are indicated by the
  abbreviations RE (radiatively efficient) and RI (radiatively
  inefficient, i.e. LERGs): the RE subsample is further divided using
  the {\it WISE} colour into narrow-line radio galaxies (N) and
  quasars (Q). SDSS DR16 quasars are also plotted; almost all of these
  are also in Q. For each line a region of statistical
  uncertainty is indicated by a shaded area indicating the $1\sigma$ range of the CDF derived from 100
  bootstrapped samples from the same dataset. The
  plots show: top left, base 10 logarithm of the core prominence,
  defined as the VLASS 3-GHz flux density divided by LoTSS total flux
  density; top right, variability index as described in the text;
  bottom left, core spectral index between VLASS and LoTSS; bottom
  right, core spectral index between VLASS and FIRST.}
\label{fig:cores}
\end{figure*}

In the study of compact or flat-spectrum features of RLAGN, the VLASS
survey offers important complementary information to LoTSS. To give an
example of how it can be used we selected a sub-sample of the RLAGN with
$L_{144} > 10^{26}$ W Hz$^{-1}$ and with largest angular size $>60$
arcsec in order to look for emission from the compact radio source
coincident with the optical host galaxy, known as the radio `core'.
The luminosity cut ensures that we will mostly be looking at FRII-type
objects \citep{Fanaroff+Riley74} and the angular size cut leverages
the fact that VLASS spatially filters structures with scale $\ga 30$
arcsec, helping to reduce contamination from extended emission. The
luminosity cut also puts us in the range where {\it WISE}
colour-colour classifications appear to be most consistent with
earlier spectroscopic work (Section \ref{sec:ccc_check}). These
two selections applied to the inclusive RLAGN sample give us 8,862
objects in total.

\begin{figure}
  \includegraphics[width=\linewidth]{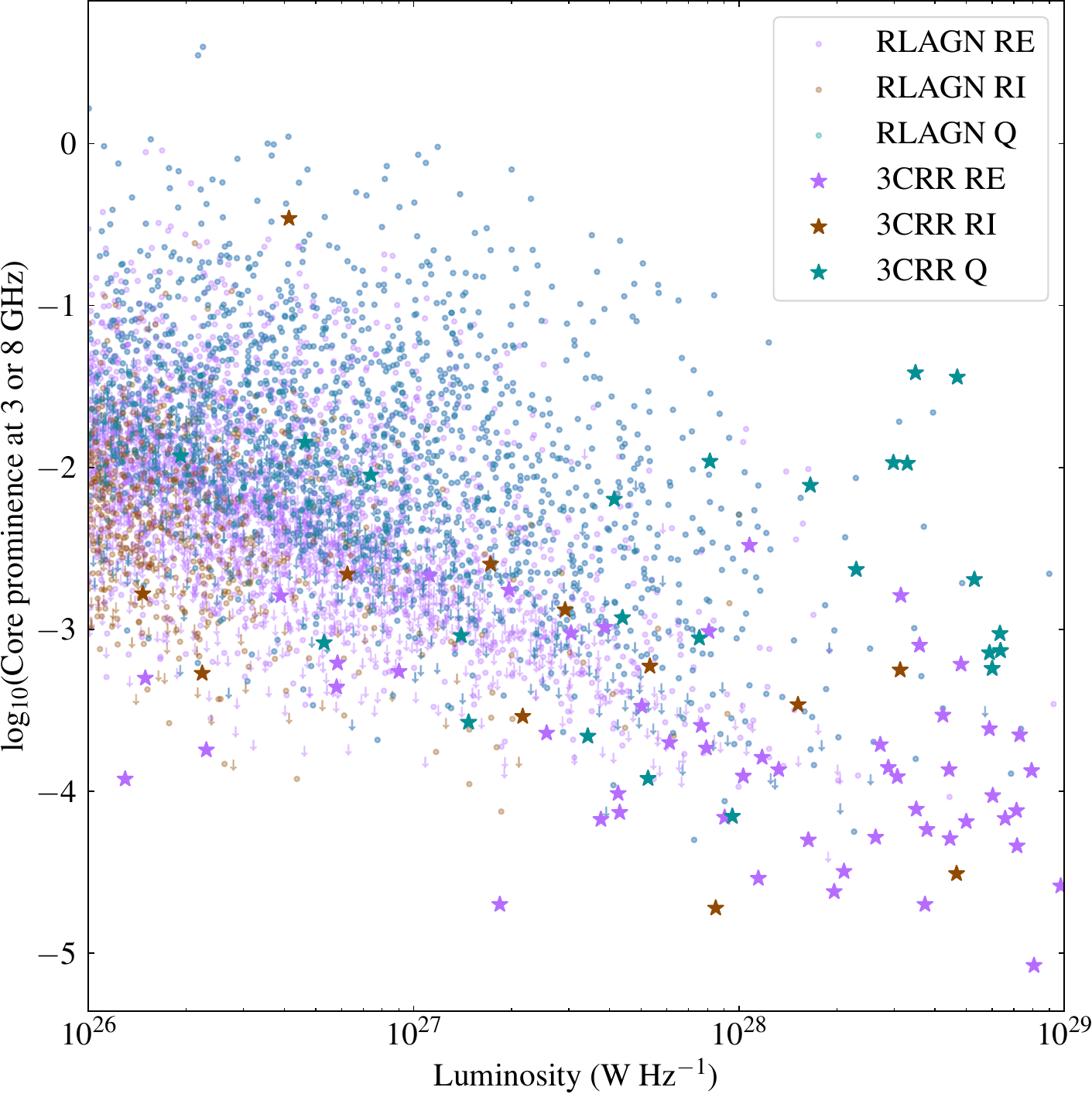}
  \caption{VLASS core prominence as a function of luminosity for
      the sample of large, luminous objects described in the text.
      $5\sigma$ upper limits are denoted with arrows. Overplotted are
      the core prominences (mostly at 8 GHz) tabulated for $z<1.0$
      3CRR FRII sources by \protect\cite{Mullin+08}. A general
      downward trend in core prominence with luminosity is seen, with
      the 3CRR objects lying in a location consistent with the
      trend.}
  \label{fig:core_luminosity}
\end{figure}

We selected all VLASS quick-look observations available in June 2024
and initially stacked all of them around the position of the radio
source to give the deepest image. The flux density of the core is then
taken to be the flux density of the brightest pixel within 2 arcsec of
the optical ID position, so long as it is detected at $>5\sigma$. This
gives a total of 5,206 detections at 3 GHz (58 per cent): the
detection threshold is dependent on the field and on the number of
VLASS images available, but the mean rms noise in the VLASS images is
76 $\mu$Jy beam$^{-1}$, so that we are sensitive to core flux
densities typically $\ga 400$ $\mu$Jy at 3 GHz. We further
cross-matched the positions of the optical IDs to the LoTSS Gaussian
catalogue and the FIRST catalogue, requiring a compact component
(deconvolved size $<6$ arcsec) within 2 arcsec of the position of the
optical ID. This gives respectively 2,462 and 1,641 core candidate
detections at 144 and 1400 MHz (28 and 18 per cent). Core prominence
for our sample can be defined by taking the ratio of the core flux
density at a given frequency to the total flux density at 144 MHz,
with the caveat that the absolute values of prominence will depend on
the frequencies used. Spectral indices can be calculated for cores
detected at more than one frequency, bearing in mind that of course we
do not have simultaneous observations at the three frequencies so that
the spectral indices will be affected by variability.

Finally, the fact that we have multiple
images for essentially all of the VLASS sources (5,198) allows us to
compute a variability index, defined as the reduced $\chi^2$ for a fit
of the mean flux to all the individual flux measurements:
\[
\chi^2_{r,VI} = \frac{1}{N-1}\sum \frac{(S_i - \mu)^2}{\sigma_i^2}
\]
where $\mu = \bar{S_i}$ and we take $\sigma_i = 0.1S_i$ to take
account of the fact that the dominant uncertainty in these images is
likely to be absolute flux calibration which it is reasonable to set
at the 10 per cent level. Apparently significant
variability at 3 GHz is detected in a non-negligible fraction of the
sources, with $\chi^2_{r,VI} >3$ for 387 sources. Some of these are
likely to be spurious due to such factors as incorrect source
identification or poor fidelity in the VLASS images: we discuss the
possible contamination fraction and show some images in Appendix
\ref{app:variable}. However, we believe that in general we are
seeing real variability in the VLASS data.

Distributions of the core prominence,
variability index and spectral index are plotted in
Fig.\ \ref{fig:cores}. There is no strong dependence of core variability or spectral properties on source luminosity or physical
size, though a downward trend of core prominence with 144-MHz
luminosity is observed (Fig.\ \ref{fig:core_luminosity}). The most striking results of this analysis
come from combining it with the RLAGN classification of Section
\ref{sec:re_selection}. 70 per cent of the sources with VLASS-detected
cores have a classification from the {\it WISE} colour selection
described in that section. Key points to note are:
\begin{enumerate}
  \item Quasars have significantly more prominent, more variable and
    flatter-spectrum cores than the sample as a whole or than
    NLRG (non-quasar RE objects).
   \item Quasar selection with the {\it WISE} colour-colour diagram
     (specifically $W1-W2>0.75$)
     selects objects that are generally less extreme than, but
     similarly positioned to, the SDSS DR16 quasars. This gives some
     confidence that the colour-based quasar selection is working as
     expected, although there may be some contamination.
   \item Radiatively inefficient objects (LERGs) have systematically
     low core prominence in this sample but behave similarly to
     NLRG in the other plots.
\end{enumerate}

The absolute values on the spectral index distribution plots should be
treated with caution since there are many non-detections and, as VLASS
is the most sensitive of the surveys to flat-spectrum cores, the spectral indices measured
here will be strongly biased towards relatively steep-spectrum cores
which can be detected both by VLASS and by one of the other two
surveys. Similarly, too much should not be read into the differences
in spectral indices for quasars and other sources since, in the
absence of a complete sample, we do not know for certain whether this
is simply an effect of quasars having more prominent cores, which
permits the detection of flatter-spectrum counterparts at other
frequencies\footnote{In order to get a sense of whether this is so for
the spectral index and variability plots, we carried out the same
analysis with VLASS core flux densities restricted to a narrow range
between 5 and 20 mJy. The results were qualitatively very similar and
a significant excess of flat-spectrum, variable quasars over NLRG was
detected, which suggests that this bias is not solely responsible for
the observed effect.}. But the overall results for quasars and
NLRG are qualitatively very consistent with the expectations
from unified models, in which quasars' cores are more strongly
Doppler-boosted and hence would be expected to be more prominent
\citep{Orr+Browne82}, more variable and (assuming a downward-curved
intrinsic spectrum) flatter-spectrum. Similar differences between the
core prominence distributions of quasars and NLRG have of course been
seen before in complete samples such as 3CRR
\citep{Hardcastle+98,Mullin+08,Marin+16} but this is by far the
largest sample, albeit an incomplete one, in which such an analysis
has been possible. The variability effect has not previously been
detected to our knowledge, although it is a straightforward
expectation from the unified model, and this illustrates the value of
the time-domain information provided by VLASS. Quantitative tests
  of unified models, and measurements of key parameters such as the
  effective beaming Lorentz factor as in \cite{Mullin+08}, would
  require us to define a complete sample, ideally with spectroscopic
  AGN classification and also without the angular size limits imposed by our
  use of the VLASS data, and we hope to be in a position to carry out
  such a study in the coming years.

The core prominences we find are quantitatively systematically
  high with respect to 3CRR samples (we show the relationship between
  our sample and the core prominence data from
  \cite{Mullin+08} in Fig.\ \ref{fig:core_luminosity})
  but this is easily understood in terms of the luminosity dependence
  of core prominence observed in the data, together with the fact that
3CRR objects generally have deeper observations available which can
probe down to lower core prominences than we can detect for our targets.
The systematically low core prominence for RI objects (LERGs) is
perhaps surprising given that studies of low-$z$ 3CRR sources have
typically found similar core prominences to those of NLRG and
quasars/BLRG \citep{Hardcastle+98}. In a complete sample, and with all
other things being equal, we would expect RI objects to be observed to lie
at all angles to the line of sight and hence to have a similar
distribution to the combined RE population. Several possible
explanations can be advanced. Firstly, we can note that highly beamed
LERGs will appear as blazars, and these will almost certainly be
under-represented in our sample, since firstly they are unlikely to
have a redshift either from spectroscopic or photometric methods, and
secondly they will be impossible to classify using the colour-colour
methods. By contrast, highly beamed RE objects (quasars) will if
anything be over-represented in our sample with respect to their
unbeamed counterparts. This, however, should only really affect a
small fraction of the overall RI population. A more subtle effect is
that the angular size selection we apply in order to obtain a clean
core prominence estimate places constraints on source age (larger
sources are in general older) and so it is possible that if LERGs in
the sample have overall lower jet power, they are typically older at
the point where they meet our luminosity and angular size selection
criteria, and hence have lower core prominence. Testing this
hypothesis would require sensitive high-resolution (sub-arcsec) radio
imaging which is capable of detecting cores even in compact sources,
in order to allow as complete as possible a sample to be used in these
tests. We do note however that \cite{Whittam+16,Whittam+22} found
  RE objects (HERGs) to have a higher rate of core detection, and
  therefore presumably prominence, than RI objects, albeit with a
  different HERG/LERG separation method than in the present paper.

\cite{Chilufya+25} present a detailed study of the
properties of extended sources in the D24 catalogue, including the
distribution of VLASS/LOFAR core prominences in objects classified
spectroscopically as LERGs and HERGs.

\section{Summary and future work}
\label{sec:summary}

Using selection methods based on the optical ID catalogue for LoTSS
DR2 (H23), the spectroscopic classifications of D24, and the {\it
  WISE} magnitudes for optically identified sources, we have carried
out RLAGN/star formation separation and generated by far the largest
catalogue of candidate radio AGN to date, with close to 600,000
objects even if the likely radio-quiet quasars are excluded. The
selection is consistent with, but we believe more robust than, our
earlier work on selection of RLAGN in LoTSS DR1 (H19), while providing a
sample $\sim 25$ times larger thanks to our deeper optical data and
more robust SF/AGN selection.

We are still some way from having a complete sample in which all radio
sources above some flux threshold have optical IDs and redshifts, and
we discuss in the paper some of the limitations of the LOFAR data,
including the limited number of size measurements and the surface
brightness limits that prevent us finding large, low-luminosity
sources that presumably exist at some level.

Based on our catalogues and the work on the LoTSS deep fields, we have
presented an empirical estimate of the total sky density of
radio-excess AGN (Section \ref{sec:howmany}). Despite the sensitivity
of our survey and the high sky density of RLAGN, we estimate
that we are only identifying around 10 per cent of all the
radio-excess AGN in the sky with surveys at this depth (principally
missing faint and/or high-redshift objects). The large number of
objects in the sample allows us to model the evolution of the
luminosity function in narrow bins, and we have shown that there
appears to be strong positive density and luminosity evolution of the
radio-excess AGN population over the cosmological time to which we are
sensitive (Section \ref{sec:lf}). Our results are different in detail
from some widely used modelling based on smaller samples
\citep[e.g.][]{Willott+01} as well as the latest results from
  deep fields \citep[e.g.][]{Kondapally+22,Wang+24}, emphasising the
  value of large samples in constraining this evolution.

Bearing in mind the clear caveats about completeness of the sample, we
have been able to demonstrate the value of combining the data with
ancillary data either present in our existing source catalogues (e.g.
the host galaxy mass estimates) or available over the same sky area
(e.g. the VLASS cores), obtaining results that would be much harder to
find in a smaller sample. Interesting outcomes include the
identification of a weak but significant relation between radio
luminosity and mean mass for luminous RLAGN, in the sense that the most powerful radio
galaxies are also on average the most massive (Section \ref{sec:masses}), and the detection for the first
time of spectral index and variability properties of the cores of
resolved sources that are consistent with the prediction of unified
models (Section \ref{sec:cores}).

LoTSS Data Release 3 (DR3), which will be the final LoTSS wide-area data
release and will cover almost all of the northern extragalactic sky,
is expected in 2025 and will have the potential to generate even
larger samples for detailed statistical analysis, with a factor $\sim
3$ more sources. The proposed
ILoTSS extension of the
survey\footnote{\url{https://lofar-surveys.org/ilotss.html}} will go
signficantly deeper than either DR2 or DR3
and have greatly improved capabilities for RLAGN/SF separation thanks to
the long international baselines of LOFAR. The
limiting factor is now, and will continue to be, the lack of wide-area
optical photometric and spectroscopic data deep enough to identify all
the radio sources, estimate their redshifts and carry out RLAGN/star
formation separation. Additional spectroscopic redshifts and
spectroscopic source classifications, coming in future from
WEAVE-LOFAR and DESI, and/or the availability of deeper optical data
over a wide area, such as will be provided by the {\it Euclid}
wide-area surveys, will allow all of these conclusions to be made more
robust and allow us to start to probe the full range of cosmic time
over which RLAGN are active. It is our intention to release further
versions of the catalogue covering the DR2 area as new data become
available.

One of our key aims in generating this catalogue was to enable a
statistical investigation of RLAGN feedback out to $z \approx 1$ as a function of
environment and cosmic time, and this will be presented in a followup
paper (Pierce et al.\ in prep.).

\section*{Acknowledgements}

MJH is grateful to Ski Antonucci for discussions which motivated the
work described in Section \ref{sec:cores}. We thank George Miley for
comments on an earlier draft of the paper. We thank an anonymous
  referee for constructive comments which allowed us to improve the paper.

MJH and JCSP thank the UK STFC for support [ST/V000624/1,
  ST/Y001249/1]. KJD acknowledges support from the STFC through an
Ernest Rutherford Fellowship [ST/W003120/1]. YG acknowledges PhD
studentship funding from the Open University. MAH acknowledges support
of STFC under a grant to the University of Cambridge [ST/Y000447/1].
BM acknowledges support from the STFC for an Ernest Rutherford
Fellowship [ST/Z510257/1]. DJBS acknowledges support from the STFC
[ST/V000624/1,ST/Y001028/1].

LOFAR is the Low Frequency Array, designed and constructed by ASTRON. It has observing, data processing, and data storage facilities in several countries, which are owned by various parties (each with their own funding sources), and which are collectively operated by the ILT foundation under a joint scientific policy. The ILT resources have benefited from the following recent major funding sources: CNRS-INSU, Observatoire de Paris and Université d'Orléans, France; BMBF, MIWF-NRW, MPG, Germany; Science Foundation Ireland (SFI), Department of Business, Enterprise and Innovation (DBEI), Ireland; NWO, The Netherlands; The Science and Technology Facilities Council, UK; Ministry of Science and Higher Education, Poland; The Istituto Nazionale di Astrofisica (INAF), Italy.

This research made use of the Dutch national e-infrastructure with support of the SURF Cooperative (e-infra 180169) and the LOFAR e-infra group. The Jülich LOFAR Long Term Archive and the German LOFAR network are both coordinated and operated by the Jülich Supercomputing Centre (JSC), and computing resources on the supercomputer JUWELS at JSC were provided by the Gauss Centre for Supercomputing e.V. (grant CHTB00) through the John von Neumann Institute for Computing (NIC).

This research made use of the University of Hertfordshire
high-performance computing facility and the LOFAR-UK computing
facility located at the University of Hertfordshire (\url{https://uhhpc.herts.ac.uk}) and supported by
STFC [ST/P000096/1], and of the Italian LOFAR IT computing
infrastructure supported and operated by INAF, and by the Physics
Department of Turin University (under an agreement with Consorzio
Interuniversitario per la Fisica Spaziale) at the C3S Supercomputing
Centre, Italy.

This research made use of {\sc Astropy}, a
community-developed core Python package for astronomy
\citep{AstropyCollaboration13} hosted at
\url{http://www.astropy.org/}, of {\sc Matplotlib} \citep{Hunter07},
of {\sc APLpy}, an open-source astronomical plotting package for
Python hosted at \url{http://aplpy.github.com/}, and of {\sc topcat}
and {\sc stilts} \citep{Taylor05}.

The National Radio Astronomy Observatory is a facility of the National
Science Foundation operated under cooperative agreement by Associated
Universities, Inc.

The Legacy Surveys consist of three individual and complementary projects: the Dark Energy Camera Legacy Survey (DECaLS; Proposal ID \#2014B-0404; PIs: David Schlegel and Arjun Dey), the Beijing-Arizona Sky Survey (BASS; NOAO Prop. ID \#2015A-0801; PIs: Zhou Xu and Xiaohui Fan), and the Mayall z-band Legacy Survey (MzLS; Prop. ID \#2016A-0453; PI: Arjun Dey). DECaLS, BASS and MzLS together include data obtained, respectively, at the Blanco telescope, Cerro Tololo Inter-American Observatory, NSF’s NOIRLab; the Bok telescope, Steward Observatory, University of Arizona; and the Mayall telescope, Kitt Peak National Observatory, NOIRLab. The Legacy Surveys project is honoured to be permitted to conduct astronomical research on Iolkam Du'ag (Kitt Peak), a mountain with particular significance to the Tohono O'odham Nation.

NOIRLab is operated by the Association of Universities for Research in Astronomy (AURA) under a cooperative agreement with the National Science Foundation.

This project used data obtained with the Dark Energy Camera (DECam), which was constructed by the Dark Energy Survey (DES) collaboration. Funding for the DES Projects has been provided by the U.S. Department of Energy, the U.S. National Science Foundation, the Ministry of Science and Education of Spain, the Science and Technology Facilities Council of the United Kingdom, the Higher Education Funding Council for England, the National Center for Supercomputing Applications at the University of Illinois at Urbana-Champaign, the Kavli Institute of Cosmological Physics at the University of Chicago, Center for Cosmology and Astro-Particle Physics at the Ohio State University, the Mitchell Institute for Fundamental Physics and Astronomy at Texas A\&M University, Financiadora de Estudos e Projetos, Fundacao Carlos Chagas Filho de Amparo, Financiadora de Estudos e Projetos, Fundacao Carlos Chagas Filho de Amparo a Pesquisa do Estado do Rio de Janeiro, Conselho Nacional de Desenvolvimento Cientifico e Tecnologico and the Ministerio da Ciencia, Tecnologia e Inovacao, the Deutsche Forschungsgemeinschaft and the Collaborating Institutions in the Dark Energy Survey. The Collaborating Institutions are Argonne National Laboratory, the University of California at Santa Cruz, the University of Cambridge, Centro de Investigaciones Energeticas, Medioambientales y Tecnologicas-Madrid, the University of Chicago, University College London, the DES-Brazil Consortium, the University of Edinburgh, the Eidgen\"ossische Technische Hochschule (ETH) Zurich, Fermi National Accelerator Laboratory, the University of Illinois at Urbana-Champaign, the Institut de Ciencies de l'Espai (IEEC/CSIC), the Institut de Fisica d'Altes Energies, Lawrence Berkeley National Laboratory, the Ludwig Maximilians Universit\"at Munchen and the associated Excellence Cluster Universe, the University of Michigan, NSF's NOIRLab, the University of Nottingham, the Ohio State University, the University of Pennsylvania, the University of Portsmouth, SLAC National Accelerator Laboratory, Stanford University, the University of Sussex, and Texas A\&M University.

BASS is a key project of the Telescope Access Program (TAP), which has been funded by the National Astronomical Observatories of China, the Chinese Academy of Sciences (the Strategic Priority Research Program “The Emergence of Cosmological Structures” Grant \# XDB09000000), and the Special Fund for Astronomy from the Ministry of Finance. The BASS is also supported by the External Cooperation Program of Chinese Academy of Sciences (Grant \# 114A11KYSB20160057), and Chinese National Natural Science Foundation (Grant \# 11433005).

This project, and the Legacy Survey project, make use of data
products from the Near-Earth Object Wide-field Infrared Survey
Explorer ({\it NEOWISE}), which is a project of the Jet Propulsion
Laboratory/California Institute of Technology. {\it NEOWISE} is funded by the National Aeronautics and Space Administration.

The Legacy Surveys imaging of the DESI footprint is supported by the Director, Office of Science, Office of High Energy Physics of the U.S. Department of Energy under Contract No. DE-AC02-05CH1123, by the National Energy Research Scientific Computing Center, a DOE Office of Science User Facility under the same contract; and by the U.S. National Science Foundation, Division of Astronomical Sciences under Contract No. AST-0950945 to NOAO.

%%%%%%%%%%%%%%%%%%%%%%%%%%%%%%%%%%%%%%%%%%%%%%%%%%
\section*{Data Availability}

Data are available through \url{https://lofar-surveys.org/}.



\bibliographystyle{mnras}
\bibliography{mjh.bib,cards.bib} 

\appendix
\section{Description of the catalogue}
\label{app:catalogue}

We release the full catalogue of 963,764 classified sources with flux
density $>1.1$ mJy and known redshift with this paper. It can be
downloaded from \url{https://lofar-surveys.org/dr2_release.html}.

The catalogue consists of the following entries for each source:
\begin{enumerate}
\item All entries from the optical ID catalogue of H23.
\item All entries from the catalogue of D24, including optical
    emission-line fluxes and probabilistic classifications, for those
    sources for which SDSS data is available. A boolean column {\tt ELC}
    indicates objects which have these entries.
\item All entries from the SDSS DR16 quasar catalogue, for those
    sources whose optical position matches a DR16 quasar. A boolean
    column {\tt DR16Q} indicates objects that have these entries.
\item Flux densities in Jy and spectral indices to allow the {\it WISE}
  absolute magnitude to be calculated ({\tt flux\_w1}, {\tt flux\_w2},
  {\tt flux\_w3}, {\tt alpha\_w1\_w2}, {\tt mu} (distance modulus),
  {\tt alpha\_w2\_w3}, and {\tt abs\_w3}.
\item Boolean flags {\tt SF\_EXCLUDE\_BROAD}, {\tt SF\_EXCLUDE}, {\tt
  RQQ\_EXCLUDE\_BROAD}, {\tt RQQ\_EXCLUDE}, {\tt AGN\_NARROW} and {\tt
  AGN\_BROAD} which correspond to the classifications summarized in
  Table \ref{tab:agnselect}, with the `broad' exclusion
  classifications corresponding to the `exclusive' classifications in
  that Table.
\end{enumerate}

The RLAGN catalogue used for the analysis in Section
\ref{sec:discussion} can be generated from this catalogue by taking
    {\tt AGN\_BROAD==True}.
  
Code used to create these catalogues and make the plots in this paper
is available on Github at \url{https://github.com/mhardcastle/agn-selection}. The data described in this paper
constitute version 1.0 of the catalogue, but we plan to update both
the DR2 optical ID catalogue and the {\it WISE}-classified RLAGN/SFG
catalogue as further redshifts (and hence, in some cases, additional
members of the sample) become available from DESI and WEAVE in
the coming years.

\section{Dealing with optical limits on the luminosity function}
\label{app:optvmax}

Construction of the luminosity functions required the calculation of
both radio and optical volumes, $V_{\rm max}$, within which a source
could be observed. For the majority of the sources, the
optical/near-IR detection limit provided the most constraining factor
for the source volume determination, and the considerations involved
in this process are discussed here.

Prior to construction of the luminosity functions, a $WISE$ Band 1 magnitude detection limit of $W1 < 21.2$ mag was imposed on the sample to ensure completeness, with the corresponding flux density determining the detection limit for a RLAGN host galaxy. For a given source, this set the maximum redshift out to which it would still be observable ($z_{\rm{max}}$) and the corresponding maximum volume of the observed sky area enclosed ($V_{\rm{max}}$), as required for the luminosity function calculations.

In addition to the reduction of flux density expected with increasing
source distance, determination of $z_{\rm max}$ required the redshift
dependence of the portion of the spectral energy distribution (SED)
observed in the $W1$ bandpass to be considered. This latter factor is
important, since the cosmological K-correction in the $W1$ spectral
region of a typical galaxy is negative for the range of redshifts
covered by our sample, meaning that higher flux-density spectral
regions would be observed at higher redshifts.

To account for this redshift dependence, we first used the
  \texttt{kcorrect} code (\citeauthor{Blanton+07}
  \citeyear{Blanton+07}; available on Github at
  \url{https://github.com/blanton144/kcorrect}) to obtain rest-frame
  SED templates for the RLAGN based on their observed DESI
  Legacy Survey $g$, $r$ and $z$ optical magnitudes and their $WISE$
  Band 1 magnitudes; $WISE$ Band 2 and 3 magnitudes were also used,
  when available, to constrain the template fit. Using these templates, the integrated fluxes in the spectral region covered by the $W1$ bandpass were then calculated at increasing redshifts from the target redshift outward. The ratios of these integrated fluxes relative to the target redshift values were then used as correction factors when calculating the predicted higher redshift source flux densities. After also accounting for the increasing cosmological distance, $z_{\rm max}$ and $V_{\rm max}$ values were then determined via comparison of these flux densities with the imposed $W1$ detection limit.

\section{Example images of `highly variable' sources}
\label{app:variable}

\begin{figure*}
  \includegraphics[width=0.32\linewidth]{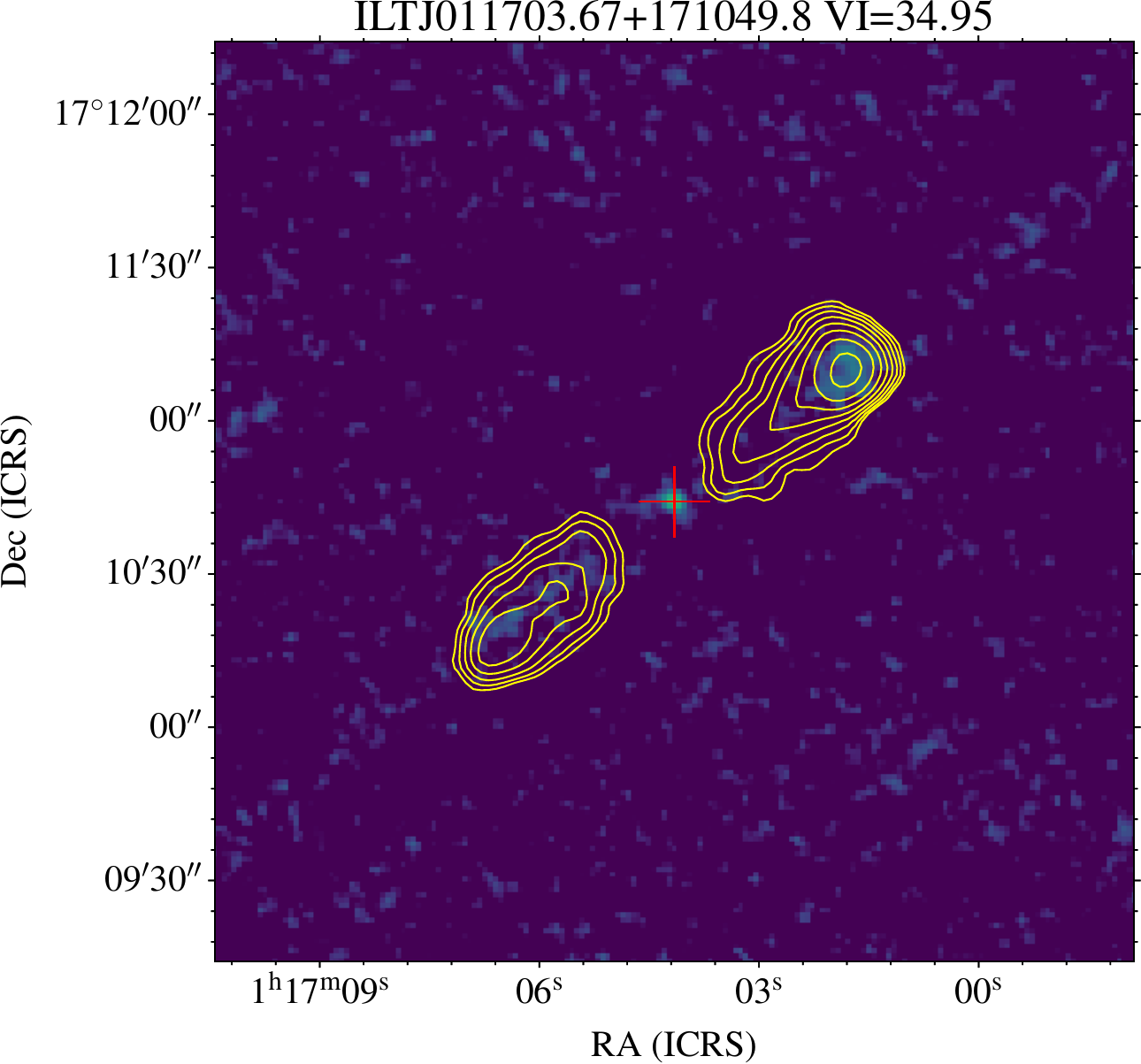}
  \includegraphics[width=0.32\linewidth]{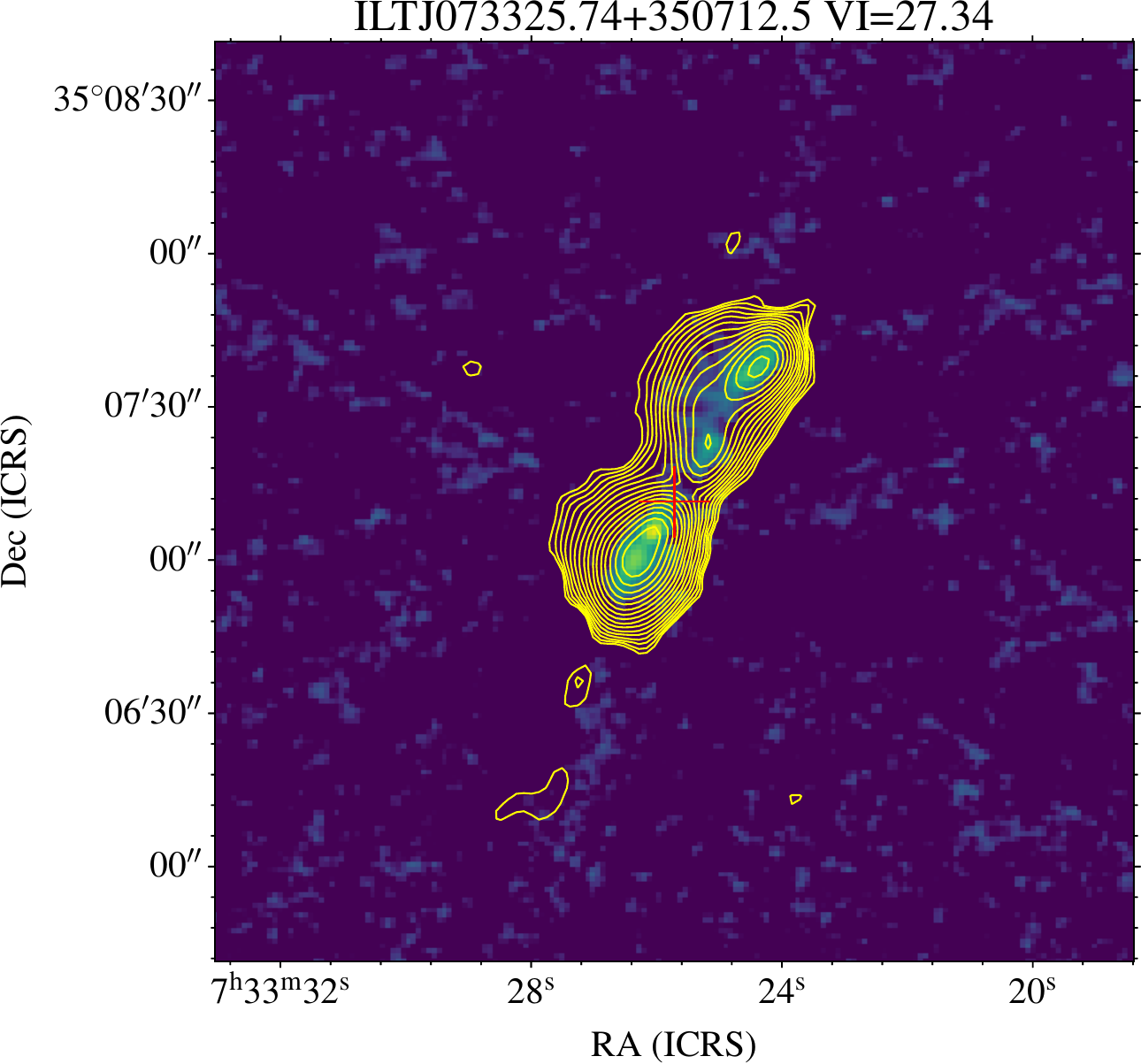}
  \includegraphics[width=0.32\linewidth]{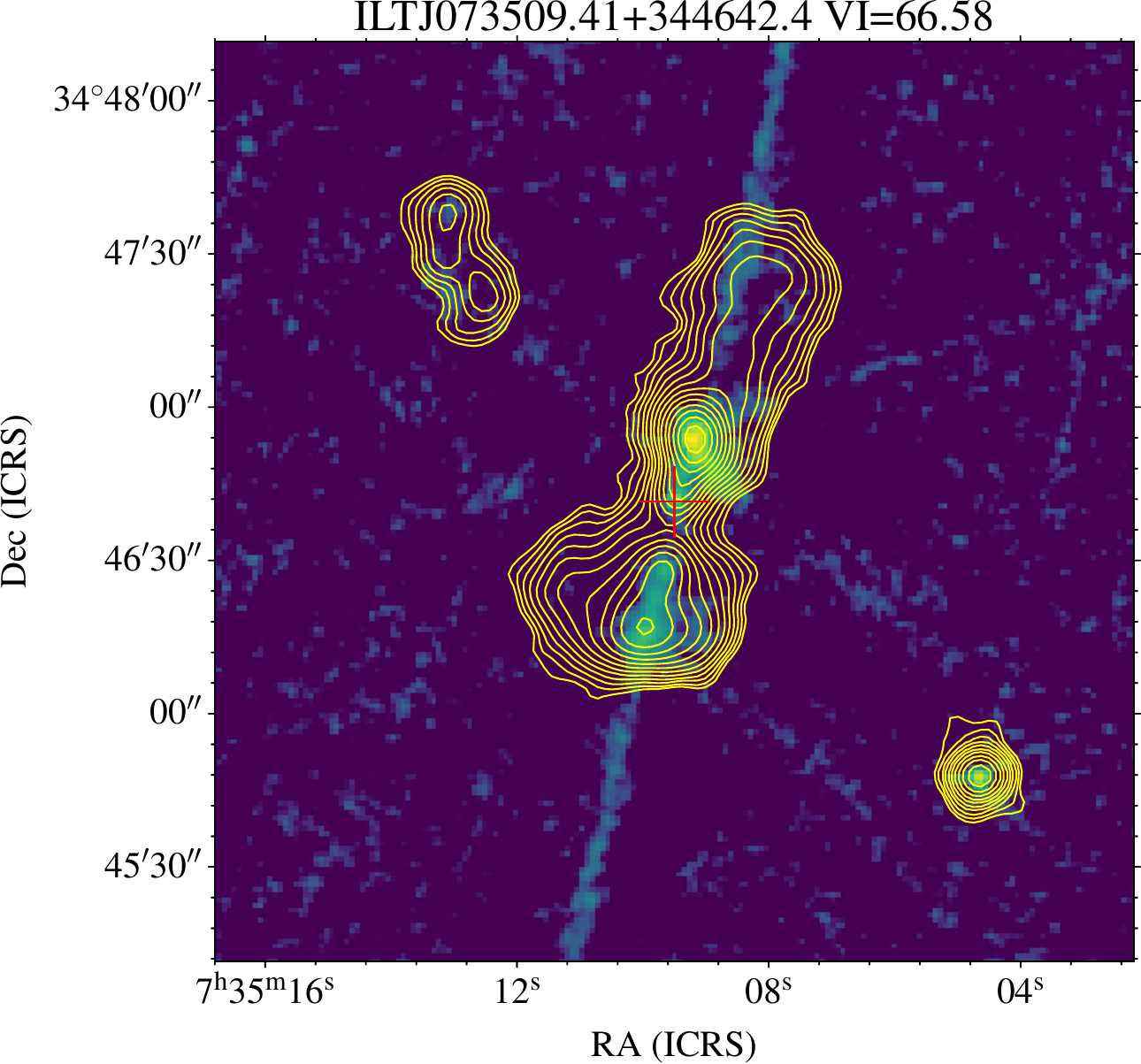}
  \includegraphics[width=0.32\linewidth]{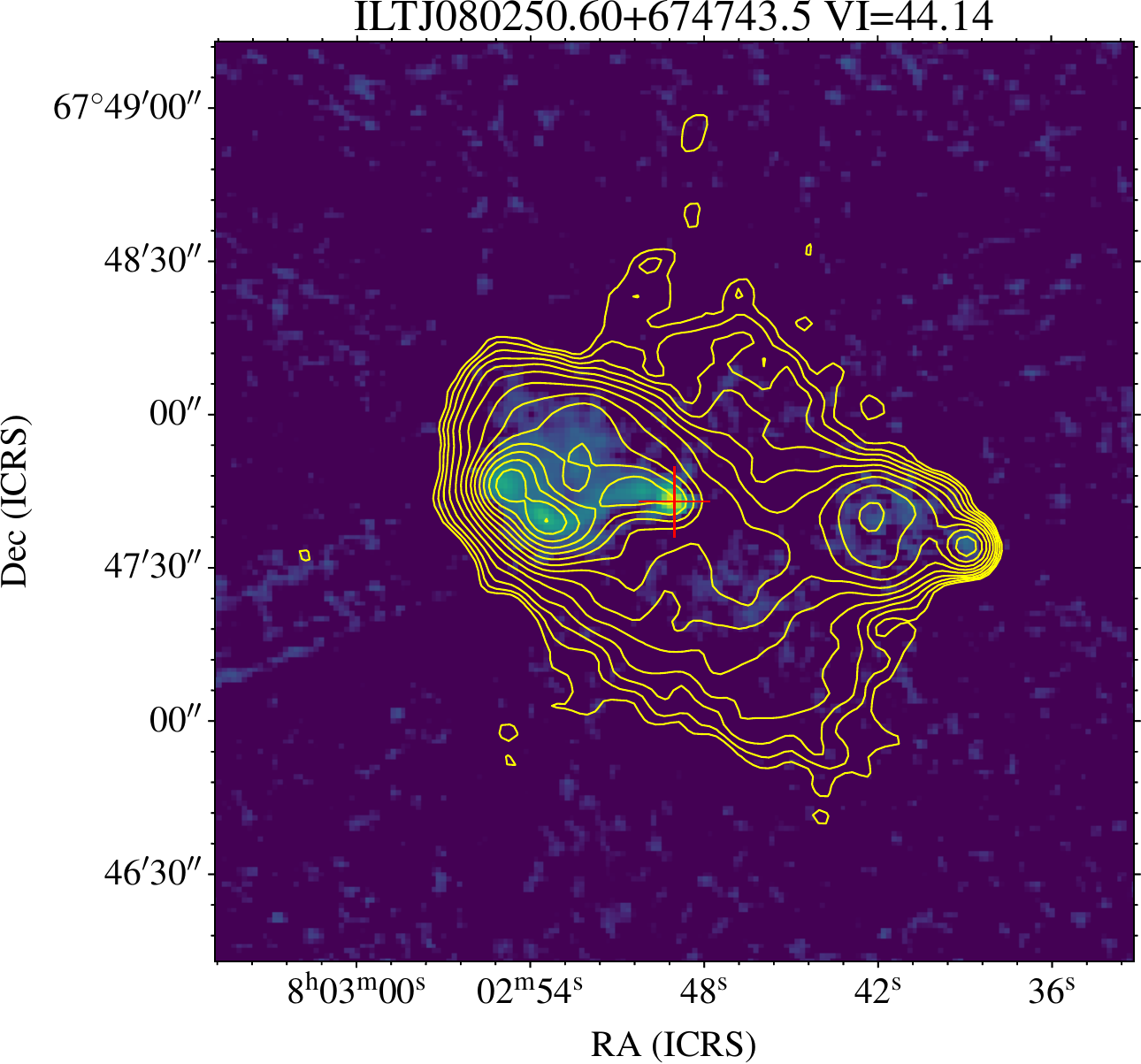}
  \includegraphics[width=0.32\linewidth]{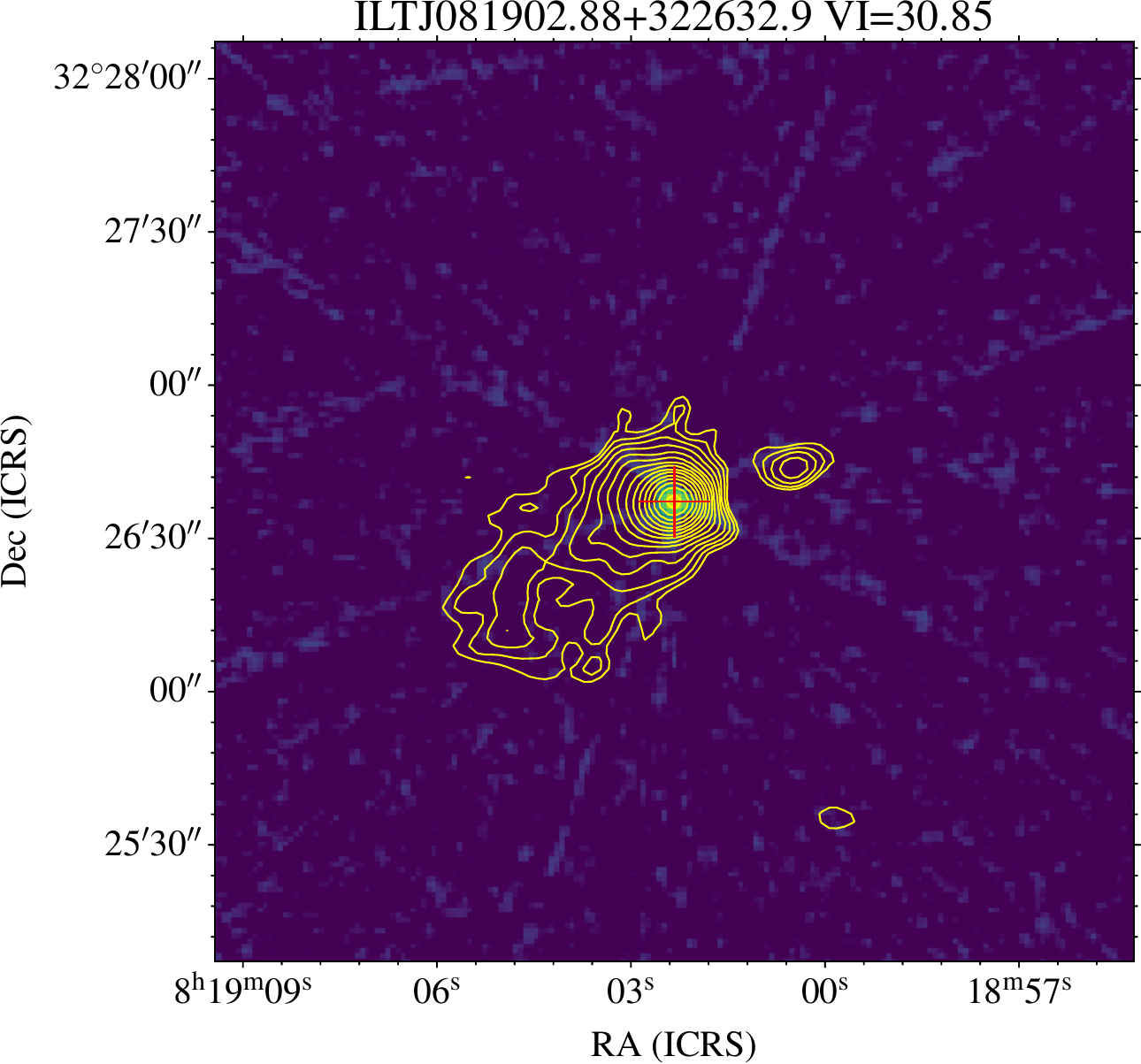}
  \includegraphics[width=0.32\linewidth]{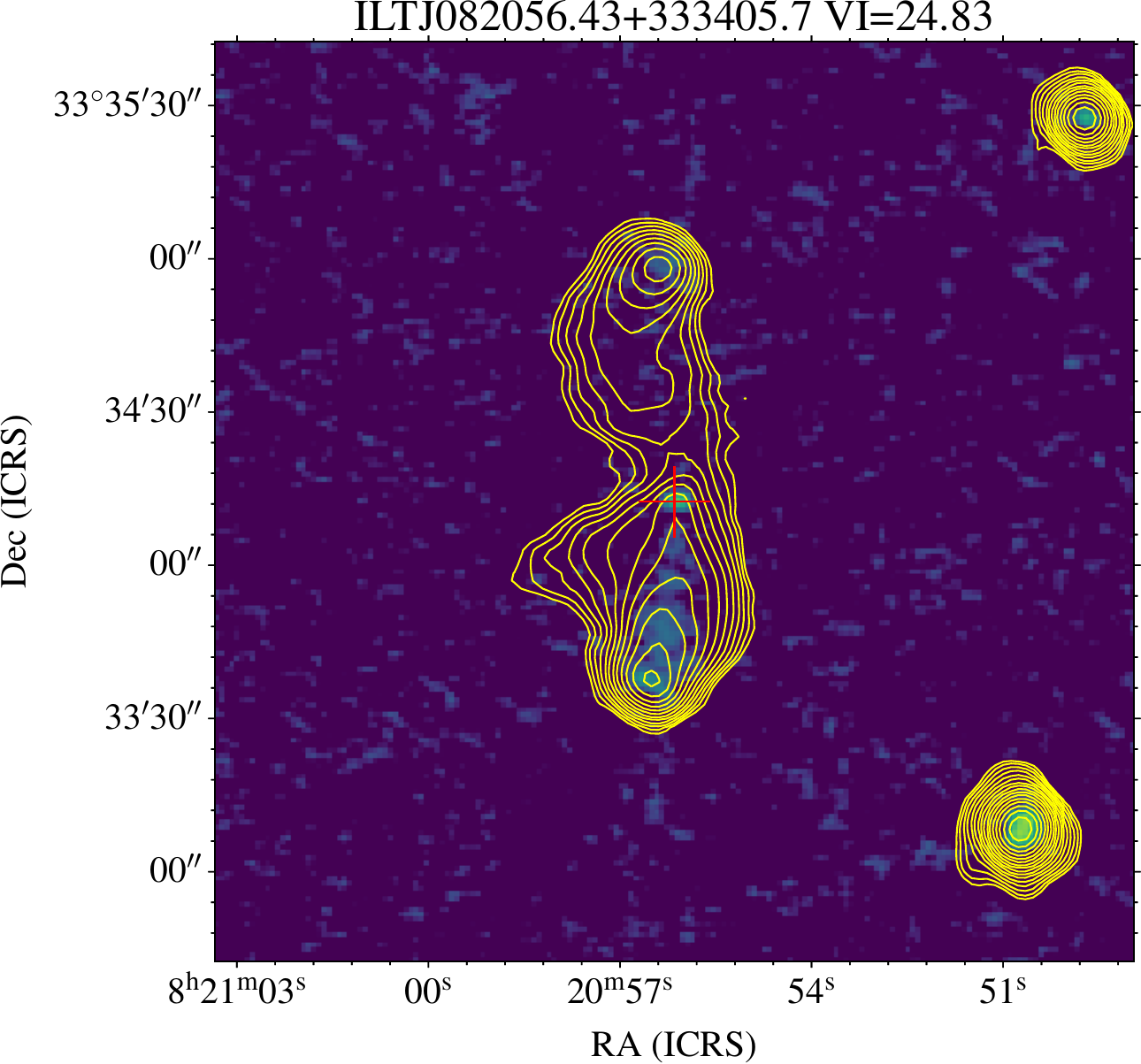}
  \includegraphics[width=0.32\linewidth]{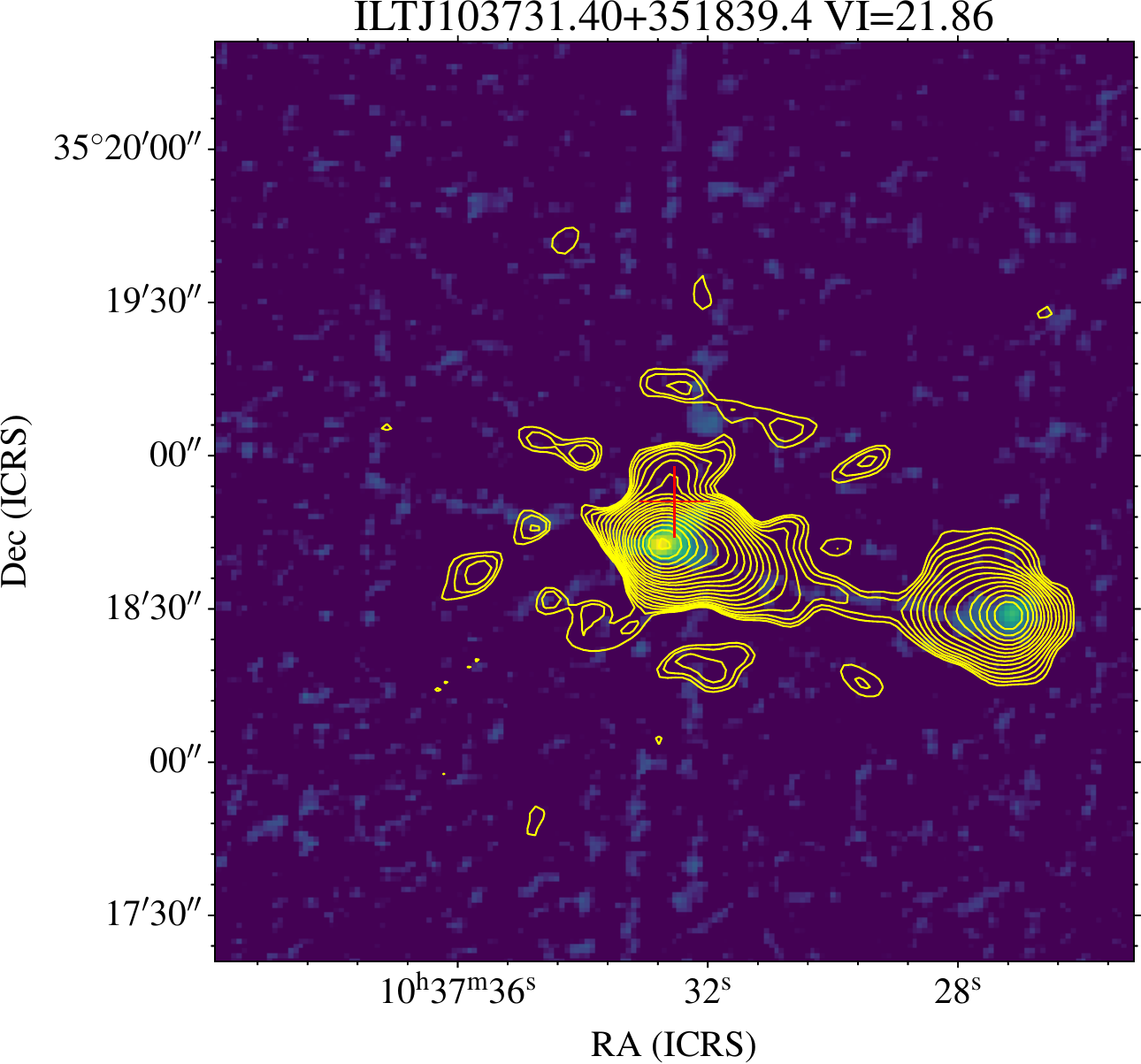}
  \includegraphics[width=0.32\linewidth]{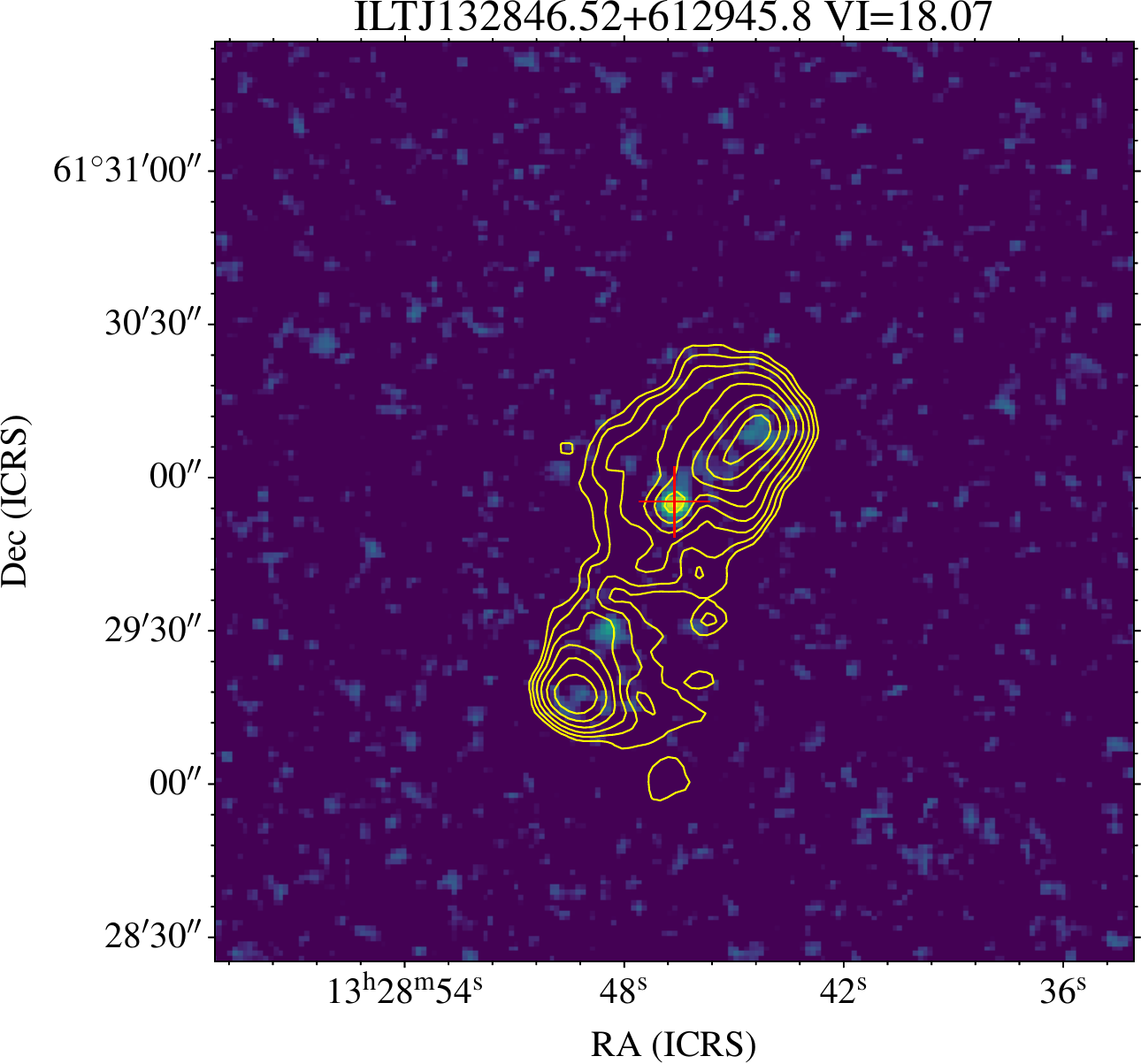}
  \includegraphics[width=0.32\linewidth]{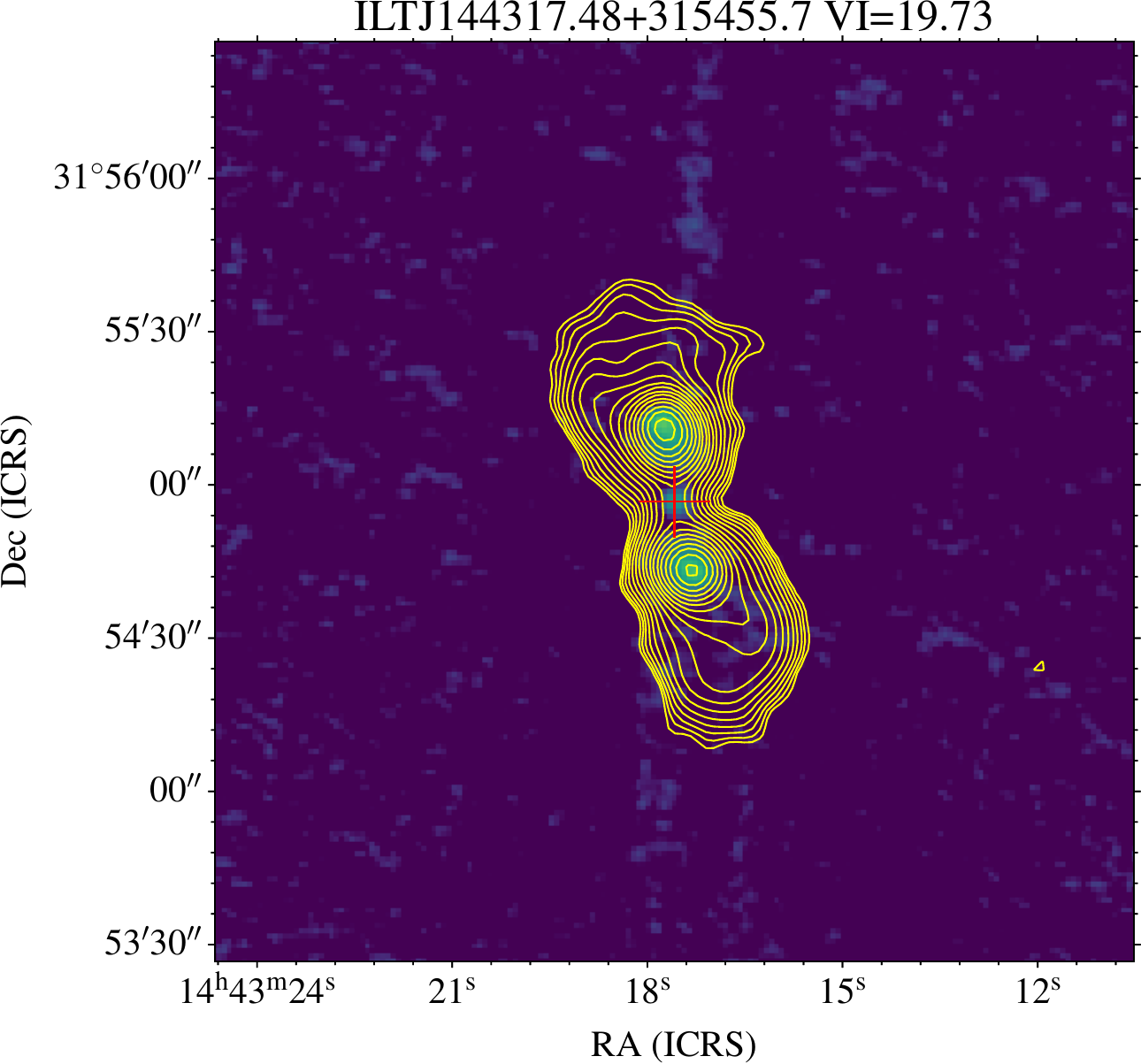}
  \includegraphics[width=0.32\linewidth]{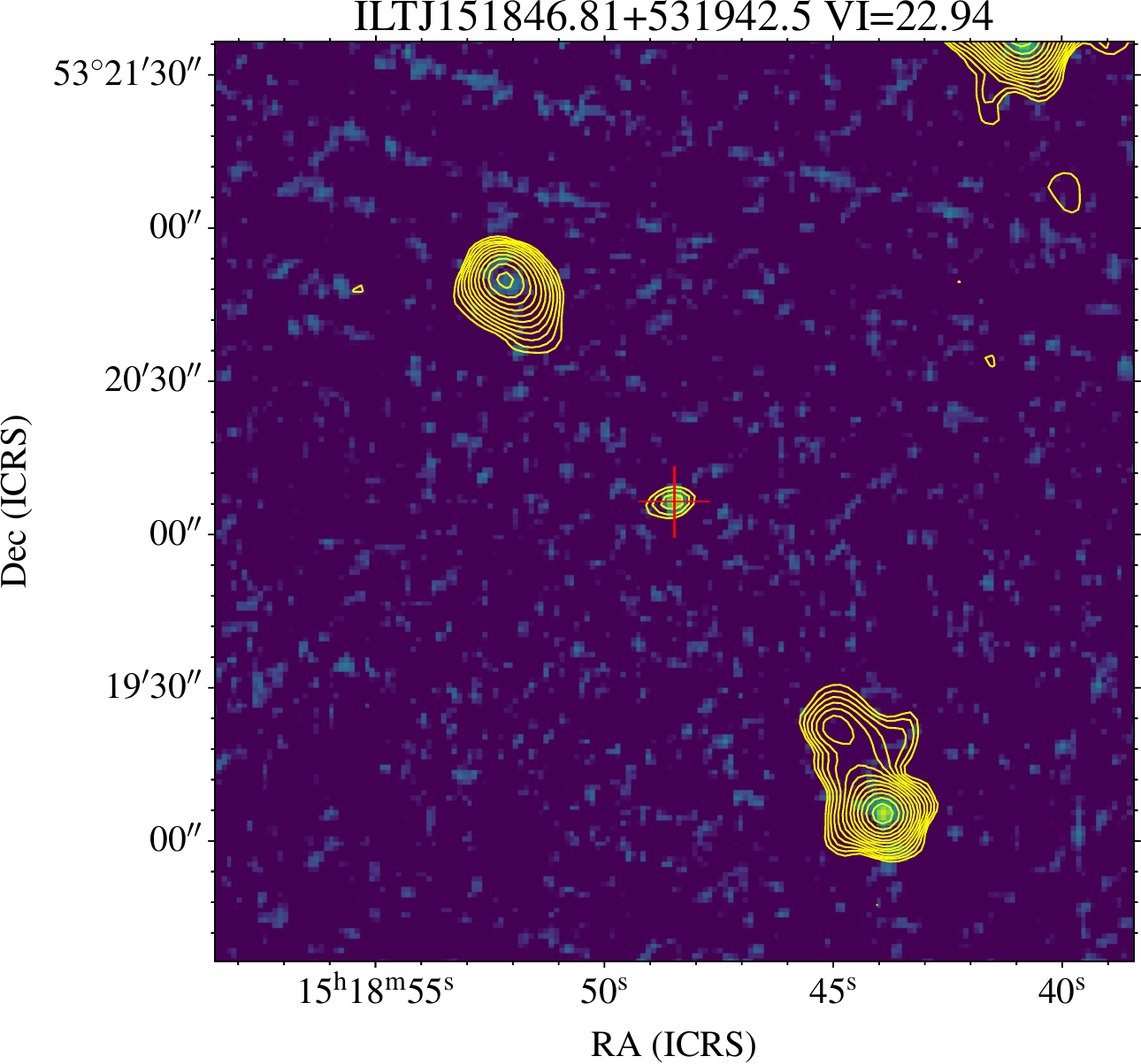}
  \includegraphics[width=0.32\linewidth]{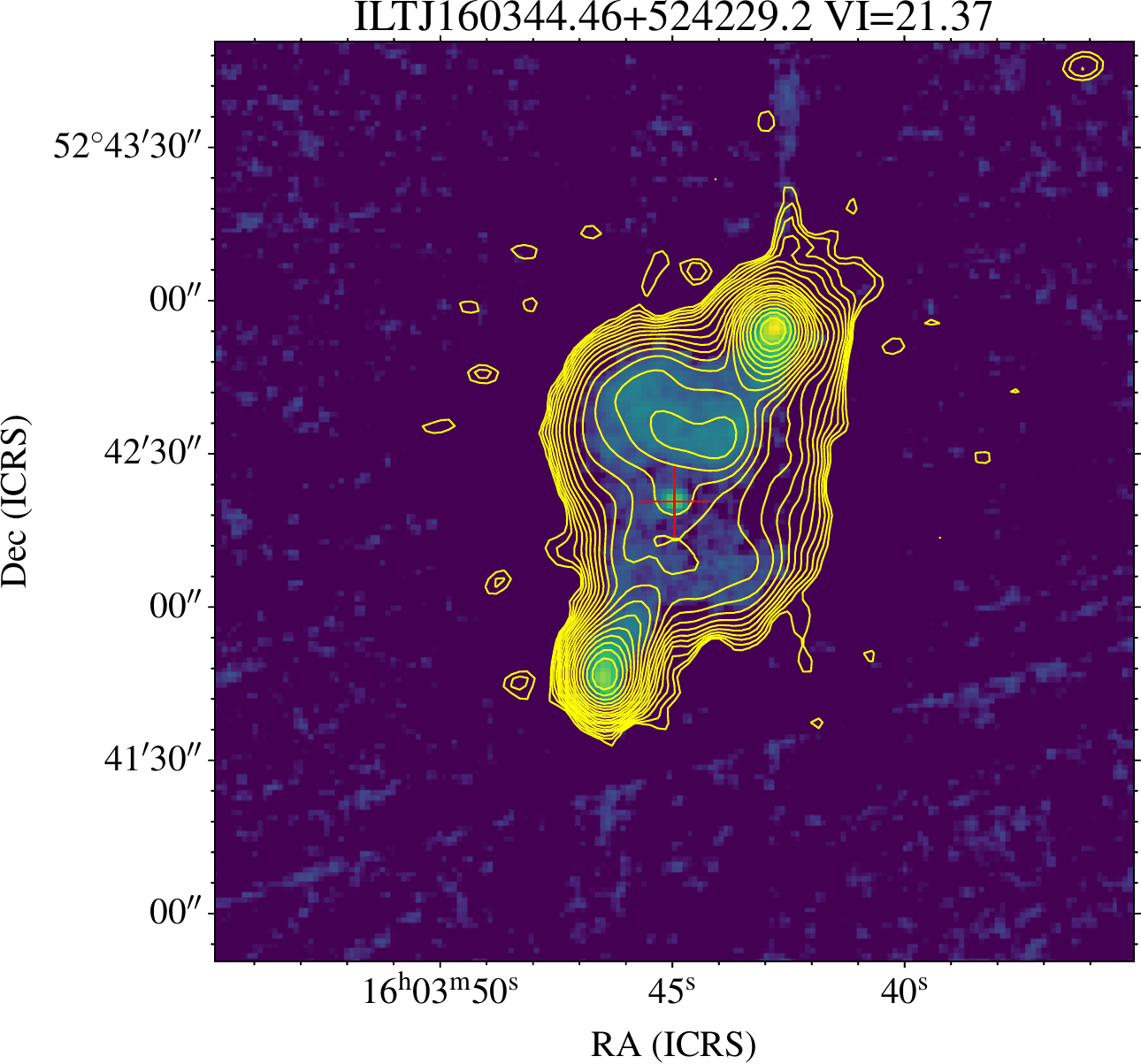}
  \includegraphics[width=0.32\linewidth]{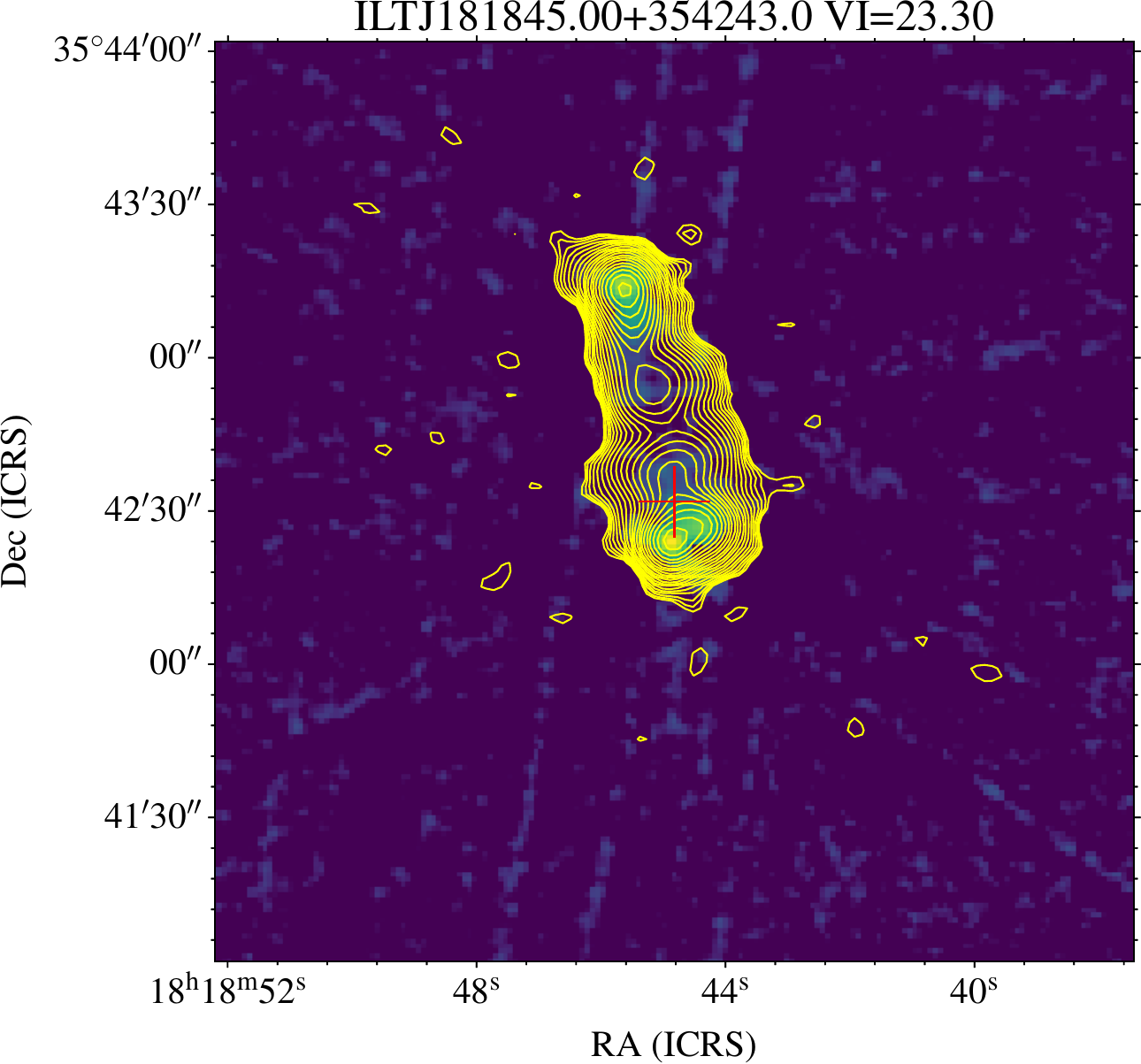}
\caption{The 12 sources with the largest variability indices as described
  in Section \ref{sec:cores}. LOFAR contours (increasing
  logarithmically in steps of 2 from $5\sigma$ or peak flux/1000,
  whichever is the larger) are overlaid on a stacked VLASS colour
  scale. The red cross indicates the optical ID position. Variability
  indices (VI) are stated in the title for each image.}
\label{fig:vi_examples}
\end{figure*}

To investigate the extreme end of the variability of the VLASS
counterparts of optical IDs discussed in Section \ref{sec:cores}, we
made cutouts of sources with the highest variability indices for
visual inspection. Examples of these are shown in
Fig.\ \ref{fig:vi_examples}. Because these are chosen to be extreme,
we would expect some of them to be spurious, and indeed we see one
example of a clearly incorrect ID in the catalogue
(ILTJ103731.40+351839.4, where the alleged optical ID position lies
outside the source and in the sidelobes of a bright hotspot), one with
a surprising though possibly correct ID position and no clear VLASS
core (ILTJ181845.00+354243.0) and one where the core position, though correct, is contaminated by a sidelobe of a bright hotspot
(ILTJ073509.41+344642.4). However, all the others seem to have a clear
VLASS detection at the optical ID position and no reason to doubt the
genuineness of the variability detection. The majority are
morphologically normal FRIIs but it is interesting that two winged
sources and one restarting object are among them. 7/12 of these
objects, and 7/9 of the ones where we can be confident of the
variability detection, are classified as quasars on the basis of their
{\it WISE} colours.

% Don't change these lines
\bsp	% typesetting comment
\label{lastpage}
\end{document}